\documentclass[a4paper,11pt]{article}
\pdfoutput=1 % if your are submitting a pdflatex (i.e. if you have
\usepackage{jheppub} % for details on the use of the package, please
\usepackage[T1]{fontenc} % if needed
\usepackage{bm}% bold math
\usepackage{url}
\DeclareMathAlphabet{\pazocal}{OMS}{zplm}{m}{n}
\usepackage{braket}
\usepackage{amsmath}
\usepackage{amssymb}
\usepackage{leftidx}
\usepackage[colorinlistoftodos]{todonotes}
\usepackage{bm}
\usepackage{slashed}
\usepackage{calrsfs}
\usepackage{cancel}
\usepackage{dirtytalk}
\usepackage{upgreek}
\usepackage{hyperref}
\title{\boldmath New conformal-like symmetry of strictly massless fermions in four-dimensional de Sitter space}

\author[a,b]{Vasileios A. Letsios}

\affiliation[a]{Department of Mathematics, University of York\\Heslington, York, YO10 5DD, UK}
\affiliation[b]{Present address: Department of Mathematics, King’s College London,\\  Strand, London WC2R 2LS, UK}

\emailAdd{vasileios.letsios@kcl.ac.uk}

\abstract{We present new infinitesimal `conformal-like' symmetries for the field equations of strictly massless spin-$s \geq 3/2$ totally symmetric tensor-spinors (i.e. gauge potentials) on 4-dimensional de Sitter spacetime ($dS_{4}$). The corresponding symmetry transformations are generated by the five {closed} conformal Killing vectors of $dS_{4}$, but they are not conventional conformal transformations. We show that the algebra generated by the ten de Sitter (dS) symmetries and the five conformal-like symmetries closes on the conformal-like algebra $so(4,2)$ up to gauge transformations of the gauge potentials. {The transformations of the gauge-invariant field strength tensor-spinors under the conformal-like symmetries are given by the product of $\gamma^{5}$ times a usual infinitesimal conformal transformation of the field strengths}. Furthermore, we demonstrate that the two sets of physical mode solutions, corresponding to the two helicities $\pm s$ of the strictly massless theories, form a direct sum of Unitary Irreducible Representations (UIRs) of the conformal-like algebra. We also fill a gap in the literature by explaining how these physical modes form a direct sum of Discrete Series UIRs of the dS algebra $so(4,1)$.}

\begin{document} 
\maketitle
\flushbottom

\section{Introduction}
 Four-dimensional de Sitter spacetime ($dS_{4}$) is believed to be a good approximation of the very early epoch of our Universe~(Inflation). Also, according to recent data indicating the accelerated expansion of space~\cite{Perlmutter_1999,SloanDigitalSky,PlanckCollab}, there is evidence to suggest that our Universe is asymptotically approaching another de Sitter phase. 
 
 The $D$-dimensional de Sitter spacetime ($dS_{D}$) is the maximally symmetric solution of the vacuum Einstein equations with positive cosmological constant $\Lambda$,
 \begin{equation}
    R_{\mu \nu}-\frac{1}{2} g_{\mu \nu} R + \Lambda  g_{\mu \nu}=0,
\end{equation}
where $g_{\mu \nu}$ is the de Sitter metric tensor, $R_{\mu \nu}$ is the Ricci tensor and $R$ is the Ricci scalar. In this paper, we use units in which $2\Lambda = (D-1)(D-2)$, while the Riemann tensor is
\begin{align}\label{Riemann}
   R_{\mu \nu \rho \sigma} = g_{\mu \rho}\, g_{\nu \sigma} - g_{\nu \rho} \,g_{\mu \sigma}. 
\end{align}

De Sitter (dS) field theories are known to exhibit characteristics with no Minkowskian analogs.
 Two such interesting characteristics of integer-spin fields on $dS_{D}$ - related to the representation theory of the dS algebra $so(D,1)$ - are: 
\begin{itemize}
    \item The existence of unitarily forbidden ranges for the mass parameters of integer-spin fields depending on both $D$ and the spin of the fields~\cite{Higuchiforb, STSHS}.
    \item The existence of exotic unitary ``partially massless'' fields for spin $s \geq 2$~\cite{DESER_NEPOM_1, DESER_NEPOM_2, STSHS, Deser_Waldron_phases, Deser_Waldron_null_propagation, Deser_Waldron_partial_masslessness}. 
\end{itemize}
For the sake of completeness, let us give here some details about these two field-theoretic characteristics. 
As discovered by Higuchi~\cite{Higuchiforb, STSHS, Yale_Thesis}, massive totally symmetric tensor fields of spin $s \geq 1$ (here by ``massive'' we mean theories that do \textbf{not} enjoy a gauge symmetry), satisfying
\begin{align}
    &\Big( \nabla^{\alpha} \nabla_{\alpha}-m^{2} +  (s-2)(s+D-3)-s \Big)h_{\mu_{1}...\mu_{s}} = 0, \nonumber\\
    & \nabla^{\alpha} h_{\alpha \mu_{2}...\mu_{s}}=0, \hspace{4mm} g^{\alpha \beta} h_{\alpha \beta \mu_{3}...\mu_{s}},
\end{align}
are always non-unitary for the following values of the mass parameter: $$m^{2} < (s-1)(s+D-4).$$ Unitary massive theories thus obey the `Higuchi bound'
\begin{align}
    m^{2} > (s-1)(s+D-4).
\end{align}
Moreover, Higuchi observed that for the following special values of the mass parameter~\cite{Higuchiforb, STSHS, Yale_Thesis}:
\begin{align}\label{pm_tuning_bosons}
    m^{2}=(\uptau -1)(2s+D-4-\uptau),\hspace{10mm}(\uptau=1,...,s),
\end{align}
the theory is unitary, while at the same time it enjoys a gauge symmetry. 
A field with mass parameter given by Eq.~(\ref{pm_tuning_bosons}) is    a gauge potential known as partially massless field of depth $\uptau$ in the modern literature~\cite{Deser_Waldron_phases, Deser_Waldron_null_propagation, Deser_Waldron_partial_masslessness}.\footnote{The partially massless spin-2 field was first discovered by Deser and Nepomechie~\cite{DESER_NEPOM_1, DESER_NEPOM_2}.}~The case with $\uptau =1$ corresponds to the theory known as strictly massless. In 4 dimensions, a strictly massless field has two propagating helicity degrees of freedom $\pm s$, while a partially massless field of depth $\uptau$ has $2 \uptau$ of them: $(\pm s, \pm (s-1), ..., \pm (s-\uptau+1))$~\cite{Deser_Waldron_phases, Deser_Waldron_null_propagation, Deser_Waldron_partial_masslessness}. In dS field theory, the strictly massless fields are the closest analogs of Minkowskian massless fields, while partially massless fields of depth $\uptau>1$ have no Minkowsian counterparts. 
{The Unitary Irreducible Representations (UIRs) of the dS algebra $so(D,1)$ corresponding to totally symmetric strictly/partially massless integer-spin fields were first discussed in Higuchi's PhD thesis~\cite{STSHS, Yale_Thesis}. More recent discussions concerning both totally symmetric and mixed symmetry integer-spin fields can be found in Ref.~\cite{Mixed_Symmetry_dS}.}

From a cosmological viewpoint, the UIRs of the dS algebra have been useful for the computation of late time correlators in the cosmological bootstrap program - see, e.g. Ref.~\cite{Bootstrap} and references therein. Also, the physical significance of studying  dS representation theory becomes clear in the ``cosmological collider'' approach, in which dS symmetries (which are slightly broken in a standard slow roll inflation scenario) play a central role in determining the outcome of calculations~\cite{Collider}.

%%%%%%%%%%%%%%%%%%%%%%%%%%%%%%%%%%%%5

$$\textbf{What about the representation theory of fermions on}~\bm{dS_{D}?}$$
Unlike the integer-spin case, the representation-theoretic properties of fermionic fields on $dS_{D}$ are not well-studied.

\noindent \textbf{Recent results.}~Although the phenomena of strict and partial masslessness also occur in the case of spin-$s \geq 3/2$ fermionic fields on $dS_{D}$~\cite{Deser_Waldron_phases, Deser_Waldron_null_propagation, Deser_Waldron_partial_masslessness}, the study of the corresponding unitarity properties was absent from the (mathematical) physics literature for a long time. Interestingly, as the author has recently shown~\cite{Letsios_announce, Letsios_in_progress}, four-dimensional dS space plays a distinguished role in the unitarity of strictly/partially massless (totally symmetric) tensor-spinors on $dS_{D}$ ($D \geq 3$). More specifically, the representations of the dS algebra $so(D,1)$, which correspond to strictly/partially massless totally symmetric tensor-spinors of spin $s \geq 3/2$, are non-unitary unless $D=4$~\cite{Letsios_announce, Letsios_in_progress}.\footnote{\textbf{Note:} For the cases with spin $s=3/2, 5/2$, these results were obtained following two different approaches: a) on the one hand, by performing a technical analysis of the representation-theoretic properties of the mode solutions on global $dS_{D}$~(this includes constructing the mode solutions explicitly, studying their transformation properties under $so(D,1)$, and investigating the existence/non-existence of dS invariant, positive definite scalar products in the space of mode solutions)~\cite{Letsios_in_progress}, and, b) on the other hand, by carefully examining the list of the dS algebra UIRs in the decomposition $so(D,1) \supset so(D)$~\cite{Letsios_announce}~and inferring (non-)unitarity from the (mis-)match between the UIR and the field-theoretic representation labels. For the cases with spin $s \geq  7/2$, the results were obtained in Ref.~\cite{Letsios_announce} motivated only by the examination of the $so(D,1)$ UIRs and the (mis-)match of the representation labels. The technical analysis of the representation-theoretic properties of the mode solutions with spin $s \geq  7/2$ is still absent from the literature for $D \neq 4$ (the $D=4$ case is studied in the present paper). Thus, if we want to be careful - as we should - the results of Ref.~\cite{Letsios_announce} for $s\geq 7/2$ may be viewed as a ``suggestion'' motivated by the examination of the $so(D,1)$ UIRs. This suggestion can be confirmed by studying the representation-theoretic properties of the spin-$s \geq 7/2$ mode solutions on $dS_{D}$, as in the spin-$s=3/2, 5/2$ cases~\cite{Letsios_in_progress}. This is something that we leave for future work.}
 %%%%%%%%%%%%%%%%%%%%%%%%%

In the present paper, we uncover a new group-theoretic feature of all strictly massless totally symmetric spin-$s \geq 3/2$ tensor-spinors on $dS_{4}$: these fermionic gauge potentials possess a conformal-like $so(4,2)$ global symmetry algebra. Moreover, we show that the mode solutions with fixed helicity, i.e. the modes forming Unitary Irreducible Representations (UIRs) of the dS algebra $so(4,1)$~\cite{Letsios_announce}, also form UIRs of the larger conformal-like $so(4,2)$ algebra. 
 
%%%%%%%%%%%%%%%%%%%%%%%%%%%%%%%%%%%%%%%%%%%%%%%%%%%%%%%%%%%%%%%%%%%%%%%%%%%%%%%%%%%%%%%%%%%%%5
\subsection{List of main results and methodology}
Here we give some information about our main results and investigations concerning the new conformal-like symmetries of strictly massless fermions on $dS_{4}$.

\begin{itemize}
    \item We present new conformal-like infinitesimal transformations~(\ref{hidden symmetry}) for strictly massless totally symmetric tensor-spinors on $dS_{4}$. These new transformations are generated by conformal Killing vectors of $dS_{4}$ and they are symmetries of the field equations [Eqs.~(\ref{Dirac_eqn_fermion_dS_sm}) and (\ref{TT_conditions_fermions_dS_sm})], i.e. they preserve the solution space of the field equations.~{In this paper, by conformal Killing vectors we mean the five genuine conformal Killing vectors of $dS_{4}$ with non-vanishing divergence - see Eq.~(\ref{V=nabla phi}).} \\ 
    
    \noindent \textbf{Note.} We call our new symmetries conformal-like, instead of just conformal, because they are not expressed in the conventional form of infinitesimal conformal transformations (i.e. they are not expressed in terms of the Lie-Lorentz derivative~(\ref{Lie_Lorentz}) with respect to conformal Killing vectors plus a conformal weight term). Also, the name ``conformal-like'' will be justified further in Section~\ref{sec_field-strength}, where the conformal-like transformation of the field strengths (i.e. curvatures) of the strictly massless tensor-spinors will be studied (see also the last bullet point in the present list of results). 

    \item The conformal-like transformations~(\ref{hidden symmetry}), together with the ten known dS transformations~(\ref{Lie_Lorentz}), generate an algebra that is isomorphic to $so(4,2)$. However, this conformal-like algebra closes up to field-dependent gauge transformations. 

    \item We fill a gap in the literature by clarifying the way in which the spin-$s\geq3/2$ physical (i.e. non-gauge) mode solutions with fixed helicity form a direct sum of Discrete Series UIRs of the dS algebra $so(4,1)$. The modes with opposite helicity correspond to different UIRs - this is also true in the case of strictly massless totally symmetric tensors~\cite{Yale_Thesis}. (Recall that a strictly massless field has only two propagating helicities $\pm s$~\cite{Deser_Waldron_phases} corresponding to two sets of physical mode solutions with opposite helicities.) 
    
    \item Then, we show that the physical mode solutions also form a direct sum of UIRs of the conformal-like $so(4,2)$ algebra. We arrive at this result by following two basic steps (which stem from the mathematical definitions of representation-theoretic irreducibility and unitarity). First, we show that the mode solutions with fixed helicity transform among themselves under all $so(4,2)$ transformations (this means under the ten dS isometries~(\ref{Lie_Lorentz}), as well as the five conformal-like symmetries~(\ref{hidden symmetry})). Then, we show that there is a $so(4,2)$-invariant, and gauge-invariant, positive definite scalar product for each set of mode solutions with fixed helicity. 

\item As the name suggests, our conformal-like symmetry transformations are \textbf{not} conventional infinitesimal conformal transformations. This is exemplified as follows. For the cases with spin $s=3/2, 5/2$, by investigating the conformal-like transformations of the field strength tensor-spinors~(i.e. curvatures) of the strictly massless fermions\footnote{The field strength tensor(-spinor), also known as ``generalised Weyl tensor(-spinor)''~(see, e.g.~\cite{Conf_Bekaert}), is invariant under gauge transformations. It plays the role of the electromagnetic tensor $F_{\mu \nu}= \partial_{[\mu} A_{\nu]}$ in the case of the $U(1)$ gauge potential $A_{\mu}$ - or, likewise, the role of the linearised Weyl tensor in the case of the spin-2 gauge potential (graviton) in linearised gravity.}, we find that these transformations correspond to the product of two transformations: an infinitesimal axial rotation (i.e. multiplication with $\gamma^{5}$) times an infinitesimal conformal transformation. For the cases with spin $s \geq 7/2$, we present a (justified) conjecture concerning the expressions for the conformal-like transformations of the field strength tensor-spinors.

\noindent  \textbf{Literature review.} We conclude this part of the Introduction with a brief literature review that is relevant to our present work.
The UIRs of $so(4,1)$ corresponding to certain fermions on $dS_{4}$ have been also discussed in Ref.~\cite{Gazeau}.~The mode solutions and the Quantum Field Theory of spin-1/2 fermions on $dS_{D}$ have been discussed in various articles, such as Refs.~\cite{Camporesi, Letsios, Schaub2022, Schaub2023, Muck, Otchik, Barut, Shishkin, Cotaescu, Cotaescu2018, Gazeau, Kanno, Allen_spinors}. The invariance of maximal-depth integer-spin partially massless theories on $dS_{4}$ under conformal transformations has been investigated in Ref.~\cite{Deser_Waldron_Conformal} - however, interestingly, a representation-theoretic study~\cite{Conf_Bekaert} suggests that the associated symmetry algebra does not correspond to the conformal algebra. Recently, I have shed some light on the unconventional conformal symmetry of the maximal-depth partially massless bosons~\cite{Letsios_unconventional} (and its relation to SUSY) on $dS_{4}$. (For recent discussions on scale vs conformal invariance for integer-spin fields on maximally symmetric spacetimes see Ref.~\cite{Hinterbichler_conformal}.) Discrete Series UIRs, which play a central role in the present paper, also exist in the case of the isometry group of $dS_{2}$. Recently, operators furnishing the discrete series UIRs of $so(2,1)$ in BF-type gauge theories on $dS_{2}$ were constructed~\cite{Discrete_charm}.
Quantum aspects of de Sitter space have been reviewed in~\cite{Anninos, Galante}.

 %\item We briefly discuss the massless spin-1/2 field on $dS_{4}$, which is known to be invariant under the familiar 15-dimensional conformal algebra $so(4,2)$. We show that the theory also enjoys fifteen conformal-like symmetries. Each conformal-like symmetry corresponds to the product of an axial rotation times a familiar infinitesimal conformal ($so(4,2)$) transformation. Then, we show that the symmetry algebra generated by the familiar conformal transformations together with the conformal-like transformations is isomorphic to $so(4,2) \bigoplus so(4,2)$. However, this is a chiral symmetry, i.e. $so(4,2)^{+} \bigoplus so(4,2)^{-}$, where $so(4,2)^{\pm}$ acts non-trivially on mode solutions with chirality $\pm 1/2$, and trivially on mode solutions with chirality $\mp 1/2$. In other words, a massless spin-1/2 field with a single, fixed, chirality has only fifteen non-trivial symmetries, corresponding to familiar $so(4,2)$ conformal transformations. But a massless theory with both chiralities enjoys a $so(4,2)^{+} \bigoplus so(4,2)^{-}$ chiral symmetry.
\end{itemize}
%%%%%%%%%%%%%%%%%%%%%%%%%%%%%%%%%%
\subsection{Outline, notation, and conventions}

   The rest of this paper is organised as follows. In Section~\ref{Sec_background}, we review the basics concerning (strictly massless) tensor-spinors on $dS_{4}$. In Section~\ref{Sec_Classification_UIRs D=4}, we review the classification of the UIRs of the dS algebra $so(4,1)$. In Section~\ref{Sec_mode solutions}, we discuss the (pure gauge and physical) mode solutions for strictly massless fermions of spin $s \geq 3/2$ on global $dS_{4}$. In particular, we use the method of separation of variables to express the physical mode solutions on global $dS_{4}$ in terms of tensor-spinor spherical harmonics on $S^{3}$ (these spherical harmonics are not constructed explicitly here). We also identify the analogs of the flat-space positive and negative frequency modes. In Section~\ref{Section dS UIR's and modes}, we discuss the way in which the (positive frequency) physical modes with fixed helicity form a direct sum of Discrete Series UIRs of $so(4,1)$. In Section~\ref{Sec_hidden}, we present our new conformal-like symmetry transformations and we show that the associated symmetry algebra (generated by both dS and conformal-like transformations) closes on $so(4,2)$ up to gauge transformations. In Section~\ref{sect_modes n uir's of so(4,2)}, we show that the physical modes that form a direct sum of $so(4,1)$ UIRs, also form a direct sum of $so(4,2)$ UIRs. In Section~\ref{sec_field-strength}, we discuss the conformal-like transformations of the gauge invariant filed strength tensor-spinors.

   There are two Appendices,~\ref{Appendix matrix elements} and \ref{Appe commutator}, in which we include technical details that were omitted in the main text.

\textbf{Notation and conventions.} We use the mostly plus metric sign convention for $dS_{4}$. Lowercase Greek tensor indices refer to components with respect to the `coordinate basis' on $dS_{4}$. Coordinate basis tensor indices on $S^{3}$ are denoted as $\tilde{\mu}_{1}, \tilde{\mu}_{2},...\,$. Lowercase Latin tensor indices refer to components with respect to the vielbein basis. Repeated indices are summer over. We denote the symmetrisation of indices with the use of round brackets, e.g. $A_{(\mu \nu)} \equiv  (A_{\mu \nu}+A_{\nu \mu})/2$, and the anti-symmetrisation with the use of square brackets, e.g. $A_{[\mu \nu]} \equiv  (A_{\mu \nu}-A_{\nu \mu})/2$.
 Spinor indices are always suppressed throughout this paper. The rank of spin-$s$ tensor-spinors on $dS_{4}$ is $r$ (i.e. $s=r+1/2$). The complex conjugate of the number $z$ is denoted as $z^{*}$. By conformal Killing vector we mean a genuine conformal Killing vector of $dS_{4}$ with non-vanishing divergence - see Eq.~(\ref{V=nabla phi}).

%%%%%%%%%%%%%%%%%%%%%%%%%%%%%%%

%%%%%%%%%%%%%%%%%%%%%%%%%%%%%%%%%%%%%%%%%%%%%%%%%%%%%%%%%%%%%%%%%%%%%%%
 \section{Background material for strictly massless fermions on \texorpdfstring{$dS_{4}$}{dS}} \label{Sec_background}

\subsection{Field equations for higher-spin fermions on \texorpdfstring{$dS_{4}$}{dS}} 
Fermions of spin $s \equiv r+1/2 \ge 3/2$ on $dS_{4}$ can be described by totally symmetric tensor-spinors $\Psi_{\mu_{1}...\mu_{r}}$ that satisfy the on-shell conditions~\cite{Deser_Waldron_null_propagation,Deser_Waldron_ArbitrarySR, Rahman}:
\begin{align}
   &\left( \slashed{\nabla}+M\right)\Psi_{\mu_{1}...\mu_{r}}=0  \label{Dirac_eqn_fermion_dS}\\
   & \nabla^{\alpha}\Psi_{\alpha \mu_{2}...\mu_{r}}=0, \hspace{4mm}  \gamma^{\alpha}\Psi_{\alpha \mu_{2}...\mu_{r}}=0, \label{TT_conditions_fermions_dS}
\end{align}
where $M$ is the mass parameter, $\gamma^{\alpha}$ are the four gamma matrices and $\slashed{\nabla}=\gamma^{\nu}\nabla_{\nu}$ is the Dirac operator. We call the conditions in Eq.~(\ref{TT_conditions_fermions_dS}) the transverse-traceless (TT) conditions.  

The `curved space gamma matrices', $\gamma^{\mu}(x)$, are defined with the use of the vierbein fields as $\gamma^{\mu}(x)= e^{\mu}_{\hspace{3mm} b}(x) \gamma^{b}$, where $\gamma^{b}$ ($b=0,1,2,3$) are the spacetime-independent gamma matrices. The gamma matrices $\gamma^{\mu}(x)$ satisfy the anti-commutation relations
\begin{align}
    \gamma^{\mu} \gamma^{\nu}+\gamma^{\nu} \gamma^{\mu}= 2 g^{\mu \nu }\,\bm{1},
\end{align}
where $\bm{1}$ is the spinorial identity matrix. The vierbein and co-vierbein fields satisfy 
\begin{align}
    e_{\mu}{\hspace{0.2mm}}^{a} \, e_{\nu}{\hspace{0.2mm}}^{b}\eta_{ab}=g_{\mu \nu}, \hspace{4mm}e^{\mu}{\hspace{0.2mm}}_{a}\,e_{\mu}{\hspace{0.2mm}}^{b}=\delta^{b}_{a},
\end{align}
where $\eta_{ab}=diag(-1,1,1,1)$.
The fifth gamma matrix $\gamma^{5}$ is determined as~\cite{Freedman}
\begin{align}\label{def_gamma5}
    \epsilon^{ \mu \nu \rho \sigma } = i \gamma^{5}\gamma^{[\mu}\gamma^{\nu}\gamma^{\rho} \gamma^{\sigma]},
\end{align}
where $\epsilon_{\mu \nu \rho \sigma}$ are the components of the $dS_{4}$ volume element. In the vierbein (i.e. orthonormal frame) basis we have $\epsilon_{0123}=+1$. The matrix $\gamma^{5}$ anti-commutes with the other four gamma matrices, and, hence, with the Dirac operator.

The derivative $\nabla_{\nu}$ acts on our totally symmetric tensor-spinors as
  \begin{align}\label{covariant_deriv_tensor_spinor_anyrank}
      \nabla_{\nu} \Psi_{\mu_{1}...\mu_{r}} =& \left(\partial_{\nu}    + \frac{1}{4} \omega_{\nu bc} \gamma^{b}\gamma^{c}\right)  \Psi_{\mu_{1}...\mu_{r}}-r \,\Gamma^{\lambda}_{\hspace{1mm}\nu (\mu_{1}} \Psi_{ \mu_{2}...\mu_{r}) \lambda} ,
  \end{align}
where $\Gamma^{\lambda}_{\hspace{1mm}\nu \mu}$ are the Christoffel symbols, while $\omega_{\nu b c  }=\omega_{\nu [b c]  } =e_{\nu}{\hspace{0.2mm}}^{a}\omega_{a b c  }$ is the spin connection. We have 
 \begin{equation}
      \partial_{\mu} e^{\rho}\hspace{0.1mm}_{b} + {\Gamma}^{\rho}_{\mu \sigma}e^{\sigma}\hspace{0.1mm}_{b} - \omega_{\mu}\hspace{0.1mm}^{c}\hspace{0.1mm}_{b}  \,e^{\rho}\hspace{0.1mm}_{c}=0.
  \end{equation}
The gamma matrices are covariantly constant, $\nabla_{\nu}\gamma^{\mu} =0$. The commutator of covariant derivatives acting on totally symmetric tensor-spinors on $dS_{4}$ is given by
\begin{align}\label{commutator of derivs}
       [{\nabla}_{\mu}, {\nabla}_{\nu}]{\Psi}_{\mu_{1}... \mu_{r}}&=\frac{1}{2}({\gamma}_{\mu}{\gamma}_{\nu}-{g}_{\mu \nu}){\Psi}_{\mu_{1} ...\mu_{r}}\nonumber\\
       &+r\, \Big({g}_{\mu(\mu_{1}} \Psi_{\mu_{2}...\mu_{r})\nu}   -{g}_{\nu(\mu_{1}} \Psi_{\mu_{2}...\mu_{r})\mu} \Big).
       \end{align}

%%%%%%%%%%%%%%%%%%%%%%%%%%%%%%%%%%%%%%%%%%%%%%%%%%%%%%%%%%%%%%%%%%%%%%%%%%%%%%%%%%%%%%%%%%%
\subsection{Basics about dS symmetries of the field equations}

%%%%%%%%%

 The dS algebra is generated by the ten Killing vectors of $dS_{4}$ satisfying $\nabla_{(   \mu}  \xi_{\nu )}=0$. The dS generators act on solutions $\Psi_{\mu_{1} ... \mu_{r}}$ in terms of the spinorial generalisation of the Lie derivative - also known as the Lie-Lorentz derivative~\cite{Kosmann, Ortin}. The Lie-Lorentz derivative acts on arbitrary tensor-spinors as follows:
\begin{align}\label{Lie_Lorentz}
 \mathbb{L}_{{\xi}}{\Psi}_{\mu_{1}...\mu_{r}}  =~&  \xi^{\nu} \nabla_{\nu} {\Psi}_{ \mu_{1}...\mu_{r}} +\nabla_{\mu_{1}} \xi^{\nu}\,{\Psi}_{ \nu \mu_{2}...\mu_{r}}+\nabla_{\mu_{2}} \xi^{\nu}\,{\Psi}_{ \mu_{1}\nu \mu_{3}...\mu_{r}}+...+\nabla_{\mu_{r}} \xi^{\nu}\,{\Psi}_{ \mu_{1} ...\mu_{r-1} \nu}
 \nonumber\\
 &+ \frac{1}{4}  \nabla_{\kappa} \xi_{\lambda}  \gamma^{\kappa \lambda}    {\Psi}_{\mu_{1}...\mu_{r}}.,
\end{align}
where $\gamma^{\kappa \lambda}  = \gamma^{[\kappa}   \gamma^{\lambda]}$.
The Lie-Lorentz derivative $\mathbb{L}_{{\xi}}{\Psi}_{\mu_{1}...\mu_{r}}$ conveniently describes the infinitesimal $so(4,1)$ transformation of $\Psi_{\mu_{1}...\mu_{r}}$ generated by the Killing vector $\xi^{\mu}$. From the properties~\cite{Ortin}:
\begin{align}\label{properties Lie-Lorentz}
    &\mathbb{L}_{\xi} \gamma^{a}=0, \nonumber\\
    &\mathbb{L}_{\xi} \nabla_{\nu} \Psi_{\mu_{1}...\mu_{r}} =  \nabla_{\nu} \mathbb{L}_{\xi} \Psi_{\mu_{1}...\mu_{r}},
\end{align}
 it follows that if $\Psi_{\mu_{1}...\mu_{r}}$ is a solution of Eqs.~(\ref{Dirac_eqn_fermion_dS}) and (\ref{TT_conditions_fermions_dS}), then so is $\mathbb{L}_{{\xi}}{\Psi}_{\mu_{1}...\mu_{r}}$. In other words, the Lie-Lorentz derivative is a symmetry of the field equations for any value of $M$. It is easy to conclude that the associated symmetry algebra is isomorphic to $so(4,1)$ as~\cite{Ortin}
\begin{align}
    [\mathbb{L}_{\xi },\mathbb{L}_{\xi '}] \Psi_{\mu_{1}...\mu_{r}} = \mathbb{L}_{[\xi , \xi']} \Psi_{\mu_{1}...\mu_{r}}
\end{align} 
for any two Killing vectors $\xi^{\mu}$ and $\xi^{' \mu}$.

The dS algebra, $so(4,1)$, has four non-compact generators (`dS boosts') and six compact ones (`dS rotations'). The compact generators generate the $so(4)$ rotational subalgebra of $so(4,1)$. For any fixed value of $M$, the mode solutions of the field equations~(\ref{Dirac_eqn_fermion_dS}) and (\ref{TT_conditions_fermions_dS}) form an infinite-dimensional representation of $so(4,1)$. The eigenvalue of the quadratic Casimir for this representation is given by~\cite{Letsios_announce}
\begin{align}\label{quadratic_Casimir}
   \mathcal{C}=\sum_{\text{dS boosts}} \mathbb{L}_{\xi}\mathbb{L}_{\xi}-\sum_{\xi \in so(4)} \mathbb{L}_{\xi}\mathbb{L}_{\xi}=&~ -{M}^{2}-\frac{9}{4}+s(s+1) ,
\end{align}
where $s = r+1/2 \geq 3/2$~\footnote{The expression~(\ref{quadratic_Casimir}) for the quadratic Casimir is also true for spin-1/2 fields.}. The unitarity of the representation depends on the value of the mass parameter $M$~\cite{Letsios_announce}. In this paper, we are interested in the strictly massless theories, which appear for special imaginary values of $M$ (see Subsection~\ref{subsection_sm_fermions_background}) - for discussions on arbitrary values of $M$ in any spacetime dimension see Ref.~\cite{Letsios_announce}.
%%%%%%%%%%%%%%
\subsection{Strictly massless fermions on \texorpdfstring{$dS_{4}$}{dS}}\label{subsection_sm_fermions_background}
For real values of $M$, Eqs.~(\ref{Dirac_eqn_fermion_dS}) and (\ref{TT_conditions_fermions_dS}) describe a unitary massive theory with $2s+1$ propagating degrees of freedom~\cite{Deser_Waldron_null_propagation,Deser_Waldron_ArbitrarySR}. The theory enjoys a gauge symmetry for each of the following imaginary tunings of $M$~\cite{Deser_Waldron_null_propagation,Deser_Waldron_ArbitrarySR}:
\begin{align}\label{values_mass_parameter_masslessness_fermion}
    M^{2}=-\left(r-\uptau+1\right)^{2}.
\end{align}
As in the bosonic case discussed in the Introduction, the value $\uptau=1$ corresponds to the strictly massless theory with two propagating helicities. Each of the values $\uptau=2,...,r$ corresponds to a partially massless field with $2 \uptau$ helicities: $(\pm s, \pm (s-1), ..., \pm (s-\uptau+1))$~\cite{Deser_Waldron_null_propagation}. 

In this paper, we are interested in the equations for strictly massless fermions, i.e. Eqs.~(\ref{Dirac_eqn_fermion_dS}) and (\ref{TT_conditions_fermions_dS}) with mass parameter given by~\cite{Deser_Waldron_null_propagation,Deser_Waldron_ArbitrarySR}
\begin{align} \label{sm_mass_parameter}
    M = \pm i\, r.
\end{align}
%%%%%%%%%%%%%%%%%%%%%%%%%%%%%%%%%%%%%%%%%%%%%%%%%%%%%5%%%%%%%%%%%%%%%%%%%%%%%%%%%%%%%%%%%%%%%%%%%%%%%%%%%
Strict masslessness occurs for either of the two signs for the mass parameter in Eq.~(\ref{sm_mass_parameter}). However, the representations of $so(4,1)$ corresponding to the `$+$' sign are equivalent to the representations corresponding to the `$-$' sign~\cite{Letsios_announce}. This is easy to understand as, if $\Psi_{\mu_{1}...\mu_{r}}$ satisfies
$$\slashed{\nabla}\Psi_{\mu_{1} ... \mu_{r}} = -M\, \Psi_{ \mu_{1}...\mu_{r}},$$
then the field $\Psi'_{\mu_{1}...\mu_{r}} \equiv \gamma^{5} \Psi_{\mu_{1} ... \mu_{r}}$ satisfies
$$\slashed{\nabla}\Psi_{\mu_{1} ... \mu_{r}}' = +M\, \Psi_{ \mu_{1}...\mu_{r}}',$$ while, also, $\gamma^{5}$ commutes with all dS transformations~(\ref{Lie_Lorentz})~\cite{Letsios_announce}. 

Based on the discussion of the previous paragraph, below we will only discuss the field with the `$+$' sign in Eq.~(\ref{sm_mass_parameter}). Thus, from now on, $\Psi_{ \mu_{1} ... \mu{r}}$ denotes the strictly massless field satisfying 
\begin{align}
   &\left( \slashed{\nabla}+ ir \right)\Psi_{ \mu_{1}...\mu_{r}}=0 , \label{Dirac_eqn_fermion_dS_sm}\\
   & \nabla^{\alpha}\Psi_{ \alpha \mu_{2}...\mu_{r}}=0, \hspace{4mm}  \gamma^{\alpha}\Psi_{ \alpha \mu_{2}...\mu_{r}}=0.\label{TT_conditions_fermions_dS_sm}
\end{align}
Equations~(\ref{Dirac_eqn_fermion_dS}) and (\ref{TT_conditions_fermions_dS}) are invariant under the following restricted gauge transformations:
\begin{align}\label{onshell_gauge}
    \delta^{res} \Psi_{ \mu_{1}...\mu_{r}} =  \nabla_{(\mu_{1}}\lambda_{ \mu_{2}...\mu_{r})}+ \frac{i}{2} \gamma_{(\mu_{1}}  \lambda_{ \mu_{2}...\mu_{r})},
\end{align}
 where the gauge functions $\lambda_{ \mu_{2}...\mu_{r}}$ are totally symmetric tensor-spinors of rank $r-1$ that satisfy
\begin{align}
   &\left( \slashed{\nabla}+ i(r+1) \right)\lambda_{ \mu_{2}...\mu_{r}}=0 , \label{EOM_for_gauge_functions_Dirac}\\
   & \nabla^{\alpha}\lambda_{ \alpha \mu_{3}...\mu_{r}}=0, \hspace{4mm}  \gamma^{\alpha}\lambda_{ \alpha \mu_{3}...\mu_{r}}=0.\label{EOM_for_gauge_functions_TT}
\end{align}
For $r=0$, i.e. in the case of the massless spin-1/2 field satisfying $\slashed{\nabla} \Psi =0$, the theory does not have gauge symmetry.
%%%%%%%%%%%%%%%%%%%%%%%%%%%%%%%%%%%%%%%%%%%%%%%%%%%%%%%%%%%%%%%%%%%%%%%%%%%%%
\section{Classification of the UIRs of the dS algebra} \label{Sec_Classification_UIRs D=4}
In this Section, we review the classification of the $so(4,1)$ UIRs in the decomposition $so(4,1) \supset$ $so(4)$~\cite{Schwarz, Ottoson}. 
An irreducible representation of $so(4)$ appears at most once in a UIR of $so(4,1)$~\cite{Dixmier}. 

Let us recall that an irreducible representation of $so(4)$ is specified by the highest weight~\cite{barut_group,Dobrev:1977qv}
\begin{equation}\label{define_highest_weight_orthogonal D=4}
    \vec{f}=(f_{1},f_{2}),
\end{equation}
where
\begin{align}
 \label{highest_weight_so(4)}   &f_{1}  \geq |f_{2}|.
\end{align}
The numbers $f_{1}$ and $f_{2}$ are both integers or half-odd integers, while $f_{2}$ can be negative. The representation $(f_{1},-f_{2})$ is the `mirror image' of $(f_{1},f_{2})$~\cite{Todorov_1978}. 

%%%%%%%%%%%%%%%%%%%%%%%%%%%%%%%%%%%%%%%%%%%%%%%%%%%%%%%%%%%%%%%%%%%%%%%%%%%%%%%%%%%%%%%%%%%%%%%%%%%%%%%%%%%%%

\noindent \textbf{UIRs of $\bm{so(4,1)}$.}  A UIR of $so(4,1)$ is specified by two numbers $\vec{F}=(F_{0},F_{1})$. The number $F_{1} \geq 0$ is an integer or half-odd integer. For the $so(4)$ representations $\vec{f}=(f_{1}, f_{2})$ contained in the UIR $\vec{F}=(F_{0},F_{1})$ we have:
\begin{align} \label{branching_rules_spin(2p,1)->spin(2p)}
  f_{1}\geq F_{1}\geq |f_{2}| . 
\end{align}

The UIRs of $so(4,1)$ are listed below~\cite{Schwarz, Ottoson}:
\begin{itemize}
    \item \textbf{Principal Series} $\bm{D}_{\textbf{prin}}\bm{(\,\vec{F}\,)}$\textbf{:}
    \begin{align}\label{Principal_UIR_D=4}
        F_{0}=-\frac{3}{2}+iy, \hspace{5mm} (y>0).
    \end{align}
     $F_{1}$ is an integer or half-odd integer.

    \item \textbf{Complementary Series} $\bm{D}_{\textbf{comp}}\bm{(\,\vec{F}\,):}$ 
    \begin{align}\label{Compl_UIR_D=4}
     -\frac{3}{2}\leq F_{0}<-\tilde{n}, \hspace{5mm}\tilde{n} \in\{0,1\}.   
    \end{align}
     If $ \tilde{n}=0$, then $F_{1}=0$, while for the $so(4)$ content we have $f_{2}=0$. If $\tilde{n} = 1$, then $F_{1}$ is a positive integer.
    
     \item \textbf{Exceptional Series} $\bm{D}_{\textbf{ex}}\bm{(\,\vec{F}\,):}$ 
     \begin{align}\label{Exceptional_UIR_D=4}
        F_{0}=-1. 
        \end{align}
    $F_{1}$ is a positive integer, while $f_{2}=0$.

 \item \textbf{Discrete Series} $\bm{D}^{\bm{\pm}}\bm{(\,\vec{F}\,):}$ 
$F_{0}$ is real, while $F_{0}$ and $F_{1}$ are both integers or half-odd integers. The following conditions have to be satisfied:
\begin{align}\label{condition_for_discrete_series_+ D=4}
   F_{1} \geq f_{2}  \geq F_{0}+2 \geq \frac{1}{2}\hspace{9mm}\text{for}~D^{+}(\,\vec{F}\,),
\end{align}
\begin{align}\label{condition_for_discrete_series_- D=4}
  -F_{1} \leq f_{2} \leq -(F_{0}+2 )\leq -\frac{1}{2}\hspace{6mm}\text{for}~D^{-}(\,\vec{F}\,).
  \end{align}
\end{itemize}

For any $so(4,1)$ UIR, $\vec{F}=(F_{0},F_{1})$, the quadratic Casimir, $C_{2}(\vec{F})$, is expressed as:
\begin{align}\label{Casimir_Spin(4,1)}
    C_{2}(\vec{F})=F_{0}\,(F_{0}+3)+F_{1}(F_{1}+1).
\end{align}

%%%%%%%%%%%%%%%%%%%%%%%%%%%%%%%%%%%%%%%%%%%%%%%%%%%%%%%%%%%%%%%%%%
\section{Mode solutions of strictly massless spin-\texorpdfstring{$(r+1/2) \geq 3/2$}{r+1/2} fermions}\label{Sec_mode solutions}
In this Section, we obtain the mode solutions of the spin-$(r+1/2)\geq 3/2$ strictly massless theories~[(\ref{Dirac_eqn_fermion_dS_sm}) and (\ref{TT_conditions_fermions_dS_sm})]. The spin-3/2 and spin-5/2 mode solutions (for arbitrary spacetime dimensions) have been already studied in Refs.~\cite{Letsios_announce, Letsios_in_progress}.

%%%%%%%%%%%%%%%%%%%%%%%%%%%%%%%%%%%%%%%%%%%%%%%%%%%%%
\subsection{Global coordinates and representation of gamma matrices}\label{subsection global coords}
In order to obtain the mode solutions of Eqs.~(\ref{Dirac_eqn_fermion_dS_sm}) and (\ref{TT_conditions_fermions_dS_sm}) we will work with the global coordinates of $dS_{4}$, where the line element is
\begin{equation} \label{dS_metric}
    ds^{2}=-dt^{2}+\cosh^{2}{t} \,d\Omega^{2}.
\end{equation}
The line element of $S^{3}$, $d\Omega^{2}$, can be parameterised as
\begin{align}\label{S^3_metric}
    d\Omega^{2} = d\theta^{2}_{3} + \sin^{2}\theta_{3} \left( d\theta_{2}^{2} + \sin^{2}{\theta_{2}} \,d\theta_{1}^{2}  \right),
\end{align}
where $ 0 \leq \theta_{j} \leq \pi$ (for $j=2,3$) and $0 \leq \theta_{1} \leq 2 \pi$. We will also use the following notation for a point on $S^{3}$: $\bm{\theta_{3}} \equiv (\theta_{3}, \theta_{2},\theta_{1})$. 

In global coordinates, the non-zero Christoffel symbols are 
\begin{align}\label{Christoffels_dS}
    &\Gamma^{t}_{\hspace{0.2mm}\theta_{i} \theta_{j}}=\cosh{t} \sinh{t} \hspace{1mm}\tilde{g}_{\theta_{i} \theta_{j}}, \hspace{2mm} \Gamma^{\theta_{i}}_{\hspace{0.2mm}\theta_{j} t} =\tanh{t}  \hspace{1mm}\tilde{g}^{\theta_{i}}_{\theta_{j}}, \nonumber \\ 
& \Gamma^{\theta_{k}}_{\hspace{0.2mm}\theta_{i} \theta_{j}}=\tilde{\Gamma}^{\theta_{k}}_{\hspace{0.2mm}\theta_{i} \theta_{j}},
\end{align}
where $\tilde{g}_{\theta_{i} \theta_{j} }$ and $\tilde{\Gamma}^{\theta_{k}}_{\hspace{0.2mm}\theta_{i} \theta_{j}}$ are the metric tensor and the Christoffel symbols, respectively, on $S^{3}$. The vierbein fields on $dS_{4}$ can be chosen to be:
\begin{equation}\label{vielbeins}
    e^{t}{\hspace{0.2mm}}_{0}=1, \hspace{5mm}  e^{\theta_{i}}{\hspace{0.2mm}}_{i}=\frac{1}{\cosh{t}} \tilde{e}^{\theta_{i}}{\hspace{0.2mm}}_{i} , \hspace{7mm}i=1,2,3,
\end{equation}
 where $\tilde{e}^{\theta_{i}}{\hspace{0.2mm}}_{i}$ are the dreibein fields on $S^{3}$. The non-zero components of the dS spin connection are given by
\begin{equation}\label{spin_connection_dS}
   \omega_{ijk} = \frac{  \tilde{\omega}_{ijk}}{ \cosh{t}  } , \hspace{6mm}  \omega_{i0k}= -\omega_{ik0}=  -\tanh{t}\hspace{1mm} \delta_{ik},\hspace{5mm}i,j,k \in \{  1,2,3\}
        \end{equation} 
where $\tilde{\omega}_{ijk}$ is the spin connection on $S^{3}$.
%%%%%%%%%%%%%%%%%%%%%%%%%%%%%%%%%%%%%%%

We will work with the following representation of gamma matrices on $dS_{4}$:
\begin{equation}\label{even_gammas D=4}
 \gamma^{0}=i \begin{pmatrix}  
   0 & \bm{1} \\
   \bm{1} & 0
    \end{pmatrix} , \hspace{5mm}
    \gamma^{j}=\begin{pmatrix}  
   0 & i\widetilde{\gamma}^{j} \\
   -i\widetilde{ \gamma}^{j} & 0
    \end{pmatrix} ,
     \end{equation} 
($ j=1,2,3$) where the lower-dimensional gamma matrices, $\widetilde{\gamma}^{j}$, satisfy 
\begin{equation} \label{Euclidean_Clifford_relns D=3}
    \{ \widetilde{\gamma}^{j}, \widetilde{\gamma}^{k}\} = 2 \delta^{jk} \bm{1}, \hspace{7mm}j,k=1,2,3.
\end{equation}
 The fifth gamma matrix~(\ref{def_gamma5}) is given by
\begin{align}\label{gamma(5)}
    \gamma^{5}=\begin{pmatrix}
    \bm{1} & 0\\
    0      & -\bm{1}
    \end{pmatrix}.
\end{align}
%%%%%%%%%%%%%%%%%%%%%%%%%%%%%%%%%%%%%%%%%%%%%%%%%%%%%%%%%%%%%%%5
\subsection{Constructing the mode solutions of the strictly massless theories} \label{subsec_constructing modes}
There are two kinds of spin-$(r+1/2)$ TT mode solutions satisfying the strictly massless field equations~[(\ref{Dirac_eqn_fermion_dS_sm}) and~(\ref{TT_conditions_fermions_dS_sm})] on $dS_{4}$:
\begin{itemize}
    \item The \textbf{`physical modes'} describing the propagating degrees of freedom of the theory.

    \item The \textbf{`pure gauge modes'} describing the gauge degrees of freedom of the theory.
\end{itemize}
 In this Subsection, we present some details for the construction of these mode solutions. 

The mode solutions on global $dS_{4}$ can be constructed using the method of separation variables. Schematically, this means that we are looking for solutions that can be expressed as a product ``function of $t$ $\times$ function of $\bm{\theta_{3}}$''. As we will see below, the functions describing the $\bm{\theta_{3}}$-dependence are tensor-spinor spherical harmonics on $S^{3}$ forming UIRs of $so(4)$. Thus, from a representation-theoretic viewpoint, the solutions on global $dS_{4}$ obtained with the method of separation of variables form $so(4,1)$ representations in the decomposition $so(4,1)  \supset so(4)$. The method of separation of variables has been applied in Refs.~\cite{STSHS, HiguchiLinearised, Yale_Thesis} for integer-spin fields, in Refs.~\cite{Camporesi, Letsios} for spin-1/2 fields and in Refs.~\cite{Letsios_announce, Letsios_in_progress} for spin-3/2 and spin-5/2 fields.

%%%%%%%%%%%%%%%%%%%%%%%%%%%%%%%%%%%%%%%%%%%%%%%%%%%%
\subsubsection{Physical spin-\texorpdfstring{$(r+1/2) \geq 3/2$}{r+/2} modes on \texorpdfstring{$dS_{4}$}{dS}} \label{SubSub_physical}
 Let us start by obtaining the physical mode solutions of Eqs.~(\ref{Dirac_eqn_fermion_dS_sm}) and (\ref{TT_conditions_fermions_dS_sm}). We first discuss the spherical eigenmodes on $S^{3}$ that describe the spatial dependence of physical modes. Then, we discuss the time dependence of physical modes and we apply the method of separation of variables.

 $$ \textbf{Spatial dependence and}~\bm{so(4)}~\textbf{content of physical modes}  $$
 The spatial dependence of the spin-$(r+1/2)$ physical mode solutions on $dS_{4}$ is expressed in terms of (totally symmetric) tensor-spinor spherical harmonics of rank $r$ on $S^{3}$.~The latter are the totally symmetric TT tensor-spinor eigenmodes of the Dirac operator on $S^{3}$ satisfying~\cite{Homma, Camporesi} 
\begin{align}
   & \tilde{\slashed{\nabla}}\tilde{\psi}^{(\ell; {m};k)}_{+ \tilde{\mu}_{1} \tilde{\mu}_{2} ... \tilde{\mu}_{{r}}}(\bm{\theta_{3}})= + i \left(\ell+\frac{3}{2}\right)  \tilde{\psi}^{(\ell; {m};k)}_{+ \tilde{\mu}_{1} \tilde{\mu}_{2} ... \tilde{\mu}_{r}}(\bm{\theta_{3}}),\hspace{5mm}(\ell=r,r+1,...)\nonumber \\
   &\tilde{\gamma}^{\tilde{\mu}_{1}}\tilde{\psi}^{(\ell; {m};k)}_{+ \tilde{\mu}_{1} \tilde{\mu}_{2} ... \tilde{\mu}_{{r}}}(\bm{\theta_{3}})=\tilde{\nabla}^{\tilde{\mu}_{1}}\tilde{\psi}^{(\ell; {m};k)}_{+ \tilde{\mu}_{1} \tilde{\mu}_{2} ... \tilde{\mu}_{{r}}}(\bm{\theta_{3}})=0 \label{tensor-spinor+eigen_S3},
\end{align}
and
\begin{align}
   & \tilde{\slashed{\nabla}}\tilde{\psi}^{(\ell; {m};k)}_{- \tilde{\mu}_{1} \tilde{\mu}_{2} ... \tilde{\mu}_{{r}}}(\bm{\theta_{3}})= - i \left(\ell+\frac{3}{2}\right)  \tilde{\psi}^{(\ell; {m};k)}_{- \tilde{\mu}_{1} \tilde{\mu}_{2} ... \tilde{\mu}_{{r}}}(\bm{\theta_{3}}),\hspace{5mm}(\ell=r,r+1,...)\nonumber \\
   &\tilde{\gamma}^{\tilde{\mu}_{1}}\tilde{\psi}^{(\ell; {m};k)}_{- \tilde{\mu}_{1} \tilde{\mu}_{2} ... \tilde{\mu}_{{r}}}(\bm{\theta_{3}})=\tilde{\nabla}^{\tilde{\mu}_{1}}\tilde{\psi}^{(\ell; {m};k)}_{-\tilde{\mu}_{1} \tilde{\mu}_{2} ... \tilde{\mu}_{{r}}}(\bm{\theta_{3}})=0 \label{tensor-spinor--eigen_S3}.
\end{align}
The subscripts `$\pm$' in $\tilde{\psi}^{(\ell; {m};k)}_{\pm \tilde{\mu}_{1} \tilde{\mu}_{2} ... \tilde{\mu}_{{r}}}$ have been used in order to indicate the sign of the eigenvalue in Eqs.~(\ref{tensor-spinor+eigen_S3}) and (\ref{tensor-spinor--eigen_S3}), while $\tilde{\gamma}_{\tilde{\mu}}, \tilde{\nabla}_{\tilde{\mu}}$ and $\tilde{\slashed{\nabla}} = \tilde{\gamma}^{\tilde{\mu}}\tilde{\nabla}_{\tilde{\mu}}$ are the gamma matrices, covariant derivative and Dirac operator, respectively, on $S^{3}$. The numbers $\ell, m$, and $k$ are representation-theoretic labels corresponding to the chain of subalgebras $so(4) \supset so(3) \supset so(2)$. In particular, the number $\ell = r, r+1,...$ is the angular momentum quantum number on $S^{3}$. The numbers ${m}$ and $k$ are the angular momentum quantum numbers on $S^{2}$ and $S^{1}$, respectively, and they are allowed to take the values: $m=r,r+1,...,\ell$ and $k=-(m+1),-m,...,0,...,m$. The explicit form of the tensor-spinors $\tilde{\psi}^{(\ell; {m};k)}_{\pm \tilde{\mu}_{1} \tilde{\mu}_{2} ... \tilde{\mu}_{{r}}}(\bm{\theta_{3}})$ is not needed for the purposes of this paper\footnote{See Refs.~\cite{Trautman_1993, Camporesi} for explicit expressions for the spinor eigenfunctions of the Dirac operator on spheres and Refs.~\cite{CHH, Letsios_in_progress} for the vector-spinor and symmetric rank-2 tensor-spinor cases. The general representation-theoretic properties of tensor-spinor spherical harmonics of arbitrary rank have been discussed in Ref.~\cite{Homma}.}.

The set of eigenmodes $\{ \tilde{\psi}^{(\ell; {m};k)}_{+\tilde{\mu}_{1} \tilde{\mu}_{2} ... \tilde{\mu}_{{r}}} \}$ (with fixed $\ell$) forms a $so(4)$ representation with highest weight~(\ref{highest_weight_so(4)}) given by~\cite{Homma}:
\begin{align}\label{so(4) +weight TT spin-r+1/2}
   \vec{f}^{+}_{r} = \left( \ell + \frac{1}{2}, r +\frac{1}{2} \right), \hspace{7mm} \ell=r,r+1,... \,.
\end{align}
Similarly, the set  $\{ \tilde{\psi}^{(\ell; {m};k)}_{-\tilde{\mu}_{1} \tilde{\mu}_{2} ... \tilde{\mu}_{{r}}} \}$ (with fixed $\ell$) forms a $so(4)$ representation with highest weight~\cite{Homma}:
\begin{align}\label{so(4) -weight TT spin-r+1/2}
   \vec{f}^{-}_{r} = \left( \ell + \frac{1}{2}, -r -\frac{1}{2} \right),\hspace{7mm} \ell=r,r+1,... \, .
\end{align}
Each of the $so(4)$ UIRs $\vec{f}^{\pm}_{r} = \left( \ell + \frac{1}{2}, \pm(r +\frac{1}{2}) \right)$ has the following content concerning its subalgebras: the $so(3)$ content corresponds to the $so(3)$ highest weight $m+\tfrac{1}{2}$ with $\ell+\tfrac{1}{2} \geq m+\tfrac{1}{2} \geq r+\tfrac{1}{2}$, while the $so(2)$ content corresponds to the $so(2)$ highest weight $k+1/2$ with $m+\tfrac{1}{2} \geq k+\tfrac{1}{2} \geq -m-\tfrac{1}{2}$.

In this paper we assume that the eigenmodes in Eqs.~(\ref{tensor-spinor+eigen_S3}) and~(\ref{tensor-spinor--eigen_S3}) are already normalised using the standard inner product on $S^{3}$:
\begin{align}\label{normlzn_S3}
    \int_{S^{3}} \sqrt{\tilde{g}}&\,d\bm{\theta_{3}} ~\tilde{g}^{\tilde{\mu}_{1}    \tilde{\nu}_{1}}\,\tilde{g}^{\tilde{\mu}_{2}    \tilde{\nu}_{2}}...\tilde{g}^{\tilde{\mu}_{r}    \tilde{\nu}_{r}}~\tilde{\psi}^{(\ell'; {m}'; k')}_{\sigma '\,\tilde{\mu}_{1} \tilde{\mu}_{2} ... \tilde{\mu}_{r}}(\bm{\theta_{3}})^{\dagger}
    ~\tilde{\psi}^{(\ell; {m};k)}_{\sigma \,\tilde{\nu}_{1} \tilde{\nu}_{2} ... \tilde{\nu}_{r}}(\bm{\theta_{3}}) \nonumber\\
    &= \delta_{\sigma \sigma'}\, \delta_{\ell \ell'} \,\delta_{{m}\,   {m}'} \delta_{k k'},
\end{align}
where $\sigma , \sigma' \in \{ +, -  \}$ and $d \bm{\theta_{3}} \equiv d{\theta_{3}}  d\theta_{2}   d\theta_{1}$, while $\tilde{g}$ is the determinant of the metric on $S^{3}$.\footnote{In Eq.~(\ref{normlzn_S3}), the eigenmodes with different values for $\sigma = \pm$ and/or $\ell = r,r+1,... $ are orthogonal to each other because they belong to different $so(4)$ representations. Similarly, eigenmodes with different values of ${m}$ and/or $k$ are orthogonal to each other because, in the decomposition $so(4)  \supset so(3) \supset so(2)$, they correspond to representations with different content concerning the chain of subalgebras $so(3) \supset so(2)$.}
%%%%%%%%%%%%%%%%%%%%%%%%%%%%%%%%

$$ \textbf{Time dependence of physical modes}  $$
The physical modes $\Psi_{\mu_{1} ... \mu_{r}} (t, \bm{\theta_{3}})$ on $dS_{4}$ are essentially TT tensor-spinors on $S^{3}$, and, thus, we have $\Psi_{t\mu_{2}...\mu_{r}} =0$, where $\mu_{2},\mu_{3},...,\mu_{r} \in \{ t,\theta_{3}, \theta_{2}, \theta_{1} \} $ - as will become clear, the TT conditions~(\ref{TT_conditions_fermions_dS_sm}) will be automatically satisfied by construction. The only non-zero components of the physical modes are the spatial components $\Psi_{\tilde{\mu}_{1}...\tilde{\mu}_{r}}$, where $\tilde{\mu_{1}},\tilde{\mu}_{2},...,\tilde{\mu}_{r} \in \{ \theta_{3}, \theta_{2}, \theta_{1} \}$. These can be determined by solving the Dirac equation~(\ref{Dirac_eqn_fermion_dS_sm}). To be specific, letting $\mu_{1}= \tilde{\mu}_{1}$, $\mu_{2} = \tilde{\mu}_{2},...,\mu_{r} = \tilde{\mu}_{r}$,~the Dirac equation~(\ref{Dirac_eqn_fermion_dS_sm}) for the physical modes is expressed as
 \begin{align}
    & \left( \frac{\partial}{\partial t}+\frac{3-2r}{2}\tanh{t} \right) \gamma^{t}\Psi_{\tilde{\mu}_{1}...\tilde{\mu}_{r}}+\frac{1}{\cosh{t}} \begin{pmatrix} 0 & i\tilde{\slashed{\nabla}} \\
    - i\tilde{\slashed{\nabla}} & 0
     \end{pmatrix}\Psi_{\tilde{\mu}_{1}...\tilde{\mu}_{r}} =  -ir \Psi_{\tilde{\mu}_{1}...\tilde{\mu}_{r}}, \label{Dirac_op_on_psi_phys_dS4}
 \end{align}
 where we have made use of the expressions for the Christoffel symbols, spin connection, vierbein fields and gamma matrices from Subsection~\ref{subsection global coords}.

Before proceeding to the construction of the modes, note that the physical modes on $dS_{4}$ are naturally split into two classes depending on their $so(4)$ representation-theoretic content - i.e. depending on whether their $\bm{\theta_{3}}$-dependence is given by the spherical eigenmodes~(\ref{tensor-spinor+eigen_S3}) or (\ref{tensor-spinor--eigen_S3}). Let us introduce the following notation:
\begin{itemize}
    \item The physical modes with $so(4)$ content given by $\vec{f}^{-}_{r}$~[Eq.~(\ref{so(4) -weight TT spin-r+1/2})] are denoted as ${\Psi}^{(phys ,\,- \ell; \,{m};k)}_{{\mu}_{1}...{\mu}_{r}}(t,\bm{\theta}_{3})$. We also refer to these modes as `physical modes with helicity $-s$' (recall that $s=r+1/2$).

    \item The physical modes with $so(4)$ content given by $\vec{f}^{+}_{r}$~[Eq.~(\ref{so(4) +weight TT spin-r+1/2})] are denoted as ${\Psi}^{(phys ,\,+ \ell; \,m;k)}_{{\mu}_{1}...{\mu}_{r}}(t,\bm{\theta}_{3})$. We also refer to these modes as `physical modes with helicity $+s$'.
\end{itemize}
Following our previous work~\cite{Letsios_announce}, we separate variables for ${\Psi}^{(phys ,\,- \ell; \,m;k)}_{{\mu}_{1}...{\mu}_{r}}(t,\bm{\theta}_{3})$ as:
\begin{equation}\label{physmodes_negative_spin_r+1/2_dS4}
  {\Psi}^{(phys ,\,- \ell; \,m;k)}_{t {\mu}_{2}...{\mu}_{r}}(t,\bm{\theta}_{3})= 0,\hspace{5mm}  {\Psi}^{(phys ,\,- \ell; \,m;k)}_{\tilde{\mu}_{1}...\tilde{\mu}_{r}}(t,\bm{\theta}_{3})= \begin{pmatrix}  \alpha^{(r)}_{\ell}(t) \, \tilde{\psi}_{-\tilde{\mu}_{1}...\tilde{\mu}_{r}}^{(\ell; m;k)} (\bm{\theta_{3}})  \\ - i  \beta^{(r)}_{\ell}(t) \, \tilde{\psi}^{(\ell; m;k)}_{-\tilde{\mu}_{1}...\tilde{\mu}_{r}} (\bm{\theta_{3}}
    )  \end{pmatrix},
\end{equation}
where $\ell=r,r+1,...$, while $\alpha^{(r)}_{\ell}(t)$ and $\beta^{(r)}_{\ell}(t)$ are functions of time that we must determine.\footnote{The functions $\alpha^{(r)}_{\ell}(t)$ and $\beta^{(r)}_{\ell}(t)$ correspond to $\varPhi^{(a)}_{M \ell}$ and $\varPsi^{(a)}_{M \ell}$, respectively, with $a=-r$ and $M=ir$ (in four spacetime dimensions) in our previous work~\cite{Letsios_announce}.}~Substituting Eq.~(\ref{physmodes_negative_spin_r+1/2_dS4}) into Eq.~(\ref{Dirac_op_on_psi_phys_dS4}), we find
 \begin{align}
    \left(\frac{d}{dt}+\frac{3-2r}{2}\tanh{t}-\frac{i\left( \ell+\frac{3}{2} \right)}{\cosh{t}}\,\right){\beta}^{(r)}_{ \ell}(t)&=- i\,r\,{\alpha}^{(r)}_{\ell}(t) \label{psia=-r_to_phi_sphere_analcont},
    \end{align}
    
\begin{align}\label{phia=-r_to_psi_sphere_analcont}
  \left(\frac{d}{d t}+\frac{3-2r}{2}\tanh{t}+\frac{i \left(\ell+\frac{3}{2} \right)}{\cosh{t}} \,\right)\alpha^{(r)}_{\ell}(t)&= \, i\,r\, \,\beta^{(r)}_{\ell}(t) .
  \end{align}
Using these two relations, and introducing the variable
\begin{align}\label{introducing x(t)=pi/2-it}
    x=\frac{\pi}{2}-it,
\end{align}
we find two second-order equations:
  \begin{align}\label{diff_eqn_for_a(t)}
    \Bigg[  ~\frac{\partial^{2}}{\partial x^{2}}&+(3-2r)\cot{x}\frac{\partial}{\partial x}+\left(\ell+\frac{3}{2}\right)\frac{\cos{x}}{\sin^{2}{x}} \nonumber\\
      &-\frac{(\ell+\frac{3}{2})^{2}-\frac{1}{4}{(3-2r)(1-2r)}}{\sin^{2}{x}} -\frac{(3-2r)^{2}}{4} \Bigg] \alpha^{(r)}_{\ell}(t)=-r^{2}\alpha^{(r)}_{\ell}(t)
  \end{align}
and
\begin{align}\label{diff_eqn_for_b(t)}
    \Bigg[  ~\frac{\partial^{2}}{\partial x^{2}}&+(3-2r)\cot{x}\frac{\partial}{\partial x}-\left(\ell+\frac{3}{2}\right)\frac{\cos{x}}{\sin^{2}{x}} \nonumber\\
      &-\frac{(\ell+\frac{3}{2})^{2}-\frac{1}{4}{(3-2r)(1-2r)}}{\sin^{2}{x}} -\frac{(3-2r)^{2}}{4} \Bigg] \beta^{(r)}_{\ell}(t)=-r^{2}\, \beta^{(r)}_{\ell}(t),
  \end{align}
  where $\cos{x}= i \sinh{t}$, $\sin{x} = \cosh{t}$ and $\cot{x} = i \tanh{t}$.\footnote{Note that the third term of the differential operator in Eq.~(\ref{diff_eqn_for_a(t)}) has an opposite sign from the third term of the differential operator in Eq.~(\ref{diff_eqn_for_b(t)}). This is the only difference between these two differential operators.} The solutions are given in terms of the Gauss hypergeometric function~\cite{gradshteyn2007} as:
\begin{align}
    \alpha^{(r)}_{\ell}(t) = & \left(\cos{\frac{x(t)}{2}}\right)^{\ell+1+r}\left(\sin{\frac{x(t)}{2}}\right)^{\ell+r}\nonumber\\
    & \times F\left(r+{2}+\ell,-r+\ell+{2};\ell+{2};\sin^{2}\frac{x(t)}{2}\right)\label{phi_a=-r, M=ir_t},
\end{align}
and
\begin{align}
    \beta^{(r)}_{\ell}(t) = &\,\frac{r}{\ell+2} \left(\cos{\frac{x(t)}{2}}\right)^{\ell+r}\left(\sin{\frac{x(t)}{2}}\right)^{\ell+r+1}\nonumber\\
    & \times F\left(r+{2}+\ell,-r+\ell+{2};\ell+{3};\sin^{2}\frac{x(t)}{2}\right)\label{psi_a=-r, M=ir_t},
\end{align}
where
\begin{align}
   &\cos{\frac{x(t)}{2}}=\left(  \sin{\frac{x(t)}{2}} \right)^{*}=\frac{\sqrt{2}}{2}\,\left(\cosh{\frac{t}{2}} + i \sinh{\frac{t}{2}} \right)    \label{cosx/2}.
       \end{align}
We have now completely determined the form of the physical modes~$
 {\Psi}^{(phys ,\,- \ell; \,m;k)}_{ {\mu}_{1}...{\mu}_{r}}(t,\bm{\theta}_{3})$ in Eq.~(\ref{physmodes_negative_spin_r+1/2_dS4}). 

Similarly, we find that the physical modes with $so(4)$ content given by $\vec{f}^{+}_{r}$~[Eq.~(\ref{so(4) +weight TT spin-r+1/2})] are expressed as
\begin{equation}\label{physmodes_positive_spin_r+1/2_dS4}
   {\Psi}^{(phys ,\,+\ell; \,m;k)}_{t {\mu}_{2}...{\mu}_{r}}(t,\bm{\theta}_{3})= 0,\hspace{5mm}    {\Psi}^{(phys ,\,+ \ell; \,m;k)}_{\tilde{\mu}_{1}...\tilde{\mu}_{r}}(t,\bm{\theta}_{3})= \begin{pmatrix} i \beta^{(r)}_{\ell}(t) \, \tilde{\psi}_{+\tilde{\mu}_{1}...\tilde{\mu}_{r}}^{(\ell; m;k)} (\bm{\theta_{3}})  \\ -   \alpha^{(r)}_{\ell}(t) \, \tilde{\psi}^{(\ell; m;k)}_{+\tilde{\mu}_{1}...\tilde{\mu}_{r}} (\bm{\theta_{3}}
    )  \end{pmatrix}.
\end{equation}
 The functions $\alpha^{(r)}_{\ell}(t)$ and $\beta^{(r)}_{\ell}(t)$ are given again by Eqs.~(\ref{phi_a=-r, M=ir_t}) and (\ref{psi_a=-r, M=ir_t}), respectively.

 The physical modes~(\ref{physmodes_negative_spin_r+1/2_dS4}) and (\ref{physmodes_positive_spin_r+1/2_dS4}) can also be obtained by analytically continuing tensor-spinor spherical harmonics on $S^{4}$ (see Refs.~\cite{Letsios, Letsios_in_progress, Letsios_announce, STSHS} for details concerning such analytic continuation techniques).

%%%%%%%%%%%%%%%%%%%%%%%%%%%%%%%%%%%%%%

$$ \textbf{Short wavelength limit of physical modes and `positive frequency' condition}  $$
Using the property~\cite{gradshteyn2007}:
\begin{align}
    F(A,B;C;z)=(1-z)^{C-A-B}\,F(C-A, C-B;C;z),
\end{align}
we find that in the limit $\ell >> 1$ (short wavelength/{high frequency} limit), the functions $\alpha^{(r)}_{\ell}(t)$~[Eq.~(\ref{phi_a=-r, M=ir_t})] and $\beta^{(r)}_{\ell}(t)$~[Eq.~(\ref{psi_a=-r, M=ir_t})] describe the time dependence of positive frequency Minkowskian modes, as
\begin{align}\label{pos freq behavour}
   & \frac{d \, \alpha^{(r)}_{\ell}(t)}{dt} \sim -\frac{i \ell}{\cosh{t}}\,\alpha^{(r)}_{\ell}(t), \nonumber \\
    & \frac{d \, \beta^{(r)}_{\ell}(t)}{dt} \sim -\frac{i \ell}{\cosh{t}}\,\beta^{(r)}_{\ell}(t) ,
\end{align}
{and thus,
\begin{align}\label{Euclidean pos freq}
    \frac{\partial}{\partial t}{\Psi}^{(phys ,\,\pm \ell; \,m;k)}_{ {\mu}_{1}...{\mu}_{r}} \sim  -\frac{i \ell}{\cosh{t}}\, {\Psi}^{(phys ,\,\pm \ell; \,m;k)}_{ {\mu}_{1}...{\mu}_{r}} \hspace{5mm}  (\text{for}~\ell>>1).
\end{align}
 Equation~(\ref{Euclidean pos freq}) holds for all (fixed) times $t$ and it has the form of the standard `positive frequency' condition for the Euclidean vacuum in global $dS_{4}$~\cite{Birrell}. Note that, as we will show in Section~\ref{Section dS UIR's and modes}, our `positive frequency' physical modes transform among themselves under dS transformations~(\ref{infinitesimal dS of physical modes}). We can express the `positive frequency' condition~(\ref{Euclidean pos freq}) in a more familiar form as follows. Introducing the conformal time parameter, $\eta$ (with $\tan{\eta} \equiv \sinh{t}$ and $ \pi/2 > \eta > -\pi/2$), the dS line element~(\ref{dS_metric}) is expressed as
 \begin{equation*}
    ds^{2}=\frac{1}{\cos^{2}{\eta}}   \left(   -d\eta^{2}+d\Omega^{2}\right),
\end{equation*}
while the `positive frequency' condition~(\ref{Euclidean pos freq}) becomes
 \begin{align}\label{Euclidean pos freq conformal time}
    \frac{\partial}{\partial \eta}{\Psi}^{(phys ,\,\pm \ell; \,m;k)}_{ {\mu}_{1}...{\mu}_{r}} \sim  -{i \ell}\,\, {\Psi}^{(phys ,\,\pm \ell; \,m;k)}_{ {\mu}_{1}...{\mu}_{r}}  \hspace{5mm}  (\text{for}~\ell>>1) .
\end{align}
% In other words, for high frequencies $\ell >>1$, the physical modes ${\Psi}^{(phys ,\,\pm \ell; \,m;k)}_{ {\mu}_{1}...{\mu}_{r}}$ behave as $ {\Psi}^{(phys ,\,\pm \ell; \,m;k)}_{ {\mu}_{1}...{\mu}_{r}}  \sim e^{-i\,\ell \,\eta }$. 
This is the standard `positive frequency' behaviour (for large $\ell$) used to define the Euclidean vacuum in the case of the scalar field in global $dS_{4}$ in Ref.~\cite{Birrell} (with the use of the conformal time parameter).}

{The condition in Eq.~(\ref{Euclidean pos freq}) is not an exact positive frequency condition in the Minkowskian sense, as indicated by the time-dependent `frequency'. However, the Euclidean vacuum - for which the notion of `particles' is defined with the help of~(\ref{Euclidean pos freq}), (\ref{Euclidean pos freq conformal time}) - is a good `no-particle' state for local observers in the high-frequency limit~\cite{Birrell}. (In global $dS_{4}$, it is this limit - and not the limit $t \rightarrow \pm \infty$ - where one can define an adiabatic vacuum with no particle production as explained in~\cite{Birrell}.)  Moreover, the Euclidean vacuum `positive frequency' condition~[(\ref{Euclidean pos freq}), (\ref{Euclidean pos freq conformal time})] is often preferred in the mathematical physics community because, as is well-known, the Euclidean vacuum in the unique dS invariant vacuum for which the 2-point functions obey the Hadamard condition, i.e. their singularity structure matches the flat-space one. For example, apart from the well-studied scalar case (discussed in, e.g.,~\cite{Birrell}), the graviton and spinor mode functions that satisfy the positive frequency condition~(\ref{Euclidean pos freq}) on global dS spacetime have been used to construct the Wightman 2-point functions with Minkowskian short-distance behaviour in Refs.~\cite{AtsushiHiguchi_2003} and \cite{Letsios}, respectively. However, there are also other positive frequency conditions giving rise to different vacua, such as the in/out vacuua in Ref.~\cite{Mottola_Anderson} (these do {not} correspond to Hadamard states).}

\noindent   \textbf{`Negative frequency' physical modes.} Apart from the physical modes~(\ref{physmodes_negative_spin_r+1/2_dS4}) and~(\ref{physmodes_positive_spin_r+1/2_dS4}), there are also physical modes that are the analogs of Minkowskian negative frequency modes. These are given by
\begin{equation}\label{physmodes_positive_spin_r+1/2_dS4_negfreq}
   {\Psi}^{(phys ,\,+\ell; \,m;k)C}_{t {\mu}_{2}...{\mu}_{r}}(t,\bm{\theta}_{3})= 0,\hspace{5mm}    {\Psi}^{(phys ,\,+ \ell; \,m;k)C}_{\tilde{\mu}_{1}...\tilde{\mu}_{r}}(t,\bm{\theta}_{3})= \begin{pmatrix}  \alpha^{(r)*}_{ \ell}(t) \, \tilde{\psi}_{+\tilde{\mu}_{1}...\tilde{\mu}_{r}}^{(\ell; m;k)} (\bm{\theta_{3}})  \\ + i   \beta^{(r)*}_{ \ell}(t) \, \tilde{\psi}^{(\ell; m;k)}_{+\tilde{\mu}_{1}...\tilde{\mu}_{r}} (\bm{\theta_{3}}
    )  \end{pmatrix}
\end{equation}
and
\begin{equation}\label{physmodes_negative_spin_r+1/2_dS4_negfreq}
   {\Psi}^{(phys ,\,-\ell; \,m;k)C}_{t {\mu}_{2}...{\mu}_{r}}(t,\bm{\theta}_{3})= 0,\hspace{5mm}    {\Psi}^{(phys ,\,- \ell; \,m;k)C}_{\tilde{\mu}_{1}...\tilde{\mu}_{r}}(t,\bm{\theta}_{3})= \begin{pmatrix}  i\beta^{(r)*}_{ \ell}(t) \, \tilde{\psi}_{-\tilde{\mu}_{1}...\tilde{\mu}_{r}}^{(\ell; m;k)} (\bm{\theta_{3}})  \\ +  \alpha^{(r)*}_{ \ell}(t) \, \tilde{\psi}^{(\ell; m;k)}_{-\tilde{\mu}_{1}...\tilde{\mu}_{r}} (\bm{\theta_{3}}
    )  \end{pmatrix}.
\end{equation}
It is straightforward to verify that these modes satisfy Eq.~(\ref{Dirac_op_on_psi_phys_dS4}), as well as the complex conjugate of Eq.~(\ref{Euclidean pos freq}).
In this paper, we do not discuss the representation-theoretic properties of the `negative frequency' modes $ {\Psi}^{(phys ,\,\pm \ell; \,m;k)C}_{ {\mu}_{1}...{\mu}_{r}}$, because they form the same $so(4,1)$ UIRs as the ones formed by the `positive frequency' modes ${\Psi}^{(phys ,\,\pm \ell; \,m;k)}_{ {\mu}_{1}...{\mu}_{r}}$.
%%%%%%%%%%%%%%%%%%%%%%%%%%%%%%%%%%%%%%%%%%%%%
%%%%%%%%%%
\subsubsection{Pure gauge spin-\texorpdfstring{$(r+1/2) \geq 3/2$}{r+/2} modes on \texorpdfstring{$dS_{4}$}{dS}} \label{SubSub_pure gauge}
The pure gauge modes of the striclty massless spin-$(r+1/2)$ equations~[(\ref{Dirac_eqn_fermion_dS_sm}) and~(\ref{TT_conditions_fermions_dS_sm})] satisfy the same conditions as the restricted gauge transformations~(\ref{onshell_gauge}). This means that the pure gauge modes are expressed as
\begin{align}\label{pure gauge modes +-}
    {\Psi}^{(pg ,\,\tilde{r},\,\pm \ell; \,\underline{m})}_{ {\mu}_{1}...{\mu}_{r}}(t,\bm{\theta}_{3}) = \left( \nabla_{(\mu_{1}} + \frac{i}{2} \gamma_{(\mu_{1}}  \right)  \lambda^{(\tilde{r},\pm\ell; \,\underline{m})}_{\mu_{2}...\mu_{r})}(t,\bm{\theta}_{3}).
\end{align}
The ``gauge-function modes'', $\lambda^{(\tilde{r},\,\pm\ell; \,\underline{m})}_{\mu_{2}...\mu_{r}}$, are totally symmetric tensor-spinors of rank $r-1$ and they satisfy Eqs.~(\ref{EOM_for_gauge_functions_Dirac}) and (\ref{EOM_for_gauge_functions_TT}). As in the case of physical modes, explicit expressions for $\lambda^{(\tilde{r},\,\pm\ell; \,\underline{m})}_{\mu_{2}...\mu_{r}}$ can be obtained using the method of separation of variables, but they are not needed for the purposes of this paper\footnote{Explicit expressions for the spin-3/2 and spin-5/2 cases can be found in Ref.~\cite{Letsios_in_progress}.}.  The two labels $\tilde{r}, \pm \ell$ in Eq.~(\ref{pure gauge modes +-}) are used to denote the $so(4)$ content of each pure gauge mode; this corresponds to the $so(4)$ highest weights
\begin{align} \label{spin content for PG MODES}
   \vec{f}^{\pm}_{\tilde{r}} = (\ell+ \tfrac{1}{2},  \pm \tilde{r} \pm \tfrac{1}{2}), \hspace{5mm} \tilde{r} \in \{ 0, 1,...,r-1\},
\end{align}
with $\ell = r,r+1,...$ [the value $\tilde{r}=r$ is excluded in Eq.~(\ref{spin content for PG MODES}) since it corresponds to the $so(4)$ content of physical modes - see Eqs.~(\ref{so(4) +weight TT spin-r+1/2}) and~(\ref{so(4) -weight TT spin-r+1/2})]. The label $\underline{m}$ represents angular momentum quantum numbers corresponding to the subalgebras $so(3) \supset so(2)$.

The pure gauge modes must have zero norm with respect to any dS invariant scalar product and be orthogonal to all physical modes~\cite{Yale_Thesis, STSHS, Letsios_in_progress, Letsios_announce, HiguchiLinearised}. Because of these properties, the pure gauge modes can be identified with zero in the solution space of the field equations~(\ref{Dirac_eqn_fermion_dS_sm}) and~(\ref{TT_conditions_fermions_dS_sm}). These properties will be demonstrated in Section~\ref{Section dS UIR's and modes} for a specific choice of dS invariant scalar product - see also Refs.~\cite{ Letsios_announce, Letsios_in_progress}.
%%%%%%%%%%%%%%%%%%%%%%%%%%%%%%%%%%%%%%%%%%5

%%%%%%%%%%%%%%%%%%%%%%%%%%%%%%%%%%%%%%%%%%%%%%%%%%%%%%%%%%%%%%%%%%%%%%%%%%%%%%%

%%%%%%%%%%%%%%%%%%%%%%%%%%%%%%%%%%%%%%%%%%%%%%%%%%%%%%%%%%%%%%%%%%%%%%%%%%%%%%%%%%
\section{The physical modes form UIRs of the dS algebra}\label{Section dS UIR's and modes}
In this Section, we explain how the `positive frequency' physical modes~(\ref{physmodes_negative_spin_r+1/2_dS4}) and~(\ref{physmodes_positive_spin_r+1/2_dS4}) of the fermionic strictly massless theories form a direct sum of Discrete Series UIRs of the dS algebra~$so(4,1)$. (Below we often use the terms `positive frequency' physical modes and physical modes interchangeably.) %From a quantum field theoretic point of view, the UIR's formed by the mode solutions can be identified with the one-particle Hilbert spaces of the corresponding free QFT's - see Sec.?? (See also Ref.~\cite{HiguchiLinearised} for the corresponding discussion concerning the graviton field on global $dS_{4}$,)
To identify the $so(4,1)$ UIRs formed by these mode solutions, we follow two basic steps:
\begin{itemize}
\item \textbf{Irreducibility:} We identify the sets of `positive frequency' physical modes that form irreducible representations of $so(4,1)$. 

This means that we need to study the infinitesimal dS transformations of the physical mode solutions. We show that the physical modes with fixed helicity $\pm s$ transform among themselves under all $so(4,1)$ transformations~(up to gauge equivalence). Thus, the physical modes form a direct sum of irreducible representations - one corresponding to the helicity $+s$ and one to $-s$. Moreover, it is already easy to see that pure gauge modes transform only into other pure gauge modes under infinitesimal dS transformations, as the Lie-Lorentz derivative~(\ref{Lie_Lorentz}) commutes with the operator $\nabla_{\mu}+\tfrac{i}{2} \gamma_{\mu}$ in Eq.~(\ref{pure gauge modes +-}), while also, it leaves invariant the conditions~(\ref{EOM_for_gauge_functions_Dirac})~and~(\ref{EOM_for_gauge_functions_TT}), which determine the restricted gauge transformations.

    \item \textbf{Unitarity:} We introduce a positive definite dS invariant and gauge-invariant scalar product for each set of `positive frequency' physical modes of fixed helicity.
    
    In particular, we start by introducing the scalar product~(\ref{axial_scalar prod}) and we show that it is dS invariant. With respect to this scalar product, the pure gauge modes are shown to be orthogonal to themselves, as well as to all physical modes~
(i.e. it is demonstrated that the pure gauge modes can be identified with zero in the solution space). Interestingly, it turns out that the scalar product~(\ref{axial_scalar prod}) is positive definite for the `positive frequency' physical modes with helicity $-s$ and negative definite for the `positive frequency' physical modes with helicity $+s$. However, as these two sets of fixed-helicity modes form different irreducible $so(4,1)$ representations, we are allowed to use a different scalar product for each set. We thus redefine the scalar product for the $+s$ modes by introducing a factor of $-1$, in order to achieve positive-definiteness. This peculiarity - i.e. having a different positive definite scalar product for physical modes with different helicity - is already known to appear in the spin-3/2 and spin-5/2 cases on even-dimensional $dS_{D}$ for $D \geq 4$~\cite{Letsios_announce, Letsios_in_progress}. See the end of this Section for more comments on this peculiarity.

    \textbf{Note.}~Although unitarity is often considered to be equivalent to the positive-definiteness of the scalar product in the Hilbert space of mode solutions, this is not a sufficient requirement. For representation-theoretic unitarity, the scalar product must be both positive definite and invariant under the symmetry algebra (or group) of interest. In this Section, the symmetries of interest correspond to the dS algebra, while, in Section~\ref{Sec_hidden}, they correspond to the conformal-like $so(4,2)$ algebra.
\end{itemize}
Once we ensure both the unitarity and irreducibility of the $so(4,1)$ representations formed by the physical modes with fixed helicity, we will recall the $so(4)$ content~[Eqs.~(\ref{so(4) +weight TT spin-r+1/2}) and (\ref{so(4) -weight TT spin-r+1/2})] of these modes, as well as the value of the field-theoretic quadratic Casimir~(\ref{quadratic_Casimir}). Then, it will be straightforward to identify the UIRs formed by the physical modes with a direct sum of Discrete Series UIRs of $so(4,1)$~[Eqs.~(\ref{condition_for_discrete_series_+ D=4}) and~(\ref{condition_for_discrete_series_- D=4})].
%%%%%%%%%%%%%%%%%%%%%%%%%%%%%%%%%%%%%%%%%%%%%%%%%%%%%%%%%%%%%%%%%5%%%%%%%%%%%%%%%%
\subsection{Infinitesimal dS transformations of physical modes and irreducibility of \texorpdfstring{$so(4,1)$}{so(4,1)} representations}~\label{Subsec_dS_transformations}
The infinitesimal dS transformations of the mode solutions can be studied with the use of the Lie-Lorentz derivative~(\ref{Lie_Lorentz}) with respect to the dS Killing vectors. Since the $so(4)$ content of the $so(4,1)$ representations formed by mode solutions is already known~(see Section~\ref{Sec_mode solutions}), we just need to study the transformation properties of our mode solutions under dS boosts. In fact, it is sufficient to focus on just one dS boost (the reason is that the Lie bracket between a boost Killing vector and a rotational one is equal to another boost Killing vector).
We choose to work with the following boost Killing vector:
\begin{align}\label{dS boost}
    X = X^{\mu} \partial_{\mu} = \cos{\theta_{3}}\, \frac{\partial}{\partial t} - \tanh{t} \sin{\theta_{3}}\, \frac{\partial}{\partial {\theta_{3}}}.
\end{align}

Our aim is to express $\mathbb{L}_{X}{\Psi}^{(phys ,\,- \ell; \,m;k)}_{{\mu}_{1}...{\mu}_{r}}(t,\bm{\theta}_{3})$ and $\mathbb{L}_{X}{\Psi}^{(phys ,\,+ \ell; \,m;k)}_{{\mu}_{1}...{\mu}_{r}}(t,\bm{\theta}_{3})$ as linear combinations of other mode solutions\footnote{In the spin-3/2 and spin-5/2 cases, the transformations $\mathbb{L}_{X}{\Psi}^{(phys ,\,- \ell; \,m;k)}_{{\mu}_{1}...{\mu}_{r}}(t,\bm{\theta}_{3})$ and $\mathbb{L}_{X}{\Psi}^{(phys ,\,+ \ell; \,m;k)}_{{\mu}_{1}...{\mu}_{r}}(t,\bm{\theta}_{3})$ have already been expressed as linear combinations of other mode solutions by direct calculation in Ref.~\cite{Letsios_in_progress}.}. There are (at least) two different ways we can follow in order to proceed:
\begin{itemize}
    \item i) Direct calculation, where in order to express $\mathbb{L}_{X}{\Psi}^{(phys ,\,\pm \ell; \,m;k)}_{{\mu}_{1}...{\mu}_{r}}$ as a linear combination of other modes, one has to use the raising and lowering differential operators for the angular momentum quantum number $\ell$, as in Refs.~\cite{STSHS, HiguchiLinearised, Letsios, Letsios_in_progress}.
\end{itemize}
or
\begin{itemize}
    \item ii) Making use of the matrix elements of $so(5)$ generators obtained by Gelfand and Tsetlin~\cite{Gelfand}. More specifically, one can use these matrix elements to find explicit expressions for the $so(5)$ transformations of tensor-spinor spherical harmonics on $S^{4}$ and then perform analytic continuation to $dS_{4}$.
\end{itemize}
 In this paper, we follow approach ii. Here we present the final expressions for $\mathbb{L}_{X}{\Psi}^{(phys ,\,\pm \ell; \,m;k)}_{{\mu}_{1}...{\mu}_{r}}$. The reader who is familiar with $so(D+1)$ representations formed by tensor-spinor spherical harmonics on $S^{D}$~\cite{Homma} can infer the results from Gelfand and Tsetlin's work~\cite{Gelfand}.~Technical details for the derivation can be found in Appendix~\ref{Appendix matrix elements}. 
 
 Without further ado, following approach ii, the infinitesimal transformation of the physical spin-$(r+1/2) \geq 3/2$ modes under the dS boost~(\ref{dS boost}) are found to be:
 \begin{align}\label{infinitesimal dS of physical modes}
     \mathbb{L}_{X}{\Psi}^{\left(phys ,\,\pm \ell; m; k\right)}_{{\mu}_{1}...{\mu}_{r}} =&-\frac{i}{2(\ell+2)}\,\sqrt{((\ell+2)^{2}-r^{2})\,(\ell-m+1)(\ell+m+3)}\,{\Psi}^{\left(phys ,\,\pm (\ell+1)\,;m;k \right)}_{{\mu}_{1}...{\mu}_{r}} \nonumber\\
     &-\frac{i(\ell+1)}{2}\,\sqrt{\frac{(\ell-m)(\ell+m+2)}{(\ell+1)^{2}-r^{2}}}\,{\Psi}^{\left(phys ,\,\pm (\ell-1)\,;{m;k}\right)}_{{\mu}_{1}...{\mu}_{r}}+(\text{pure gauge}),
     \end{align}
  where the term `(pure gauge)' is proportional to the pure gauge mode ${\Psi}^{(pg ,\,\tilde{r}={r}-1,\,\pm \ell; \,{m};k)}_{ {\mu}_{1}...{\mu}_{r}}$ [see Eq.~(\ref{pure gauge modes +-})].

  \noindent \textbf{Conclusion.}
  From the transformation properties~(\ref{infinitesimal dS of physical modes}), we conclude that the modes $\{  {\Psi}^{\left(phys ,\,- \ell; m; k\right)}_{{\mu}_{1}...{\mu}_{r}}\}$ and $\{  {\Psi}^{\left(phys ,\,+ \ell; m; k\right)}_{{\mu}_{1}...{\mu}_{r}}\}$ separately form irreducible representations of $so(4,1)$ (up to gauge equivalence).
  
  In the next Subsection, by making a choice of a dS invariant scalar product, we will explicitly show that all pure gauge modes have zero associated norm. Thus, the Lie-Lorentz derivatives~(\ref{Lie_Lorentz}) essentially act on equivalence classes of physical modes, i.e. if for any two physical modes, $\Psi^{(1)}_{\mu_{1}...\mu_{r}}$ and $\Psi^{(2)}_{\mu_{1}...\mu_{r}}$, the difference $\Psi^{(1)}_{\mu_{1}...\mu_{r}} -\Psi^{(2)}_{\mu_{1}...\mu_{r}}$ is a linear combination of pure gauge modes, then $\Psi^{(1)}_{\mu_{1}...\mu_{r}}$ and $\Psi^{(2)}_{\mu_{1}...\mu_{r}}$ belong to the same equivalence class.
%%%%
%%%%%%%%%%%%%%%%%%%%%%5
\subsection{dS invariant scalar product and unitarity of \texorpdfstring{$so(4,1)$}{so(4,1)} representations}\label{subsec_ds scalar prod and unitarity}
 For any two (physical or pure gauge) solutions $\Psi^{(1)}_{\mu_{1}...\mu_{r}}, \Psi^{(2)}_{\mu_{1}...\mu_{r}}$ of Eqs.~(\ref{Dirac_eqn_fermion_dS_sm}) and (\ref{TT_conditions_fermions_dS_sm}), define the (axial) vector current $ {J}^{\mu}( \Psi^{(1)}, \Psi^{(2)})$ as
\begin{equation}\label{axial_current}
      {J}^{\mu}( \Psi^{(1)}, \Psi^{(2)})=-i \,\overline{\Psi}^{(1)}_{\nu_{1}...\nu_{r}} \gamma^{5} \gamma^{\mu}  \Psi^{(2)\nu_{1}...\nu_{r}},
    \end{equation}
    where $\overline{\Psi}^{(1)}_{\nu_{1}...\nu_{r} }= i {\Psi}^{(1)\dagger}_{\nu_{1}...\nu_{r}} \gamma^{0}$.
This is covariantly conserved, $\nabla^{\mu} {J}_{\mu}( \Psi^{(1)}, \Psi^{(2)})=0$~\footnote{I would like to thank Atsushi Higuchi for pointing out that this current is conserved.}. Thus, it immediately follows that the scalar product
\begin{align}\label{axial_scalar prod}
 \braket{ \Psi^{(1)}| \Psi^{(2)}}   &= \,\int_{S^{3}} \sqrt{-{g}} \,d\bm{\theta_{3}}\,{J}^{0}( \Psi^{(1)}, \Psi^{(2)}) \nonumber\\
 &= \cosh^{3}{t}\,\int_{S^{3}} \sqrt{\tilde{g}} \,d\bm{\theta_{3}}\,{\Psi}^{(1)\dagger}_{\nu_{1}...\nu_{r}}(t,\bm{\theta_{3}}) \,\gamma^{5}\,  \Psi^{(2)\nu_{1}...\nu_{r}}(t,\bm{\theta_{3}})
\end{align}
is time-independent, where $\cosh^{3}{t} \sqrt{\tilde{g}} = \sqrt{-g}$, while $g$ is the determinant of the dS metric.

\noindent \textbf{dS invariance of the scalar product.} The dS invariance of the scalar product~(\ref{axial_scalar prod}) can be demonstrated as follows. Let $\delta_{\xi}{J}^{\mu}$ be the change of the current~(\ref{axial_current}) under the infinitesimal dS transformation generated by a dS Killing vector $\xi^{\mu}$. {Then, we have
\begin{align}
    \delta_{\xi} {J} ^{\mu}( \Psi^{(1)}, \Psi^{(2)})&={J}^{\mu}( \mathbb{L}_{\xi}\Psi^{(1)}, \Psi^{(2)})+{J}^{\mu}( \Psi^{(1)}, \mathbb{L}_{\xi}\Psi^{(2)}) \nonumber\\
    &=\nabla_{\nu} \left(  \xi^{\nu} \,{J}^{\mu}( \Psi^{(1)}, \Psi^{(2)})- \xi^{\mu}\,{J}^{\nu}( \Psi^{(1)}, \Psi^{(2)})   \right) ,\\
        \delta_{\xi} {J}^{t}( \Psi^{(1)}, \Psi^{(2)})&= \frac{1}{\sqrt{-g}}\partial_{\tilde{\nu}}\Bigg[ \sqrt{-g}\left(  \xi^{\tilde{\nu}} \,{J}^{t}( \Psi^{(1)}, \Psi^{(2)})- \xi^{t}\,{J}^{\tilde{\nu}}( \Psi^{(1)}, \Psi^{(2)})   \right) \Bigg],
\end{align}
 where we have used that $\nabla_{\nu}{J}^{\nu}=\nabla_{\nu}\xi^{\nu}=0$}. As $\delta_{\xi} {J} ^{\mu}$ is equal to the divergence of an anti-symmetric tensor, the following integral vanishes:
\begin{equation}\label{delta_J_0}
   \delta_{\xi} \braket{ \Psi^{(1)}| \Psi^{(2)}} = \int_{S^{3}} \sqrt{-{g}} \,d\bm{\theta_{3}}\, \delta_{\xi} {J} ^{t}( \Psi^{(1)}, \Psi^{(2)})= 0.
\end{equation}  
In other words, the value of the scalar product~(\ref{axial_scalar prod}) does not change under infinitesimal dS transformations.
This directly implies that
\begin{equation}\label{dS invariance of axial inner}
  \braket{ \mathbb{L}_{\xi}\Psi^{(1)}| \Psi^{(2)}} +  \braket{ \Psi^{(1)}| \mathbb{L}_{\xi}\Psi^{(2)}} =0  ,
\end{equation}
for any dS Killing vector $\xi$.

\noindent \textbf{Gauge invariance of the scalar product.} Let us show that, with respect to the scalar product~(\ref{axial_scalar prod}), all pure gauge modes~(\ref{pure gauge modes +-}) are orthogonal to themselves, as well as to all physical modes. In particular, letting $\Psi^{(2)}_{\mu_{1}...\mu_{r}}$ be a pure gauge mode~(\ref{pure gauge modes +-}) - i.e. $\Psi^{(2)}_{\mu_{1}...\mu_{r}}=\Psi^{(pg)}_{\mu_{1}...\mu_{r}}= (\nabla_{(\mu_{1}}+ \frac{i}{2} \gamma_{(\mu_{1}}) \lambda_{\mu_{2}...\mu_{r})}$, where we have omitted the quantum number labels for convenience - the current~(\ref{axial_current}) can be expressed as
\begin{align}\label{gauge invariance of current}
    J^{\mu}( \Psi^{(1)}, \Psi^{(pg)})=2i\, \nabla_{\lambda}\left( \overline{\Psi}^{(1)\nu_{2}\nu_{3}...\nu_{r}[\lambda}\gamma^{\mu]}\gamma^{5} \, \lambda_{\nu_{2}\nu_{3}...\nu_{r}}   \right),
\end{align}
where $\Psi^{(1)}$ is any physical or pure gauge mode. As $J^{\mu}( \Psi^{(1)}, \Psi^{(pg)})$ in Eq.~(\ref{gauge invariance of current}) is equal to the divergence of an anti-symmetric tensor, the scalar product between any pure gauge mode and any other mode is always zero. Also, this directly implies that the scalar product~(\ref{axial_scalar prod}) is invariant under restricted gauge transformations~(\ref{onshell_gauge}).

\noindent \textbf{Positive-definiteness.} Let us now calculate the norm of the physical `positive frequency' mode solutions with respect to the scalar product~(\ref{axial_scalar prod}). Substituting the expressions for the physical modes, ~(\ref{physmodes_negative_spin_r+1/2_dS4}) and (\ref{physmodes_positive_spin_r+1/2_dS4}), into the scalar product~(\ref{axial_scalar prod}), we find
\begin{align}
    \braket{{\Psi}^{\left(phys ,\,\sigma \ell; m; k\right)}|{\Psi}^{\left(phys ,\,\sigma' \ell'; m'; k'\right)}}=&(-\sigma)\times\, \cosh^{3-2r}{t}\nonumber\\
    &\times \left( |\alpha^{(r)}_{\ell}(t)|^{2}-|\beta^{(r)}_{\ell}(t)|^{2}   \right) \delta_{\sigma \sigma'}\delta_{\ell \ell'}\delta_{m m'}\delta_{k k'},
\end{align}
where $\sigma, \sigma' \in \{ -,+ \}$, while we have made use of the normalisation condition~(\ref{normlzn_S3}) of the tensor-spinor spherical harmonics on $S^{3}$. This expression is time-independent and its value has been calculated in equation~(8.26) of Ref.~\cite{Letsios_in_progress}. The result is~\cite{Letsios_in_progress}
\begin{align}\label{norm of physical modes}
    \braket{{\Psi}^{\left(phys ,\,\sigma \ell; m; k\right)}|{\Psi}^{\left(phys ,\,\sigma' \ell'; m'; k'\right)}}=(-\sigma)\times\,2^{3-2r}\frac{|\Gamma(\ell+2)|^{2}}{\Gamma(\ell+2+r)\Gamma(\ell+2-r)}\delta_{\sigma \sigma'}\delta_{\ell \ell'}\delta_{m m'}\delta_{k k'}.
\end{align}
According to this equation, the `positive frequency' physical modes with helicity $-s$, $\{ {\Psi}^{\left(phys ,\,- \ell; m; k\right)}_{\mu_{1}...\mu_{r}} \}$, form a UIR of $so(4,1)$ with positive definite scalar product given by Eq.~(\ref{axial_scalar prod}), while the `positive frequency' physical modes with helicity $+s$, $\{ {\Psi}^{\left(phys ,\,+ \ell; m; k\right)}_{\mu_{1}...\mu_{r}}\}$, form a UIR of $so(4,1)$ with positive definite scalar product given by the negative of Eq.~(\ref{axial_scalar prod}).

{For the sake of completeness, we note that the norm of the physical `negative frequency' modes [Eqs.~(\ref{physmodes_positive_spin_r+1/2_dS4_negfreq}) and~(\ref{physmodes_negative_spin_r+1/2_dS4_negfreq})] is given by
\begin{align}\label{norm of physical modes neg freq}
    \braket{{\Psi}^{\left(phys ,\,\sigma \ell; m; k\right)C}|{\Psi}^{\left(phys ,\,\sigma' \ell'; m'; k'\right)C}}=(+\sigma)\times\,2^{3-2r}\frac{|\Gamma(\ell+2)|^{2}}{\Gamma(\ell+2+r)\Gamma(\ell+2-r)}\delta_{\sigma \sigma'}\delta_{\ell \ell'}\delta_{m m'}\delta_{k k'}.
\end{align}}

%%%%%%%%%%%%%%%%%%%%%%%%%%%%%%%%%%%%%%%%%%%%%%%%%%%%%%%%%%%%%%%%%%%%%%%%%%%%%%%%%%%%%%%
\subsection{Some comments on why the scalar product~(\ref{axial_scalar prod}) is not positive definite for both helicities}
 {As we mentioned at the beginning of this Section, the physical `positive frequency' modes with helicity $-s$ have a different positive definite scalar product from the physical `positive frequency' modes with helicity $+s$ [recall our definition of `positive frequency' conditions~(\ref{Euclidean pos freq})]. The two scalar products differ in their definition by a factor of $-1$: the scalar product for the $-s$ modes is given by~(\ref{axial_scalar prod}), while the scalar product for the $+s$ modes is given by the negative of~(\ref{axial_scalar prod}). This means that if we use the same scalar product~(\ref{axial_scalar prod}) for both helicities, then the physical `positive frequency' modes with helicity $-s$ will have positive definite norm, while the physical `positive frequency' modes with helicity $+s$ will have negative definite norm [this becomes clear from Eq.~(\ref{norm of physical modes})]. However, from the representation theory point of view, we are allowed to use a different scalar product for each set of fixed-helicity `positive frequency' modes because each set separately forms an irreducible $so(4,1)$ representation. This peculiarity of having a different positive definite scalar product for each fixed-helicity set of `positive frequency' modes stems from the appearance of $\gamma^{5}$ in the scalar product~(\ref{axial_scalar prod}). $\gamma^{5}$ is necessary to achieve dS invariance of the scalar product in the case of tensor-spinors with an imaginary mass parameter~\cite{Letsios_in_progress}. If we remove $\gamma^{5}$ from the scalar product~(\ref{axial_scalar prod}), then we will obtain an inner product which is always positive definite but is \textbf{not} dS invariant for tensor-spinors with imaginary mass parameters. However, the scalar product with $\gamma^{5}$ removed is dS invariant for tensor-spinors with real mass parameters.~\cite{Letsios_in_progress}}  

{\noindent \textbf{Comparing to the Klein-Gordon case.} At first sight, the aforementioned indefiniteness of the norm for strictly massless spin-$s \geq 3/2$ fermions might remind us of the indefiniteness of the Klein-Gordon norm in the case of the scalar field. The mode functions of the scalar field on global $dS_{4}$~(\ref{dS_metric}) can be found in, e.g., Refs.~\cite{STSHS, Mottola_Anderson}. However, the indefiniteness of the Klein-Gordon norm is of a different kind. The scalar field mode functions on global $dS_{4}$ that satisfy a generalised positive frequency condition at short wavelengths as in Eq.~(\ref{Euclidean pos freq}) [these are called CTBD (Chernikov-Tagirov-Bunch-Davies) mode functions in Ref.~\cite{Mottola_Anderson}] have positive definite norm associated with the Klein-Gordon scalar product\footnote{{Recall that the Klein-Gordon scalar product is dS invariant~\cite{STSHS}.}}. The complex conjugate mode functions satisfy generalised negative frequency conditions [corresponding to the complex conjugate of Eq.~(\ref{Euclidean pos freq})] and have negative definite norm associated with the Klein-Gordon scalar product. \textbf{To sum up}, in the case of the scalar field - unlike in the case of strictly massless spin-$s \geq 3/2$ fermions - the positive-definiteness or negative-definiteness of the dS invariant norm can be used to distinguish between modes with generalised positive or negative frequency behaviour. On the other hand, the special feature of the strictly massless spin-$s \geq 3/2$ theories is that indefiniteness of the norm appears among the set of `positive frequency' modes [see Eq.~(\ref{norm of physical modes})], as well as among the set of `negative frequency' modes [see Eq.~(\ref{norm of physical modes neg freq})]. This happens because the sign of the norm is related to the helicity and not to the positive or negative frequency behaviour.  }

{\noindent \textbf{Comparing to the massless spin-1/2 case.} It is worth recalling that in the spin-$1/2$ (massive and massless) case, the standard dS invariant Dirac norm is always positive definite and \textbf{is not} related to the `positive frequency' and `negative frequency' behaviour of the modes - see, e.g., Ref.~\cite{Letsios}. Below, to get some more insight into the peculiarities of the strictly massless spin-$s \geq 3/2$ fermions, we briefly discuss the massless spin-1/2 field on global $dS_{4}$, which is the simplest strictly massless fermion on $dS_{4}$. We will use this example to demonstrate that the appearance of $\gamma^{5}$ in the scalar product leads to the indefiniteness of the norm among the `positive frequency' modes, as well as among the `negative frequency' modes, as in the case of strictly massless spin-$s \geq 3/2$ fields. }

$$ \textbf{The massless spin-1/2 field}  $$
{The massless spin-1/2 field on $dS_{4}$ satisfies the massless Dirac equation 
$$\slashed{\nabla} \Psi = 0.$$ We will show that the indefiniteness of the norm~(\ref{norm of physical modes}) for the `positive frequency' modes of the strictly massless spin-$s \geq 3/2$ theories can also appear in the massless spin-1/2 case by making an unconventional choice of a dS invariant scalar product. (The unconventional choice corresponds to the scalar product~(\ref{axial_scalar prod}) for spin-1/2 fields, i.e. with $r=0$, while the conventional choice corresponds to the Dirac inner product discussed below.) }

{The mode functions of the massless Dirac equation on global $dS_{4}$~(\ref{dS_metric}) can be found in Ref.~\cite{Letsios}. Both `positive frequency' and `negative frequency' modes (of any helicity $\pm 1/2$) have positive definite norm associated with the standard dS invariant Dirac inner product~\cite{Letsios}~\footnote{Unlike the Klein-Gordon norm, the Dirac norm is always positive definite (in both dS and Minkowski spacetimes).}. The Dirac inner product is defined as
\begin{align*}
 \braket{ \Psi^{(1)}| \Psi^{(2)}}_{Dirac}   
 &= \cosh^{3}{t}\,\int_{S^{3}} \sqrt{\tilde{g}} \,d\bm{\theta_{3}}\,{\Psi}^{(1)\dagger}(t,\bm{\theta_{3}}) \,  \Psi^{(2)}(t,\bm{\theta_{3}}),
\end{align*}
where $\Psi^{(1)}$ and $\Psi^{(2)}$ are any two solutions of the massless Dirac equation.
By inserting $\gamma^{5}$ in the Dirac inner product, we can define the modified/unconventional scalar product (as in Eq.~(\ref{axial_scalar prod})):
\begin{align*}
 \braket{ \Psi^{(1)}| \Psi^{(2)}}_{mod}   
 &= \cosh^{3}{t}\,\int_{S^{3}} \sqrt{\tilde{g}} \,d\bm{\theta_{3}}\,{\Psi}^{(1)\dagger}(t,\bm{\theta_{3}}) \, \gamma^{5} \Psi^{(2)}(t,\bm{\theta_{3}}).
\end{align*}
 (This is just the conserved charge corresponding to the axial current  $i\,\overline{\Psi^{(1)}} \, \gamma^{\mu} \gamma^{5} \Psi^{(2)}$.) We will show that the `positive frequency' modes with helicity $-1/2$ ($+1/2$) have positive (negative) norm with respect to the modified/unconventional scalar product, as in the case of the strictly massless fermions of spin $s \geq 3/2$~(\ref{norm of physical modes}). }
 
\noindent { \textbf{Note.} In the case of the massless spin-1/2 field, both the Dirac and the modified/unconventional scalar products are dS invariant\footnote{{The dS invariance of the Dirac inner product follows from the covariant conservation of the Dirac current $i\,\overline{\Psi^{(1)}} \gamma^{\mu} \Psi^{(2)}$~\cite{Letsios}. Similarly, the dS invariance of the modified/unconventional scalar product follows from the covariant conservation of axial current $i\,\overline{\Psi^{(1)}} \gamma^{\mu} \gamma^{5} \Psi^{(2)}$.}}. As is well-known, the conventional choice is the (always positive definite) Dirac inner product. On the other hand, in the case of strictly massless spin-$s \geq 3/2$ fermions, there seems to be only one dS invariant choice for the scalar product; the scalar product that includes $\gamma^{5}$ given by Eq.~(\ref{axial_scalar prod}).}

\noindent {\textbf{Spin-$1/2$ mode functions.} The `positive frequency' modes with helicities $+1/2$ and $-1/2$ are given by~\cite{Letsios}
\begin{align*}
     {\Psi}^{(+ \ell; \,m;k)}(t,\bm{\theta}_{3})= \alpha^{(0)}_{\ell}(t) \begin{pmatrix} 0  \\ \tilde{\psi}^{(\ell; m;k)}_{+} (\bm{\theta_{3}}
    )  \end{pmatrix},~
\end{align*}
and
\begin{align*}
     {\Psi}^{(- \ell; \,m;k)}(t,\bm{\theta}_{3})= \alpha^{(0)}_{\ell}(t)\begin{pmatrix} \tilde{\psi}^{(\ell; m;k)}_{-} (\bm{\theta_{3}})  \\ 0
      \end{pmatrix},
\end{align*}
respectively, where $\alpha^{(0)}_{\ell}(t)$ is found by letting $r=0$ in Eq.~(\ref{phi_a=-r, M=ir_t}) as
$$\alpha^{(0)}_{\ell}(t) = \frac{(\tan{\frac{x(t)}{2}})^{\ell}}{(\cos{\frac{x(t)}{2}})^{3}}, $$
while $\ell = 0,1,...$, $m=0,...,\ell$, and $k=-m-1,...,0,...,m$. The spinor spherical harmonics $\tilde{\psi}^{(\ell; m;k)}_{\pm } (\bm{\theta_{3}})$ on $S^{3}$ satisfy Eqs.~(\ref{tensor-spinor+eigen_S3}) and (\ref{tensor-spinor--eigen_S3}) with $r=0$ (see also Ref.~\cite{Camporesi}). In the limit $\ell >>1$, the `positive frequency' modes satisfy the generalised positive frequency condition~\cite{Letsios}
\begin{align*}
   & \frac{\partial \, {\Psi}^{(\pm \ell; \,m;k)}(t,\bm{\theta}_{3})}{\partial t} \sim -\frac{i \ell}{\cosh{t}}\,{\Psi}^{(\pm \ell; \,m;k)}(t,\bm{\theta}_{3}),
\end{align*}
as in the higher-spin case~(\ref{Euclidean pos freq}). The two sets of mode functions, $\{{\Psi}^{(+ \ell; \,m;k)}\}$ and $\{{\Psi}^{(- \ell; \,m;k)}\}$, separately form UIRs of $so(4,1)$~\cite{Letsios_announce, Letsios}. Thus, it is group-theoretically allowed to have a different scalar product for each set of mode functions.}
%%%%%%%%%%%%%%%%%

{The `negative frequency' modes with helicities $+1/2$ and $-1/2$ are given by~\cite{Letsios}
\begin{align*}
     {\Psi}^{(- \ell; \,m;k)C}(t,\bm{\theta}_{3})= \left( \alpha^{(0)}_{\ell}(t) \right)^{*} \begin{pmatrix} 0  \\ \tilde{\psi}^{(\ell; m;k)}_{-} (\bm{\theta_{3}}
    )  \end{pmatrix},~
\end{align*}
\begin{align*}
     {\Psi}^{(+ \ell; \,m;k)C}(t,\bm{\theta}_{3})= \left( \alpha^{(0)}_{\ell}(t) \right)^{*}\begin{pmatrix} \tilde{\psi}^{(\ell; m;k)}_{+} (\bm{\theta_{3}})  \\ 0
      \end{pmatrix}.
\end{align*}
In the limit $\ell >>1$, the `negative frequency' modes satisfy the generalised negative frequency condition~\cite{Letsios}
\begin{align*}
   & \frac{\partial \, {\Psi}^{(\pm \ell; \,m;k)C}(t,\bm{\theta}_{3})}{\partial t} \sim +\frac{i \ell}{\cosh{t}}\,{\Psi}^{(\pm \ell; \,m;k)C}(t,\bm{\theta}_{3}).
\end{align*}}

\noindent   {\textbf{dS invariant norm of spin-1/2 mode functions.} As we mentioned earlier, there are two choices of dS invariant scalar products for the massless spin-1/2 field. Using the conventional Dirac inner product we find~\cite{Letsios} 
\begin{align*}
    \braket{{\Psi}^{\left(\sigma \ell; m; k\right)}|{\Psi}^{\left(\sigma' \ell'; m'; k'\right)}}_{Dirac}= 
 \braket{{\Psi}^{\left(\sigma \ell; m; k\right)C}|{\Psi}^{\left(\sigma' \ell'; m'; k'\right)C}}_{Dirac}  =~2^{3}\,\delta_{\sigma \sigma'}\delta_{\ell \ell'}\delta_{m m'}\delta_{k k'},
\end{align*}
while using the modified scalar product we find
\begin{align*}
    &\braket{{\Psi}^{\left(\sigma \ell; m; k\right)}|{\Psi}^{\left(\sigma' \ell'; m'; k'\right)}}_{mod}=~(-\sigma) \times \,2^{3}\,\delta_{\sigma \sigma'}\delta_{\ell \ell'}\delta_{m m'}\delta_{k k'}, \\
 &\braket{{\Psi}^{\left(\sigma \ell; m; k\right)C}|{\Psi}^{\left(\sigma' \ell'; m'; k'\right)C}}_{mod} =~(+\sigma) \times  \,2^{3}\,\delta_{\sigma \sigma'}\delta_{\ell \ell'}\delta_{m m'}\delta_{k k'}, 
\end{align*}
where we have used the fact that the massless spin-1/2 mode functions are eigenfunctions of $\gamma^{5}$ [Eq.~(\ref{gamma(5)})]:
$$ \gamma^{5}{\Psi}^{\left(\sigma \ell; m; k\right)} = (-\sigma)\,{\Psi}^{\left(\sigma \ell; m; k\right)}, ~~~~\gamma^{5}{\Psi}^{\left(\sigma \ell; m; k\right)C} = (+\sigma)\,{\Psi}^{\left(\sigma \ell; m; k\right)C}. $$
It is clear that using the modified scalar product we have indefiniteness of the norm among the `positive frequency' and among the `negative frequency' massless spin-1/2 modes, as in the case of strictly massless spin-$s \geq 3/2$ fermions. However, in the spin-1/2 case, this is easily explained because the mode functions are eigenfunctions of $\gamma^{5}$ with eigenvalues $\pm 1$ that determine the sign of the norm. In the spin-$s \geq 3/2$ cases, the mode functions are not eigenfunctions of $\gamma^{5}$ (see Section~\ref{Sec_mode solutions}), but still, it is natural to blame the appearance of $\gamma^{5}$ for the indefiniteness of the norm (recall that $\gamma^{5}$ is needed to ensure the dS invariance of the scalar product in the presence of an imaginary mass parameter~\cite{Letsios_in_progress}).}
%%%%%%%%%%%%%%%%%%%%%%%%%%%%%%%%%%%%%%%%%%%%%%%%%%%%%%%%%%%%%%%%%%%%%%%%%%%%%%%%%%%%%%%%%%%%%%%%%%%%%%%%%%%%%%%%55
\subsection{Identifying the dS algebra UIRs} \label{subsec
_identifying ds uir}
The analysis presented in the previous Subsections has demonstrated that the physical modes, $\{ {\Psi}^{\left(phys ,\,+ \ell; m; k\right)}_{\mu_{1}...\mu_{r}}\}$ and $\{ {\Psi}^{\left(phys ,\,- \ell; m; k\right)}_{\mu_{1}...\mu_{r}} \}$, of the strictly massless theories separately form UIRs of $so(4,1)$. It can be understood that we have a direct sum of Discrete Series UIRs~(\ref{condition_for_discrete_series_+ D=4}) and (\ref{condition_for_discrete_series_- D=4}) as follows. Combining the $so(4)$ content of physical modes [Eqs.~(\ref{so(4) +weight TT spin-r+1/2}) and (\ref{so(4) -weight TT spin-r+1/2})] with the branching rules~(\ref{branching_rules_spin(2p,1)->spin(2p)}), we find that both $\{ {\Psi}^{\left(phys ,\,+ \ell; m; k\right)}_{\mu_{1}...\mu_{r}} \}$ and $\{ {\Psi}^{\left(phys ,\,- \ell; m; k\right)}_{\mu_{1}...\mu_{r}} \}$ correspond to UIRs with $F_{1} = r+1/2$ (see Section~\ref{Sec_Classification_UIRs D=4}). Then, comparing the field-theoretic quadratic Casimir~(\ref{quadratic_Casimir}) (with $M=ir$) with the representation-theoretic one~(\ref{Casimir_Spin(4,1)}), we find the following ``field theory - representation theory dictionary'':
\begin{itemize}
    \item The set of physical modes with helicity $+s$, $\{ {\Psi}^{\left(phys ,\,+ \ell; m; k\right)}_{\mu_{1}...\mu_{r}} \}$, forms the Discrete Series UIR $D^{+}(F_{0} ,F_{1}) =D^{+}(r-\tfrac{3}{2} , r+\tfrac{1}{2})$ [Eq.~(\ref{condition_for_discrete_series_+ D=4})] of $so(4,1)$. The $so(4)$ content is given by Eq.~(\ref{so(4) +weight TT spin-r+1/2}). The positive definite norm is given by the negative of Eq.~(\ref{norm of physical modes}) (with $\sigma=+$).

    \item The set of physical modes with helicity $-s$, $\{ {\Psi}^{\left(phys ,\,- \ell; m; k\right)}_{\mu_{1}...\mu_{r}} \}$, forms the Discrete Series UIR $D^{-}(F_{0} ,F_{1}) =D^{-}(r-\tfrac{3}{2} , r+\tfrac{1}{2})$ [Eq.~(\ref{condition_for_discrete_series_- D=4})] of $so(4,1)$. The $so(4)$ content is given by Eq.~(\ref{so(4) -weight TT spin-r+1/2}). The positive definite norm is given by Eq.~(\ref{norm of physical modes}) (with $\sigma=-$).
\end{itemize}
Thus, the set of all physical mode solutions for the strictly massless spin-$(r+1/2) \geq 3/2$ theory, satisfying Eqs.~(\ref{Dirac_eqn_fermion_dS_sm}) and (\ref{TT_conditions_fermions_dS_sm}), corresponds to the direct sum of Discrete Series UIRs $D^{-}(r-\tfrac{3}{2} , r+\tfrac{1}{2}) \bigoplus D^{+}(r-\tfrac{3}{2} , r+\tfrac{1}{2})$~\footnote{This is also true for the massless spin-1/2 field on $dS_{4}$, i.e. for $r=0$~\cite{Letsios_announce}.}. This is in agreement with the ``field theory - representation theory dictionary'' suggested previously by us~\cite{Letsios_announce}.
%%%%%%%%%%%%%%%%%%%%%%%%%%%%%%%%%%%%%%%%%%%%%%%%%%%%%%%%%%%%%%%%%%%%%%%%%%%%%%%%%%%%%%%%%%%%%%%%%%%%%%%%%%%%%%%%%%%%%%%%%%%%%%%%%%%%%%%%%
\section{Conformal-like symmetries for strictly massless fermions}\label{Sec_hidden}
In this Section, we present our main results, i.e. we present and study new conformal-like symmetries for strictly massless spin-$s \geq 3/2$ fermions on $dS_{4}$. 

%%%%%%%%%%%%%%%%%%%%%%%%%%%%%
\noindent \textbf{Conformal Killing vectors of $\bm{dS_{4}}$.}~For later convenience, let us review the basics concerning the conformal Killing vectors on $dS_{4}$. The five conformal Killing vectors of $dS_{4}$ satisfy 
\begin{align}\label{CKV equation}
  \nabla_{\mu}  V_{\nu} + \nabla_{\nu}  V_{\mu}= g_{\mu \nu}\frac{\nabla^{\alpha}    V_{\alpha}}{2}
\end{align}
with $\nabla^{\alpha}    V_{\alpha} \neq 0$. (The ten dS Killing vectors, $\xi^{\mu}$, satisfy the same equation, but they are divergence-free.) The 15-dimensional Lie algebra generated by the dS Killing vectors and the conformal Killing vectors is isomorphic to $so(4,2)$. The Lie bracket between a dS Killing vector and a conformal Killing vector is equal to a conformal Killing vector, while the Lie bracket between two conformal Killing vectors closes on $so(4,1)$. These facts can be understood from the $so(4,2)$ commutation relations:
    \begin{align}\label{abstract so(4,2) com relnts}
   [M_{A'B'},M_{C'D'} ] = (\eta'_{B'C'} M_{A'D'} +\eta'_{A'
   D'}M_{B'C'})- (A'\leftrightarrow B') ,
\end{align}
 with $A',B',C',D' = -1,0,...,4$, where $M_{A'B'}= -M_{B'A'}$ and $\eta'_{A'B'}=diag(-1,-1,1,1,1,1)$ (with $\eta'_{-1-1}=\eta'_{00}=-1$). The generators $M_{-1A'}$, with $A'=0,...,4$, can be identified with the five conformal Killing vectors of $dS_{4}$, while the generators $M_{A'B'}$, with $A',B'= 0,...,4$, can be identified with the ten dS Killing vectors.

Each of the five conformal Killing vectors of $dS_{4}$, denoted for convenience as $V^{(0)\mu}, V^{(1)\mu},$ $..., V^{(4)\mu}$, can be expressed as a gradient of a scalar function~\footnote{I would like to thank Atsushi Higuchi for pointing this out.}
\begin{align}\label{V=nabla phi}
    V^{(A)}_{\mu}= \nabla_{\mu} \phi_{V^{(A)}}.
\end{align}
 Each of the five scalar functions $\phi_{V^{(A)}}$ ($A=0,1,...,4$) satisfies
\begin{align}
   & \nabla_{\mu} \nabla_{\nu}\phi_{V^{(A)}} =  - g_{\mu \nu}\phi_{V^{(A)}}\label{properties of phi 1} ,
\end{align}
i.e. $\nabla_{\mu} V^{(A)}_{\nu}=-g_{\mu \nu} \phi_{V^{(A)}}$.
 The scalar functions satisfying Eq.~(\ref{properties of phi 1}) can be found by analytically continuing the scalar functions that are annihilated by the operator $\nabla_{\mu} \nabla_{\nu}+ g_{\mu \nu}$ on $S^{4}$~\footnote{It is known that such functions on $S^{4}$ exist~\cite{Allen}. More specifically, they correspond to scalar spherical harmonics on $S^{4}$~\cite{STSHS}.}. If we embed $dS_{4}$ in 5-dimensional Minkowski space as $-(X^{0})^{\, \,  2} + \sum_{j = 1}^{4} (X^{j})^{ \, \, 2}=1$, then the five scalar functions $\phi_{V^{(A)}}$ are $\phi_{V^{(A)}} = X^{A}$ (this equality holds up to a proportionality constant, which we ignore in the present paper). In the case of global coordinates~(\ref{dS_metric}) we have
\begin{align}\label{global coords embedding}
    X^{0}&=\sinh{t}\nonumber \\
   X^{4} &=\cosh{t} \,\cos{\theta_{3}}\nonumber \\
     X^{3} &=\cosh{t} \,\sin{\theta_{3}}\, \cos{\theta_{2}}\,  \nonumber \\
      X^{2} &=\cosh{t}\, \sin{\theta_{3}}\, \sin{\theta_{2}}\,  \cos{\theta_{1}} \nonumber \\
       X^{1} &=\cosh{t}\, \sin{\theta_{3}}\, \sin{\theta_{2}} \, \sin{\theta_{1}} .
\end{align}

Below we will often drop the label `$(A)$' from $V^{(A)\mu}$ and $\phi_{V^{(A)}}$. Thus, from now on, we will denote conformal Killing vectors of $dS_{4}$ as $V^{\mu} = \nabla^{\mu}  \phi_{V}$ or $W^{\mu} = \nabla^{\mu}  \phi_{W}$, unless otherwise stated. 

%%%%%%%%%%%%%%%%%%%%%%%%%%%
    
    \subsection{Conformal-like symmetry transformations}\label{subsec_hidden transf}
    The main new result of the present paper is:
    
    \begin{itemize}
        \item If $\Psi_{\mu_{1}...\mu_{r}}$ is a strictly massless tensor-spinor satisfying Eqs.~(\ref{Dirac_eqn_fermion_dS_sm}) and (\ref{TT_conditions_fermions_dS_sm}), then these equations are also satisfied by $T_{V} \Psi_{\mu_{1}...\mu_{r}}$ defined as
\begin{align}\label{hidden symmetry}
    T_{V}\Psi_{\mu_{1}...\mu_{r}} 
    \equiv & ~\gamma^{5}\Big( V^{\rho}\nabla_{\rho}\Psi_{\mu_{1}...\mu_{r}}+ i\, r\, V^{\rho}\gamma_{\rho} \Psi_{\mu_{1} ... \mu_{r}}  -i\,r\, V^{\rho}\gamma_{(\mu_{1}} \Psi_{\mu_{2}...\mu_{r}) \rho} - \frac{3}{2} \phi_{V} \, \Psi_{\mu_{1}...\mu_{r}}\Big ) \nonumber\\ 
   & -\frac{2r}{2r+1} \left( \nabla_{(\mu_{1}} +\frac{i}{2}\gamma_{(\mu_{1}}\right) \gamma^{5}\Psi_{\mu_{2}...\mu_{r})\rho}V^{\rho},
\end{align}
for any conformal Killing vector $V^{\mu}= \nabla^{\mu} \phi_{V}$. The latter satisfies $\nabla_{\mu} V^{\mu}=-4 \phi_{V}$ [see Eq.~(\ref{properties of phi 1})]. Equation~(\ref{hidden symmetry}) describes the new conformal-like infinitesimal symmetry transformations for strictly massless fermions generated by conformal Killing vectors on $dS_{4}$. 
    \end{itemize}
The differential operator $T_{V}$ maps solutions of Eqs.~(\ref{Dirac_eqn_fermion_dS_sm}) and (\ref{TT_conditions_fermions_dS_sm}) into other solutions, i.e. $T_{V}$ corresponds to a symmetry of these equations.

\noindent \textbf{Note.} The term in the last line of Eq.~(\ref{hidden symmetry}) does \textbf{not} correspond to a restricted gauge transformation~(\ref{onshell_gauge}). This can be understood by observing that the gauge function, $\lambda_{\mu_{2}...\mu_{r}}$, in the restricted gauge transformation~(\ref{onshell_gauge}) satisfies Eq.~(\ref{EOM_for_gauge_functions_Dirac}), while $\gamma^{5}\Psi_{\mu_{2}...\mu_{r}\rho}V^{\rho}$ in the last line of Eq.~(\ref{hidden symmetry}) does not; it satisfies the following equation instead~\footnote{Although the term in the second line of Eq.~(\ref{hidden symmetry}) is not a restricted gauge transformation, it still corresponds to an ``off-shell'' gauge transformation - by ``off-shell'' gauge transformation we mean any gauge transformation that leaves invariant the Lagrangian for strictly massless fermions (the restricted gauge transformations~(\ref{onshell_gauge}) correspond to a special case of the ``off-shell'' transformations). Hermitian and gauge-invariant Lagrangians for strictly massless fermions on $AdS_{4}$ have been constructed in Ref.~\cite{Fang-Fronsdal} (see also Ref.~\cite{Rahman 2}). By analytically continuing $AdS_{4}$ to $dS_{4}$, i.e. by replacing the AdS radius as $\mathcal{R}_{AdS} \rightarrow i \mathcal{R}_{dS}$, where $\mathcal{R}_{dS}$ is the dS radius ($\mathcal{R}_{dS}=1$ in our units), one can extend the Lagrangians for strictly massless fermions on $AdS_{4}$~\cite{Fang-Fronsdal} to non-hermitian Lagrangians on $dS_{4}$. The field equations derived from these non-hermitian Lagrangians on $dS_{4}$ are invariant under ``off-shell'' gauge transformations that have the form $\delta \Psi_{\mu_{1} ... \mu_{r}}  = (\nabla_{(\mu_{1}}+ \frac{i}{2} \gamma_{(\mu_{1}} ) \chi_{\mu_{2} ...\mu_{r})}$, where $\chi_{\mu_{2} ... \mu_{r}}$ is a totally symmetric tensor-spinor with $\gamma^{\mu_{2}} \chi_{\mu_{2} ... \mu_{r}} = 0$. If one specialises to the TT gauge, these field equations reduce to Eqs.~(\ref{Dirac_eqn_fermion_dS_sm}) and (\ref{TT_conditions_fermions_dS_sm}), while the initial ``off-shell'' gauge invariance reduces to the restricted gauge invariance with gauge transformations given by~(\ref{onshell_gauge}).}: $$\slashed{\nabla}\gamma^{5}\Psi_{\mu_{2}...\mu_{r}\rho}V^{\rho} = i r\, \gamma^{5}\Psi_{\mu_{2}...\mu_{r}\rho}V^{\rho}.$$

In order to prove that the conformal-like transformation~(\ref{hidden symmetry}) corresponds to a symmetry we need to show that $T_{V}\Psi_{\mu_{1}...\mu_{r}}$ satisfies the same field equations as $\Psi_{\mu_{1}...\mu_{r}}$, i.e. Eqs.~(\ref{Dirac_eqn_fermion_dS_sm}) and (\ref{TT_conditions_fermions_dS_sm}). It is convenient to define the totally symmetric tensor-spinors $\Delta_{V} \Psi_{\mu_{1}...\mu_{r}}$ and $P_{V} \Psi_{\mu_{1} ... \mu_{r}}$ as
\begin{align}\label{A}
 \Delta_{V} \Psi_{\mu_{1}...\mu_{r}} \equiv ~\gamma^{5}\Big( V^{\rho}\nabla_{\rho}\Psi_{\mu_{1}...\mu_{r}}+ i\, r\, V^{\rho}\gamma_{\rho} \Psi_{\mu_{1} ... \mu_{r}} 
   & -i\,r\, V^{\rho}\gamma_{(\mu_{1}} \Psi_{\mu_{2}...\mu_{r}) \rho} - \frac{3}{2} \phi_{V} \, \Psi_{\mu_{1}...\mu_{r}}\Big )
\end{align}
and
\begin{align}\label{B}
P_{V} \Psi_{\mu_{1}...\mu_{r}}\equiv -\frac{2r}{2r+1} \left( \nabla_{(\mu_{1}} +\frac{i}{2}\gamma_{(\mu_{1}}\right) \gamma^{5}\Psi_{\mu_{2}...\mu_{r})\rho}V^{\rho},
\end{align}
such that 
\begin{align} \label{hidden= A+B}
T_{V}\Psi_{\mu_{1}...\mu_{r}}=\Delta_{V} \Psi_{\mu_{1}...\mu_{r}} +P_{V} \Psi_{\mu_{1}...\mu_{r}}   .
\end{align}
We observe that $\Delta_{V} \Psi_{\mu_{1}...\mu_{r}} $ and $P_{V} \Psi_{\mu_{1}...\mu_{r}} $ have opposite gamma traces
\begin{align}\label{gamma_traces_A_and_B}
    \gamma^{\alpha}\Delta_{V} \Psi_{\alpha \mu_{2}...\mu_{r}} = -\gamma^{\alpha}P_{V} \Psi_{\alpha \mu_{2}...\mu_{r}} = 2 i \gamma^{5}\Psi_{\mu_{2}...\mu_{r}\rho}V^{\rho} .
\end{align}
Thus, the gamma-tracelessness property of the conformal-like transformation~(\ref{hidden symmetry}), $$\gamma^{\alpha}\, T_{V}\Psi_{\alpha \mu_{2}...\mu_{r}}=0,$$ is straightforwardly shown. 

Now let us show that, if $\Psi_{\mu_{1}...\mu_{r}}$ satisfies Eq.~(\ref{Dirac_eqn_fermion_dS_sm}), then so does   $T_{V} \Psi_{\mu_{1}...\mu_{r}}$. In other words, we will show that $T_{V} \Psi_{\mu_{1}...\mu_{r}}$ is an eigenfunction of the Dirac operator with eigenvalue $-i r$. Acting with the Dirac operator on $\Delta_{V}  \Psi_{\mu_{1}...\mu_{r}}$ and $P_{V}\Psi_{\mu_{1}...\mu_{r}}$, we find
\begin{align}\label{Dirac_op on A}
   \left( \slashed{\nabla} + i r \right) \Delta_{V}  \Psi_{\mu_{1}...\mu_{r}}= \,r \left( \nabla_{(\mu_{1}}- \frac{i}{2}\gamma_{(\mu_{1}}   \right) \gamma^{\alpha} \Delta_{V} \Psi_{\mu_{2}...\mu_{r}) \alpha}
\end{align}
and
\begin{align}\label{Dirac_op on B}
   \left( \slashed{\nabla} + i r \right) P_{V}  \Psi_{\mu_{1}...\mu_{r}} = \,r \left( \nabla_{(\mu_{1}}- \frac{i}{2}\gamma_{(\mu_{1}}   \right) \gamma^{\alpha}P_{V} \Psi_{\mu_{2}...\mu_{r}) \alpha},
\end{align}
respectively, where we have used Eq.~(\ref{commutator of derivs}). Adding Eqs.~(\ref{Dirac_op on A}) and~(\ref{Dirac_op on B}) by parts, and making use of Eqs.~(\ref{hidden= A+B}) and (\ref{gamma_traces_A_and_B}), we find
$$  \left( \slashed{\nabla} + ir  \right) T_{V}\Psi_{\mu_{1} ... \mu_{r}}  = 0, $$
as required. Finally, contracting this equation with $\gamma^{\mu_{1}}$, and using the gamma-traceleness property of $T_{V} \Psi_{\mu_{1} ... \mu_{r}}$, we find that $T_{V} \Psi_{\mu_{1} ... \mu_{r}}$ is also divergence-free.
%%%%%%%%%%%%%%%%%%%%%%%%%%%%%%%%%%%%%%%%%%%%%%%%%%%%%%%%%%%%%%%%%%%%%%%%%%%%%%%%%%%%%%%%%%%%%%%%%%%%%%%%%%%%%%%%%%%%%%%%%%%%%%%%%%%%%%%%%%%%%%%%%%%%%%%%%%%%%%%%%%%%%%%%%%%%%%%%%%%%

To conclude, we have proved that the conformal-like transformation $T_{V}\Psi_{\mu_{1} ... \mu_{r}}$ [Eq.~(\ref{hidden symmetry})] satisfies
\begin{align}
   &\left( \slashed{\nabla}+ ir \right)T_{V}\Psi_{ \mu_{1}...\mu_{r}}=0  ,\label{Dirac_eqn_fermion_dS_sm_TV}\\
   & \nabla^{\alpha}\,T_{V}\Psi_{ \alpha \mu_{2}...\mu_{r}}=0, \hspace{4mm}  \gamma^{\alpha}\,T_{V}\Psi_{ \alpha \mu_{2}...\mu_{r}}=0\label{TT_conditions_fermions_dS_sm_TV}
\end{align}
for any conformal Killing vector $V^{\mu}$ and for all spins $s \geq 3/2$. In other words, the operator $T_{V}$~[Eq.~(\ref{hidden symmetry})] is a symmetry of the field equations~(\ref{Dirac_eqn_fermion_dS_sm}) and (\ref{TT_conditions_fermions_dS_sm}) for strictly massless fermions.
%%%%%%%%%%%%%%%%%%%%%%%%%%%%%%%%%%%%%555%%%%%%%%%%%%%%%%%%%%%%%%%%%%%%%%%%%%%%55%%%%%%%%%%%%%%%%%%%%%%%%%%%%%%%%%%%%%%%%
\subsection{Conformal-like \texorpdfstring{$so(4,2)$}{so(4,2)} algebra generated by the dS symmetries and the conformal-like symmetries}\label{subSec_hidden algebra}
In order to understand the structure of the algebra generated by the dS transformations~(\ref{Lie_Lorentz}) and the conformal-like transformations~(\ref{hidden symmetry}) we need to study the corresponding Lie brackets (i.e. commutators). Below, $V^{\mu} = \nabla^{\mu}  \phi_{V}$ and $W^{\mu} = \nabla^{\mu}  \phi_{W}$ denote any two conformal Killing vectors of $dS_{4}$ [see Eq.~(\ref{V=nabla phi})]. 

\noindent \textbf{Commutator between dS and conformal-like transformations.} After a straightforward calculation, the commutator between a dS transformation~(\ref{Lie_Lorentz}) and a conformal-like transformation~(\ref{hidden symmetry}) is found to be
\begin{align}\label{[Lie, hidden]}
    [\mathbb{L}_{\xi} , T_{V}] \Psi_{\mu_{1} ... \mu_{r}}&= \mathbb{L}_{\xi}  T_{V}\Psi_{\mu_{1} ... \mu_{r}}-T_{V}\mathbb{L}_{\xi}\Psi_{\mu_{1} ... \mu_{r}} \nonumber \\ &=T_{[\xi,V]}\Psi_{\mu_{1} ... \mu_{r}},
\end{align}
where $[\xi, V]$ is the Lie bracket between the Killing vector $\xi$ and the conformal Killing vector $V$, i.e. $[\xi, V]^{\mu}= \mathcal{L}_{\xi}V^{\mu}$ ($\mathcal{L}_{\xi}$ is the usual Lie derivative with respect to $\xi$).

\noindent \textbf{Commutator between two conformal-like transformations.} The calculation of the commutator between two conformal-like transformations, $[T_{W}, T_{V}]\Psi_{\mu_{1}...\mu_{r}}$, is quite long. Thus, here we present the final result and we refer the reader to Appendix~\ref{Appe commutator} for some details of the calculation. The result is
\begin{align}\label{[hidden, hidden]}
    [T_{W} , T_{V}] \Psi_{\mu_{1} ... \mu_{r}}&= \mathbb{L}_{[W,V]}\Psi_{\mu_{1} ... \mu_{r}}+\left(\nabla_{(\mu_{1}}+ \frac{i}{2} \gamma_{(\mu_{1}}\right)  L_{ \mu_{2}...\mu_{r})},
\end{align}
where $[W,V]^{\mu}= \mathcal{L}_{W}V^{\mu}= \phi_{W} V^{\mu} - \phi_{V} W^{\mu}  $ is a Killing vector. The second term on the right-hand side of Eq.~(\ref{[hidden, hidden]}) is a restricted gauge transformation of the form~(\ref{onshell_gauge}), where
\begin{align}\label{gauge parametr in [hid,hid]}
    L_{ \mu_{2}...\mu_{r}}=\frac{4r}{(2r+1)^{2}}\left((\nabla^{\lambda}-\frac{i}{2}  \,\gamma^{\lambda})\Psi^{\rho}_{\hspace{1mm}\mu_{2}...\mu_{r}}\,\,\nabla_{\lambda}[W,V]_{\rho}-(r+1) \,\Psi^{\rho}_{\hspace{1mm}\mu_{2}...\mu_{r}} \,[W,V]_{\rho}   \right).
\end{align}
(We have verified that $L_{ \mu_{2}...\mu_{r}}$ satisfies Eqs.~(\ref{EOM_for_gauge_functions_Dirac}) and (\ref{EOM_for_gauge_functions_TT}).)

\noindent \textbf{Structure of the conformal-like algebra}. To conclude, the structure of the conformal-like algebra generated by the ten dS transformations~(\ref{Lie_Lorentz}) and the five conformal-like transformations~(\ref{hidden symmetry}) is determined by the following commutation relations:
\begin{subequations}
\begin{equation}\label{full hidden algebra 1}
   [\mathbb{L}_{\xi} , \mathbb{L}_{\xi'}] \Psi_{\mu_{1} ... \mu_{r}} =\mathbb{L}_{[\xi,\xi']}\Psi_{\mu_{1} ... \mu_{r}},
\end{equation}    
\begin{equation}\label{full hidden algebra 2}
    [\mathbb{L}_{\xi} , T_{V}] \Psi_{\mu_{1} ... \mu_{r}} =T_{[\xi,V]}\Psi_{\mu_{1} ... \mu_{r}},
\end{equation}
\begin{equation}\label{full hidden algebra 3}
     [T_{W} , T_{V}] \Psi_{\mu_{1} ... \mu_{r}}= \mathbb{L}_{[W,V]}\Psi_{\mu_{1} ... \mu_{r}}+\left(\nabla_{(\mu_{1}}+ \frac{i}{2} \gamma_{(\mu_{1}}\right)  L_{ \mu_{2}...\mu_{r})},
\end{equation}
\end{subequations}
where $L_{\mu_{2}...\mu_{r}}$ is given by~(\ref{gauge parametr in [hid,hid]}), $\xi^{\mu}$ and $\xi'^{\mu}$ are any two dS Killing vectors, while $W^{\mu}=\nabla^{\mu}\phi_{W}$ and $V^{\mu}=\nabla^{\mu}\phi_{V}$ are any two conformal Killing vectors. The commutation relations~(\ref{full hidden algebra 1})-(\ref{full hidden algebra 3}) coincide with the $so(4,2)$ commutation relations~(\ref{abstract so(4,2) com relnts}) up to the restricted gauge transformation in Eq.~(\ref{full hidden algebra 3}).

Our results demonstrate that there is a representation of $so(4,2)$ (which closes up to field-dependent gauge transformations) acting on the solution space of Eqs.~(\ref{Dirac_eqn_fermion_dS_sm}) and (\ref{TT_conditions_fermions_dS_sm}). In the following Subsection, we will show that the physical modes, which have been shown to form a direct sum of $so(4,1)$ UIRs (see Section~\ref{Section dS UIR's and modes}), also form a direct sum of $so(4,2)$ UIRs.    

\begin{itemize}
    \item \textbf{Note.} One might think that the closure of the conformal-like algebra up to (field-dependent) gauge transformations is a consequence of the term in the second line of Eq.~(\ref{hidden symmetry}). In order to argue that this is \textbf{not} the case, let us focus on the strictly massless spin-3/2 field and depart from the TT gauge:

Consider the full Rarita-Schwinger (RS) equation for the strictly massless spin-3/2 field (gravitino) on $dS_{4}$~\cite{Freedman}
\begin{align}\label{RS equation}
    \gamma^{\mu \rho \sigma} (\nabla_{\rho} + \frac{i}{2} \gamma_{\rho}) \psi_{\sigma}=0,
\end{align}
where $\gamma^{\mu \rho \sigma}  = \gamma^{[\mu}  
 \gamma^{\rho} \gamma^{\sigma]}$.
This equation is invariant under ``off-shell'' gauge transformations 
\begin{align}\label{offshell gauge spin-3/2}
    \delta \psi_{\mu}= (\nabla_{\mu} + \frac{i}{2}  \gamma_{\mu}) \epsilon,
\end{align}
where $\epsilon$ is an arbitrary spinor.
If we choose to work in the TT gauge, then the RS equation reduces to Eqs.~(\ref{Dirac_eqn_fermion_dS_sm}) and (\ref{TT_conditions_fermions_dS_sm}), which have a smaller gauge invariance corresponding to restricted gauge transformations~(\ref{onshell_gauge}). After a straightforward calculation, we find that the RS equation~(\ref{RS equation}) enjoys the conformal-like symmetry\footnote{The expression in Eq.~(\ref{hidden in non-TT 3/2}) corresponds just to the first part~(\ref{A}) of the conformal-like transformation in the TT gauge [Eq.~(\ref{hidden symmetry})].}
\begin{align}\label{hidden in non-TT 3/2}
 \Delta_{V} \psi_{\mu} = ~\gamma^{5}\Big( V^{\rho}\nabla_{\rho}\psi_{\mu}+ i\, \, V^{\rho}\gamma_{\rho} \psi_{\mu} 
   & -i\, V^{\rho}\gamma_{\mu} \psi_{ \rho} - \frac{3}{2} \phi_{V} \, \psi_{\mu}\Big ).
\end{align}
 In other words, if $\psi_{\mu}$ satisfies the RS equation, then so does $\Delta_{V} \psi_{\mu}$. Because of the ``off-shell'' gauge symmetry~(\ref{offshell gauge spin-3/2}), Eq.~(\ref{hidden in non-TT 3/2}) does not include a part corresponding to the second line of Eq.~(\ref{hidden symmetry}). Then, the commutator between two conformal-like transformations~(\ref{hidden in non-TT 3/2}) is found to be
\begin{equation}
     [\Delta_{W} , \Delta_{V}] \psi_{\mu}= \mathbb{L}_{[W,V]}\psi_{\mu}-2i\,\left(\nabla_{\mu}+ \frac{i}{2} \gamma_{\mu}\right)  \gamma^{\lambda} \psi^{\rho}\,\nabla_{\lambda}[W,V]_{\rho},
\end{equation}
 where we notice the appearance of an ``off-shell'' gauge transformation (which is \textbf{not} a restricted gauge transformation~(\ref{onshell_gauge})) on the right-hand side. The rest of the structure of the symmetry algebra is determined by the same commutation relations as in Eqs.~(\ref{full hidden algebra 1}) and (\ref{full hidden algebra 2}) (with $T_{V}$ replaced by $\Delta_{V}$). 
 
 \noindent   \textbf{Conclusion.} As in the TT gauge, the full RS equation~(\ref{RS equation}) enjoys a conformal-like $so(4,2)$ symmetry and the algebra closes up to ``off-shell'' gauge transformations~(\ref{offshell gauge spin-3/2}) that do not correspond to restricted gauge transformations~(\ref{onshell_gauge}). However, in the TT case~(\ref{full hidden algebra 3}), the algebra closes up to restricted gauge transformations.
    
\end{itemize}

%%%%%%%%%%%%%%%%%%ga
\section{The physical modes also form UIRs of the conformal-like algebra} \label{sect_modes n uir's of so(4,2)}
In this Section, we show that the `positive frequency' physical modes (\ref{physmodes_negative_spin_r+1/2_dS4}) and (\ref{physmodes_positive_spin_r+1/2_dS4}) of the strictly massless spin-$s \geq 3/2$ fermionic theories form UIRs of the conformal-like $so(4,2)$ algebra. To be specific:
\begin{itemize}
    \item The \textbf{irreducibility} of the $so(4,2)$ representations will be demonstrated by showing that the physical modes with fixed helicity transform among themselves under the infinitesimal conformal-like transformations~(\ref{hidden symmetry}). In particular, the physical modes with helicity $+s$~[Eq.~(\ref{physmodes_positive_spin_r+1/2_dS4})], and the ones with helicity $-s$~[Eq.~(\ref{physmodes_negative_spin_r+1/2_dS4})], will be shown to separately form irreducible representations of $so(4,2)$. (Recall that we have already shown that these modes form a direct sum of UIRs of the dS algebra $so(4,1)$ - see Section~\ref{Section dS UIR's and modes}.)

    \item As for showing the \textbf{unitarity} of the two aforementioned irreducible $so(4,2)$ representations, we work as follows. First, we recall from Section~\ref{Section dS UIR's and modes} that the physical modes with helicity $\mp s$ form a $so(4,1)$ UIR with dS invariant and positive definite scalar product given by $(\pm 1) \times$(\ref{axial_scalar prod}). Then, since a positive definite and $so(4,1)$-invariant scalar product is known, it is sufficient to show that this scalar product is also invariant under the conformal-like symmetries~(\ref{hidden symmetry}).
\end{itemize}

%%%%%%%%%%%%%%%%%%%%%%%%%%%%%%%%%%%%%%%%%%%%%%%%
\subsection{Conformal-like transformations of physical modes and irreducibility of \texorpdfstring{$so(4,2)$}{so(4,1)} representations}~\label{Subsec_hidden_trnsfrmns of phys modes}
Let us start with the simple observation that, according to Eq.~(\ref{abstract so(4,2) com relnts}), the Lie bracket between a conformal Killing vector and a dS Killing vector is equal to a conformal Killing vector. Similarly, the commutator $[\mathbb{L}_{\xi}, T_{V}] \Psi_{\mu_{1}...\mu_{r}}$ in Eq.~(\ref{full hidden algebra 2}) is equal to a conformal-like symmetry transformation. Thus, as the $so(4,1)$ representation-theoretic properties of the physical modes are known (see Section~\ref{Section dS UIR's and modes}), it is sufficient to study just one of the five conformal-like transformations~(\ref{hidden symmetry}) for our physical modes. Then, the transformation properties of the physical modes under the rest of the conformal-like transformations can be found using the commutation relations~(\ref{full hidden algebra 2}).

Let us now choose to work with the conformal Killing vector $V^{(0)\mu}$ [Eq.~(\ref{V=nabla phi})] given by
\begin{align}
    V^{(0)}_{\mu}= \nabla_{\mu} \sinh{t},
\end{align}
i.e. $(V^{(0)}_{t},V^{(0)}_{\theta_{3}}, V^{(0)}_{\theta_{2}}, V^{(0)}_{\theta_{1}}) = (\cosh{t},0,0,0)$. The conformal-like transformation~(\ref{hidden symmetry}) generated by $V^{(0)\mu}$ is expressed as
\begin{align}
    T_{V^{(0)}}\Psi_{\mu_{1}...\mu_{r}}&=-\gamma^{5}\,\cosh{t} \nonumber \\ & \times\Bigg( \frac{\partial}{\partial t}  +\left(-r+\frac{3}{2}\right)\tanh{t} -ir \,\gamma^{t}\Bigg)\Psi_{\mu_{1}...\mu_{r}}.
\end{align}
Specialising to the physical modes (\ref{physmodes_negative_spin_r+1/2_dS4}) and (\ref{physmodes_positive_spin_r+1/2_dS4}), and making use of Eqs.~(\ref{even_gammas D=4}), (\ref{psia=-r_to_phi_sphere_analcont}) and (\ref{phia=-r_to_psi_sphere_analcont}), we readily find
\begin{align}
    T_{V^{(0)}}\Psi^{(phys ,\,- \ell; \,m;k)}_{\mu_{1}...\mu_{r}}&=+ i \left(\ell +\frac{3}{2}   \right)\Psi^{(phys ,\,-\ell; \,m;k)}_{\mu_{1}...\mu_{r}}
\end{align}
and
\begin{align}
    T_{V^{(0)}}\Psi^{(phys ,\,+ \ell; \,m;k)}_{\mu_{1}...\mu_{r}}&=- i \left(\ell +\frac{3}{2}   \right)\Psi^{(phys ,\,+\ell; \,m;k)}_{\mu_{1}...\mu_{r}}.
\end{align}
From these equations (and from the discussion at the beginning of this Subsection), it follows that the two sets of modes, $\{ \Psi^{(phys ,\,+ \ell; \,m;k)}_{\mu_{1}...\mu_{r}}\}$ and $\{ \Psi^{(phys ,\,- \ell; \,m;k)}_{\mu_{1}...\mu_{r}}\}$, separately form irreducible representations of the conformal-like $so(4,2)$ algebra.
%%%%%%%%%%%%%%%%%%%%%%%%%%%%%%%%%%%%%%%%%%%%%%%%%%%%%%%%%%%%%%%%%%%%%%%%%%%%%%%%%%%%%%%%%%%%%%%%%%%%%%%%%%%%%%%%%%%%%%%%%%%%%
\subsection{\texorpdfstring{$so(4,2)$}{so(4,1)}-invariant scalar product and unitarity}\label{subsec_so(4,2) scalar prod and unitarity}

In the previous Subsection, we showed that the physical modes form a direct sum of irreducible representations of the conformal-like algebra. The only remaining step for showing that this is a direct sum of $so(4,2)$ UIRs is to ensure the existence of a $so(4,2)$-invariant and positive definite scalar product. 

Let us show that the dS invariant scalar product~(\ref{axial_scalar prod}) is also invariant under the conformal-like symmetries~(\ref{hidden symmetry}) - and, thus, under the whole conformal-like $so(4,2)$ algebra (recall that this scalar product is also invariant under restricted gauge transformations). Let $\Psi^{(1)}_{\mu_{1}...\mu_{r}}$ and $\Psi^{(2)}_{\mu_{1}...\mu_{r}}$ be any two solutions of Eqs.~(\ref{Dirac_eqn_fermion_dS_sm}) and (\ref{TT_conditions_fermions_dS_sm}). We consider the change $$\delta_{V}J^{\mu}(\Psi^{(1)}, \Psi^{(2)})=J^{\mu}(T_{V}\Psi^{(1)}, \Psi^{(2)}) + J^{\mu}(\Psi^{(1)}, T_{V}\Psi^{(2)})$$ of the vector current~(\ref{axial_current}) under the conformal-like transformations~(\ref{hidden symmetry}).
After a straightforward calculation, we find
\begin{align}
    \delta_{V}J^{\mu}(\Psi^{(1)}, \Psi^{(2)})=-i\,\nabla_{\lambda}E^{\lambda \mu},
\end{align}
where $E^{\lambda  \mu}$ is an anti-symmetric tensor given by:
\begin{align}
    &\frac{1}{2}E^{\lambda  \mu}\nonumber \\
    &=-\overline{\Psi}^{(1)}_{\nu_{1}...\nu_{r}}  V^{[\lambda}\gamma^{\mu]}  \Psi^{(2)\nu_{1}...\nu_{r}}+\frac{2r}{2r+1} V^{\rho}\left( \overline{\Psi}^{(1)}_{\nu_{2}...\nu_{r}\rho}  \gamma^{[\mu}\Psi^{(2)\lambda ]\nu_{2}...\nu_{r}}+  \overline{\Psi}^{(1)\nu_{2}...\nu_{r}[\lambda} \gamma^{\mu]}\Psi^{(2)}_ {\nu_{2}...\nu_{r} \rho}  \right).
\end{align}
This ensures that the dS invariant scalar product~(\ref{axial_scalar prod}) is also invariant under infinitesimal conformal-like transformations, as
$$\delta_{V}\braket{ \Psi^{(1)}| \Psi^{(2)}}= \int_{S^{3}} \sqrt{-{g}} \,d\bm{\theta_{3}}\,\delta_{V}J^{0}(\Psi^{(1)}, \Psi^{(2)})  =0.$$

Based on the discussions in the previous paragraph (and in the previous Subsection), we conclude the following:
\begin{itemize}
    \item The set of physical modes with helicity $+s$, $\{ {\Psi}^{\left(phys ,\,+ \ell; m; k\right)}_{\mu_{1}...\mu_{r}} \}$, forms a UIR of $so(4,2)$ with positive definite norm given by the negative of Eq.~(\ref{norm of physical modes}) (with $\sigma=+$).

    \item The set of physical modes with helicity $-s$, $\{ {\Psi}^{\left(phys ,\,- \ell; m; k\right)}_{\mu_{1}...\mu_{r}} \}$, forms a UIR of $so(4,2)$ with positive definite norm given by Eq.~(\ref{norm of physical modes}) (with $\sigma=-$).
\end{itemize}

%%%%%%%%%%%%%%%%%%%%%%%%%%%%%%%%%%%%%%%%%%%%%%%%%%%%%%%%%%%%%%%%%%%%%%%%%%%%%%%%%%%%%%%%%%%%%%%%%%%%%%%%%%%%%%%%%%
\section{Conformal-like transformations of field strength tensor-spinors}\label{sec_field-strength}
In order to gain some insight into the interpretation of the conformal-like transformations $T_{V} \Psi_{\mu_{1}...\mu_{r}}$~(\ref{hidden symmetry}), we study the corresponding transformations of the field strength tensor-spinors (i.e. curvatures). In particular, we study the transformations of the spin-$s=3/2,5/2$ field strengths explicitly, while in the spin-$s \geq 7/2$ cases we make a conjecture for the expressions of the transformations.
%%%%%%%%%%%%%%%%%%%%%%%%%%%%%%%%%%%%%%%%%%%%%%%%%%%%%%%%%%%%%%%%%%%%%%%%%%%%%%%%%%%%%%%%%%%%%%%%%%%%%%%%%%%%%%%%%%%%%%%%%%%%%%%%%%%%%%%%%%%%%%
\subsection{Spin-3/2 field strength tensor-spinor}\label{Subsec induced spin-3/2}

 The field strength tensor-spinor for the strictly massless spin-3/2 field is
\begin{align} \label{field-strength 3/2}
    \mathbb{F}_{\mu_{1} \nu_{1}}=-  \mathbb{F}_{\nu_{1} \mu_{1}} = \left( \nabla_{[ \mu_{1}} + \frac{i}{2} \gamma_{[\mu_{1}}  \right)   \Psi_{\nu_{1}]}.
\end{align}
For later convenience, we will denote this as  $\mathbb{F}_{\mu_{1} \nu_{1}}(\Psi)$.
The field strength $ \mathbb{F}_{\mu_{1} \nu_{1}}(\Psi)$ is invariant under not only restricted gauge transformations~(\ref{onshell_gauge}) but also ``off-shell'' gauge transformations~(\ref{offshell gauge spin-3/2}).

\noindent \textbf{Useful properties.} Let us discuss some of the properties of $\mathbb{F}_{\mu_{1} \nu_{1}}(\Psi)$ that will be useful in studying its conformal-like transformation. Using the field equations~(\ref{Dirac_eqn_fermion_dS_sm}) and (\ref{TT_conditions_fermions_dS_sm}) for $\Psi_{\nu}$, we find 
\begin{align}
    \gamma^{\mu_{1}}\mathbb{F}_{\mu_{1} \nu_{1}}(\Psi)= \nabla^{\mu_{1}}\mathbb{F}_{\mu_{1} \nu_{1}}(\Psi)=0.
\end{align}
The dual field strength tensor-spinor is defined as
\begin{align}\label{define_dual_spin3/2}
  ^{*} \mathbb{F}_{\mu_{1} \nu_{1}} (\Psi) \equiv \frac{1}{2}\epsilon_{\mu_{1} \nu_{1}}^{\hspace{6mm}\kappa \lambda} \,  \mathbb{F}_{\kappa \lambda }(\Psi).
\end{align}
Expressing $\epsilon_{\mu_{1} \nu_{1}}^{\hspace{6mm}\kappa \lambda}$ in Eq.~(\ref{define_dual_spin3/2}) in terms of gamma matrices [see Eq.~(\ref{def_gamma5})], and using the gamma-tracelessness of $ \mathbb{F}_{\mu_{1} \nu_{1}}(\Psi)$, we find
\begin{align}\label{dual_property_spin3/2}
  ^{*} \mathbb{F}_{\mu_{1} \nu_{1}}(\Psi) = - i \gamma^{5}  \mathbb{F}_{\mu_{1} \nu_{1}}(\Psi).
\end{align}
Also, a straightforward calculation shows that the following identity holds:
\begin{align}
    \nabla_{[\rho}\mathbb{F}_{\mu_{1} \nu_{1}]}(\Psi) + \frac{i}{2}\gamma_{[\rho}  \mathbb{F}_{\mu_{1} \nu_{1}]}(\Psi)=0.
\end{align}
It is easy to show that each of the two terms in this equation is zero by observing that\footnote{\textbf{Proof of Eq.~(\ref{Bianchi_for_dual_3/2}).} In order to prove Eq.~(\ref{Bianchi_for_dual_3/2}), we contract $\nabla_{[\rho}\,^{*}\mathbb{F}_{\mu_{1} \nu_{1}]}(\Psi)$ with $\epsilon_{\alpha \beta}^{\hspace{4mm} \mu_{1} \nu_{1}}$ and we use the definition~(\ref{define_dual_spin3/2}) of the dual field strength. Then, using well-known identities for $\epsilon_{\alpha \beta}^{\hspace{4mm} \mu_{1} \nu_{1}}$, while also using the divergence-freedom of the field strength, we can show that $\epsilon_{\alpha \beta}^{\hspace{4mm} \sigma \kappa}\,\nabla_{[\rho}\,^{*}\mathbb{F}_{\sigma \kappa]} (\Psi)=0.$
Finally, contracting this equation with $\epsilon_{\mu_{1} \nu_{1}}^{\hspace{6mm} \alpha \beta}$, we arrive at Eq.~(\ref{Bianchi_for_dual_3/2}). \textbf{End of proof.}}
\begin{align}\label{Bianchi_for_dual_3/2}
   \nabla_{[\rho}\,^{*}\mathbb{F}_{\mu_{1} \nu_{1}]}(\Psi) =0. 
\end{align}
It immediately follows from  Eqs.~(\ref{dual_property_spin3/2})-(\ref{Bianchi_for_dual_3/2}) that
\begin{align}\label{Bianchies_for_normal_3/2}
   \nabla_{[\rho}\mathbb{F}_{\mu_{1} \nu_{1}]} (\Psi)=\gamma_{[\rho}\mathbb{F}_{\mu_{1} \nu_{1}]}(\Psi) =0. 
\end{align}
%%%%%%%%%%%%%%%%%%%%%%%%%%%%%%%%%%%%%%%%%%%%%%%%%%%%%%%%%%%%%%%%%%%%%%%%%%%%%%%%%%%%%%%%%%%%%%%%%

\noindent \textbf{Conformal-like transformation.} After a straightforward calculation, the conformal-like transformation of the field strength, $\mathbb{F}_{\mu_{1} \nu_{1}}(T_{V}\Psi)$, is expressed as
\begin{align}
    \mathbb{F}_{\mu_{1} \nu_{1}}(T_{V}\Psi)&=  \mathbb{F}_{\mu_{1} \nu_{1}}(\Delta_{V}\Psi)\nonumber\\
    &=\gamma^{5}\left( V^{\rho}\nabla_{\rho}-\frac{5}{2}\phi_{V}   \right) \mathbb{F}_{\mu_{1} \nu_{1}}(\Psi)+3\,i\gamma^{5}\,V^{\rho} \,\gamma_{[\rho}\mathbb{F}_{\mu_{1} \nu_{1}]}(\Psi),
\end{align}
where in the first line we have used $T_{V} \Psi_{\mu}=\Delta_{V} \Psi_{\mu} +P_{V} \Psi_{\mu}$ [see Eq.~(\ref{hidden= A+B})] and $\mathbb{F}_{\mu_{1} \nu_{1}}(P_{V}\Psi)=0$ (the latter follows from the gauge-invariance of the field strength). Then, using Eq.~(\ref{Bianchies_for_normal_3/2}), we find
\begin{align}
    \mathbb{F}_{\mu_{1} \nu_{1}}(T_{V}\Psi)=\gamma^{5}\left( V^{\rho}\nabla_{\rho}-\frac{5}{2}\phi_{V}   \right) \mathbb{F}_{\mu_{1} \nu_{1}}(\Psi),
\end{align}
or equivalently
\begin{align}\label{induced final 3/2}
    \mathbb{F}_{\mu_{1} \nu_{1}}(T_{V}\Psi)=\gamma^{5}\left(\mathbb{L}_{V}+\frac{\nabla_{\kappa} V^{\kappa}}{8}\right) \mathbb{F}_{\mu_{1} \nu_{1}}(\Psi),
\end{align}
where $\mathbb{L}_{V}$ is the Lie-Lorentz derivative~(\ref{Lie_Lorentz}) with respect to the conformal Killing vector $V$~(\ref{V=nabla phi})\footnote{The infinitesimal Lorentz transformation term $\nabla_{\alpha} V_{\beta} \gamma^{\alpha \beta}/4$ in the Lie-Lorentz derivative $\mathbb{L}_{V}$ in Eq.~(\ref{induced final 3/2}) vanishes because, according to Eq.~(\ref{V=nabla phi}), $\nabla_{[\alpha} V_{\beta]}=0$.}.

\noindent \textbf{Conclusion.} The expression~(\ref{induced final 3/2}) makes clear that the conformal-like transformation of the spin-3/2 field strength tensor-spinor corresponds to the product of two transformations: an infinitesimal axial rotation (i.e. multiplication with $\gamma^{5}$) times an infinitesimal conformal transformation (i.e. Lie-Lorentz derivative plus a conformal weight term).

%%%%%%%%%%%%%%%%%%%%%%%%%%%%5555%%%%%%%%%%%%%%%%%%%%%%%%%%%%%%%%%%%%%%%%%%%%%%%%%%%%%%%
\subsection{Spin-5/2 field strength tensor-spinor}\label{Subsec induced spin-5/2}

 The field strength tensor-spinor for the strictly massless spin-5/2 field is a rank-4 tensor-spinor given by
\begin{align}\label{field-strength 5/2}
    \mathbb{F}_{\mu_{1} \nu_{1} \mu_{2} \nu_{2}  }&(\Psi)\nonumber\\
    =&\frac{1}{2} \left( \nabla_{ \mu_{2}}\nabla_{[ \mu_{1}} + \frac{3}{4} g_{\mu_{2} 
 [\mu_{1}}-\frac{1}{4}\gamma_{\mu_{2} [\mu_{1}} +\frac{i}{2}\nabla_{\mu_{2}}\gamma_{[\mu_{1}}+\frac{i}{2}\gamma_{\mu_{2}}  \nabla_{[\mu_{1}} \right)   \Psi_{\nu_{1}]\nu_{2}}-(\mu_{2}   \leftrightarrow  \nu_{2}).
\end{align}
This is symmetric under the exchange of pairs of indices
\begin{align}
  \mathbb{F}_{\mu_{2} \nu_{2} \mu_{1} \nu_{1}  }(\Psi) =     \mathbb{F}_{\mu_{1} \nu_{1} \mu_{2} \nu_{2}  }(\Psi).
\end{align}
It is also anti-symmetric in its first two and last two indices
\begin{align}
 \mathbb{F}_{\mu_{1} \nu_{1} \mu_{2} \nu_{2}  } (\Psi)  =\mathbb{F}_{[\mu_{1} \nu_{1}] \mu_{2} \nu_{2}  }(\Psi)=\mathbb{F}_{\mu_{1} \nu_{1} [\mu_{2} \nu_{2}]  }(\Psi),
\end{align}
and satisfies the identity
\begin{align}
 \mathbb{F}_{\mu \alpha \beta \gamma } (\Psi)  + \mathbb{F}_{\mu \gamma   \alpha \beta } (\Psi) +\mathbb{F}_{\mu \beta \gamma  \alpha } (\Psi) =0.
\end{align}
As in the spin-3/2 case, the field strength is invariant under not only restricted gauge transformations~(\ref{onshell_gauge}) but also gauge transformations of the following form:
\begin{align}
    \delta \Psi_{\mu \nu } =\left( \nabla_{(\mu}  +\frac{i}{2} \gamma_{(\mu} \right)\epsilon_{\nu)}
\end{align}
(i.e. $ \mathbb{F}_{\mu_{1} \nu_{1} \mu_{2} \nu_{2}  } (\delta \Psi)=0$), where $\epsilon_{\nu}$ is an arbitrary vector-spinor.

Working as in the spin-3/2 case, we can show that the spin-5/2 field strength~(\ref{field-strength 5/2}) is gamma-traceless and divergence-free with respect to all of its indices, and it also satisfies the identities
\begin{align}\label{Bianchies_for_normal_5/2}
\nabla_{[\rho}\mathbb{F}_{\mu_{1} \nu_{1}] \mu_{2} \nu_{2}} (\Psi)=\gamma_{[\rho}\mathbb{F}_{\mu_{1} \nu_{1}] \mu_{2} \nu_{2}  }(\Psi) =0. 
\end{align}
\noindent \textbf{Conformal-like transformation.} Let us find the conformal-like transformation of the field strength, $\mathbb{F}_{\mu_{1} \nu_{1} \mu_{2}  \nu_{2}}(T_{V}\Psi)$. The calculation is similar to the spin-3/2 case, but quite longer. The result is 
\begin{align}
    \mathbb{F}_{\mu_{1} \nu_{1} \mu_{2}   \nu_{2}}(T_{V}\Psi)=\gamma^{5}\left( V^{\rho}\nabla_{\rho}-\frac{7}{2}\phi_{V}   \right) \mathbb{F}_{\mu_{1} \nu_{1} \mu_{2}   \nu_{2}}(\Psi),
\end{align}
or equivalently
\begin{align}\label{induced final 5/2}
    \mathbb{F}_{\mu_{1} \nu_{1} \mu_{2} \nu_{2}  }(T_{V}\Psi)=\gamma^{5}\left(\mathbb{L}_{V}-\frac{\nabla_{\kappa} V^{\kappa}}{8}\right) \mathbb{F}_{\mu_{1} \nu_{1} \mu_{2} \nu_{2}  }(\Psi).
\end{align}

\noindent \textbf{Conclusion.} As in the spin-3/2 case~(\ref{induced final 3/2}), the expression~(\ref{induced final 5/2}) makes clear that the conformal-like transformation of the spin-5/2 field strength corresponds to the product: infinitesimal axial rotation times infinitesimal conformal transformation.

%%%%%%%%%%%%%%%%%%%%%%%%%%%%%%%%%%%%%%%%%%%%%%%%%%%%%%%%%%%%%%%%%5%%%%%%%%%%%%%%%%%%%%%%%%%%%%%%%%%%%%%%%%%%%%%%%%%%%%%%%%%%%%%%%%%%%%%%%%%%%%%%%%%%%%%%
\subsection{A conjecture for the spin-\texorpdfstring{$(r+1/2) \geq 7/2$}{7/2} field strength tensor-spinors}\label{Subsec induced spin->=7/2}

%%%%%%%%%%%%%%%%%%%%%%%%%%%%%%%%%%%%%%%%%%%%%%%%%%%%%%%%%%%%%%%%%%%%%%%%%%%%%%%%%%%%%%%%%%%%%%%%%%%%%%%%%%%%%%%%
(Here we do not present explicit expressions for the field strength tensor-spinors $\mathbb{F}_{\mu_{1}\nu_{1}....\mu_{r} \nu_{r}}(\Psi)$ of the strictly massless spin-$(r+1/2) \geq 7/2$ fields.) We define the field strength $\mathbb{F}_{\mu_{1}\nu_{1}....\mu_{r} \nu_{r}}(\Psi)$ as the gauge-invariant rank-$2r$ tensor-spinor that satisfies 
\begin{align}\label{div gam trace field strength conj}
    \gamma^{\mu_{1}}\mathbb{F}_{\mu_{1} \nu_{1}...\mu_{r}   \nu_{r}}(\Psi)= \nabla^{\mu_{1}}\mathbb{F}_{\mu_{1} \nu_{1}...\mu_{r}\nu_{r}}(\Psi)=0,
\end{align}
and it is also anti-symmetric under the exchange of the indices $\mu_{l} \leftrightarrow  \nu_{l}$ for $l=1,...,r$. It is also symmetric under the exchange of any two pairs of indices as in the following example:
\begin{align}
   \mathbb{F}_{\textcolor{red}{\mu_{1}\nu_{1}} \mu_{2} \nu_{2}....\mu_{r} \nu_{r}}(\Psi)= \mathbb{F}_{\mu_{2}\nu_{2} \textcolor{red}{\mu_{1} \nu_{1}}....\mu_{r} \nu_{r}}(\Psi) = \mathbb{F}_{\mu_{r}\nu_{r} \mu_{2} \nu_{2}....\mu_{r-1} \nu_{r-1}   \textcolor{red}{\mu_{1} \nu_{1}}} \hspace{4mm}\text{and so forth,}  
\end{align}
while it also satisfies the identities
\begin{align}
  \mathbb{F}_{{[\mu_{1}\nu_{1}} \mu_{2}] \nu_{2}....\mu_{r} \nu_{r}}(\Psi)=0  
\end{align}
and
\begin{align} \label{alg bianchi conjecture}
     \nabla_{[\rho}\mathbb{F}_{{\mu_{1}\nu_{1}}] \mu_{2} \nu_{2}....\mu_{r} \nu_{r}}(\Psi)=   \gamma_{[\rho}\mathbb{F}_{{\mu_{1}\nu_{1}}] \mu_{2} \nu_{2}....\mu_{r} \nu_{r}}(\Psi)   =0 . 
\end{align}

\noindent \textbf{Conjecture.} The conformal-like transformation of the spin-$(r+1/2) \geq 7/2$ field strength tensor-spinor is given by
\begin{align}
    \mathbb{F}_{\mu_{1} \nu_{1} ...\mu_{r}   \nu_{r}}(T_{V}\Psi)=\gamma^{5}\left( V^{\rho}\nabla_{\rho}-\left(r+\frac{3}{2}\right)\phi_{V}   \right) \mathbb{F}_{\mu_{1} \nu_{1}... \mu_{r}   \nu_{r}}(\Psi),
\end{align}
or equivalently
\begin{align}\label{induced final r+1/2 conj}
    \mathbb{F}_{\mu_{1} \nu_{1}... \mu_{r} \nu_{r}  }(T_{V}\Psi)=\gamma^{5}\left(\mathbb{L}_{V}- (2r-3)\frac{\nabla_{\kappa} V^{\kappa}}{8}\right) \mathbb{F}_{\mu_{1} \nu_{1}... \mu_{r} \nu_{r}  }(\Psi).
\end{align}
This conjecture has been verified for $r=1$ in Subsection~\ref{Subsec induced spin-3/2} and for $r=2$ in Subsection~\ref{Subsec induced spin-5/2}. Our conjecture is further justified by observing that $\mathbb{F}_{\mu_{1} \nu_{1}... \mu_{r} \nu_{r}  }(T_{V}\Psi)$ [Eq.~(\ref{induced final r+1/2 conj})] satisfies Eqs.~(\ref{div gam trace field strength conj})-(\ref{alg bianchi conjecture}).

%%%%%%%%%%%%%%%%%%%%%%%%%%%%%%%%%%%%%%%%%%%%%%%%%%%%%%%%%%%%%%%%%%%%%%%%%%%%%%%%%%5
\section{Summary and Discussions}
In this paper, we uncovered new conformal-like symmetries~(\ref{hidden symmetry}) for the field equations [(\ref{Dirac_eqn_fermion_dS_sm}) and~(\ref{TT_conditions_fermions_dS_sm})] of strictly massless fermionic potentials of spin $s \geq 3/2$ on $dS_{4}$. The associated symmetry algebra closes on $so(4,2)$ up to gauge transformations [see Eqs.~(\ref{full hidden algebra 1})-(\ref{full hidden algebra 3})]. We also showed that the physical (positive frequency) mode solutions~(\ref{physmodes_negative_spin_r+1/2_dS4}) and (\ref{physmodes_positive_spin_r+1/2_dS4}) form a direct sum of UIRs of the conformal-like $so(4,2)$ algebra. As for the interpretation of the conformal-like symmetries, we found that, at the level of the field strength tensor-spinors, each conformal-like transformation is expressed as a product of two transformations: an infinitesimal axial rotation and an infinitesimal conformal transformation (this was shown explicitly for the spin-$s=3/2, 5/2$ cases and conjectured for the cases with $s \geq 7/2$ - see Section~\ref{sec_field-strength}). 

   $$\textbf{Flat-space limit}$$ 
   
\noindent Let us discuss the flat-space limit of the conformal-like symmetries (i.e. the limit of zero cosmological constant). We will start by observing that the flat-space limit of the five conformal Killing vectors~(\ref{V=nabla phi}) of $dS_{4}$ gives rise to the four translation Killing vectors and the generator of dilatations of Minkowski spacetime (rather than the five conformal Killing vectors - i.e. four special conformal transformations and one dilatation - of Minkowski spacetime, as one might expect). {Using this observation, we will show that the flat-space limits of the conformal-like symmetries~(\ref{hidden symmetry}) are `trivial', in the sense that they correspond to known symmetries of the flat-space theories: usual infinitesimal spacetime translations and scale transformations. (At the level of the potentials, massless higher-spin fields on Minkowski spacetime do not enjoy the full $so(4,2)$ symmetry of infinitesimal conformal transformations~\cite{Conf_Bekaert}. They are only invariant under usual $iso(3,1)$ isometries and scale transformations. However, massless higher-spin fields enjoy the full $so(4,2)$ symmetry at the level of the gauge-invariant field strengths. See Ref.~\cite{Conf_Bekaert} and references therein for group-theoretic discussions. Also, note that the 2-point function for the gauge-invariant field strength of the massless spin-3/2 field on Minkowski spacetime has the expected form for conformal primaries~\cite{RS_sphere}.)} 

{To take the flat-space limit, let us recover the dS radius, $\mathcal{R}_{dS}$, such that~(\ref{dS_metric}) is written as $$ds^{2} ={\mathcal{R}_{dS}^{2}}\Bigg(-dt^{2} +\cosh^{2}{t} \left[d\theta^{2}_{3} + \sin^{2}\theta_{3} \left( d\theta_{2}^{2} + \sin^{2}{\theta_{2}} \,d\theta_{1}^{2}  \right) \right]  \Bigg),$$
and Eqs.~(\ref{Dirac_eqn_fermion_dS_sm}) and (\ref{TT_conditions_fermions_dS_sm}) as
\begin{align*}
   &\left( \slashed{\nabla}+ \frac{ir}{\mathcal{R}_{dS}} \right)\Psi_{ \mu_{1}...\mu_{r}}=0 , \\
   & \nabla^{\alpha}\Psi_{ \alpha \mu_{2}...\mu_{r}}=0, \hspace{4mm}  \gamma^{\alpha}\Psi_{ \alpha \mu_{2}...\mu_{r}}=0.
\end{align*}
{Also, recovering $\mathcal{R}_{dS}$,  the conformal-like transformations~(\ref{hidden symmetry}) are written as
\begin{align*}
    T_{V}\Psi_{\mu_{1}...\mu_{r}} 
    \equiv & ~\gamma^{5}\Big( V^{\rho}\nabla_{\rho}\Psi_{\mu_{1}...\mu_{r}}+ \frac{i\, r}{\mathcal{R}_{dS}}\, V^{\rho}\gamma_{\rho} \Psi_{\mu_{1} ... \mu_{r}}  -\frac{i\,r}{\mathcal{R}_{dS}} V^{\rho}\gamma_{(\mu_{1}} \Psi_{\mu_{2}...\mu_{r}) \rho} + \frac{3}{8} \nabla^{\alpha}{V}_{\alpha} \, \Psi_{\mu_{1}...\mu_{r}}\Big ) \nonumber\\ 
   & -\frac{2r}{2r+1} \left( \nabla_{(\mu_{1}} +\frac{i}{2 \, \mathcal{R}_{dS}}\gamma_{(\mu_{1}}\right) \gamma^{5}\Psi_{\mu_{2}...\mu_{r})\rho}V^{\rho}.
\end{align*}}
{Now let us define $t \equiv  T/ \mathcal{R}_{dS}$ and $\theta_{3} \equiv \varrho / \mathcal{R}_{dS}$. Letting $\mathcal{R}_{dS} \rightarrow \infty$, we can obtain the Minkowskian line element as
$$ds^{2}|_{\mathcal{R}_{dS} \rightarrow  \infty} =-dT^{2} +d\varrho^{2} + \varrho^{2} \left( d\theta_{2}^{2} + \sin^{2}{\theta_{2}} \,d\theta_{1}^{2}  \right) = -(dx^{0})^{2}+\sum_{j=1}^{3} (dx^{j})^{2},$$
where $x^{0}=T$, $x^{1} = \varrho \sin{\theta_{2}}   \sin{\theta_{1}}$, $x^{2} = \varrho \sin{\theta_{2}}   \cos{\theta_{1}}$ and $x^{3} = \varrho \cos{\theta_{2}}$. The flat-space version of the field equations for massless higher-spin fermions is 
\begin{align}\label{flat space eqn}
   &\slashed{\partial}\Psi_{ \mu_{1}...\mu_{r}}=0 \nonumber, \\
   & \partial^{\alpha}\Psi_{ \alpha \mu_{2}...\mu_{r}}=0, \hspace{4mm}  \gamma^{\alpha}\Psi_{ \alpha \mu_{2}...\mu_{r}}=0.
\end{align} 
Now, let us find the flat-space limit of the dS conformal Killing vectors~(\ref{V=nabla phi}) (by re-scaling them appropriately). In particular, the four translation Killing vectors, $\delta_{0}^{\mu}, \delta_{1}^{\mu}, \delta_{2}^{\mu}$ and $\delta_{3}^{\mu}$, of Minkowski spacetime are obtained from the following limits:
\begin{align}
  &  \left( -\mathcal{R}_{dS} V^{(0)}_{\mu} \right)|_{\mathcal{R}_{dS} \rightarrow  \infty} =-\delta^{0}_{\mu}=-\partial_{\mu}x^{0},~~~\left( \mathcal{R}_{dS} V^{(1)}_{\mu} \right)|_{\mathcal{R}_{dS} \rightarrow  \infty} = \delta^{1}_{\mu}=\partial_{\mu}x^{1}, \nonumber\\
    &\left( \mathcal{R}_{dS} V^{(2)}_{\mu} \right)|_{\mathcal{R}_{dS} \rightarrow  \infty} = \delta^{2}_{\mu}=\partial_{\mu}x^{2},~~~~~\left( \mathcal{R}_{dS} V^{(3)}_{\mu} \right)|_{\mathcal{R}_{dS} \rightarrow  \infty} = \delta^{3}_{\mu}=\partial_{\mu}x^{3},
\end{align}
while the dilatation conformal Killing vector $x^{\mu}$ is obtained as
\begin{align}
  &  \left(- \mathcal{R}^{2}_{dS} V^{(4)}_{\mu} \right)|_{\mathcal{R}_{dS} \rightarrow  \infty} = \frac{1}{2}\partial_{\mu}\left( -T^{2}+\varrho^{2}   \right) =  \frac{1}{2}\partial_{\mu}\left( x^{\alpha}x_{\alpha}  \right) =x_{\mu},
\end{align}
where, here, $\mu$ takes the values $\{0,1,2,3\}$ corresponding to the standard Minkowski coordinates $x_{0}=-x^{0}, x^{1}, x^{2}, x^{3}$. It is natural that the flat-space limit of the {five exact} dS conformal Killing vectors~(\ref{V=nabla phi}) corresponds to the {five exact} $so(4,2)$ generators on Minkowski spacetime (recall that, unlike the generators of spacetime translations and dilatations, the Killing vectors of Lorentz transformations and the conformal Killing vectors of special conformal transformations are not exact).} 
}   

{Following the aforementioned limiting procedure, it is straightforward to show that the five de Sitterian conformal-like symmetries~(\ref{hidden symmetry}) reduce to five not interesting (i.e. known) flat-space symmetries as:
\begin{align*}
    T_{V^{(A)}}\Psi_{\mu_{1}...\mu_{r}} 
   \rightarrow  ~\gamma^{5} \partial_{A}\Psi_{\mu_{1}...\mu_{r}}  -\frac{2r}{2r+1} \partial_{(\mu_{1}}\left(\gamma^{5}\Psi_{\mu_{2}...\mu_{r})\rho}\delta^{\rho}_{A}\right),~~~\text{for}~A=0,1,2,3,
\end{align*}
and
\begin{align*}
    T_{V^{(4)}}\Psi_{\mu_{1}...\mu_{r}} 
   \rightarrow  ~\gamma^{5} \left(x^{\rho}\partial_{\rho}\Psi_{\mu_{1}...\mu_{r}} + \frac{3}{8}\, \partial_{\rho}x^{\rho}\,\Psi_{\mu_{1}...\mu_{r}}\right)   -\frac{2r}{2r+1} \partial_{(\mu_{1}}\left(\gamma^{5}\Psi_{\mu_{2}...\mu_{r})\rho}x^{\rho}\right).
\end{align*}
These are known flat-space symmetries of Eqs.~(\ref{flat space eqn}) - they correspond to infinitesimal translations and scale transformations (times $\gamma^{5}$) accompanied by gauge transformations. We can re-express these symmetries in a more familiar form using (the flat-space version of) the Lie-Lorentz derivative~(\ref{Lie_Lorentz}) and dropping the gauge transformations\footnote{{The gauge transformations in the dS conformal-like symmetries~(\ref{hidden symmetry}) were necessary to ensure that the transformed tensor-spinor remains in the TT gauge. However, in the case of the flat-space symmetries, the corresponding gauge transformations are not necessary.}}, as
\begin{align}\label{flat space extra symmetry}
    T^{flat}_{w}{\Psi}_{\mu_{1}...\mu_{r}} \equiv  &\gamma^{5} \left(\mathbb{L}_{{w}}{\Psi}_{\mu_{1}...\mu_{r}} +\frac{3-2r}{8} \, \partial^{\alpha}w_{\alpha}\Psi_{\mu_{1}...\mu_{r}} \right),
\end{align}
where $w^{\rho}    \in \{ \delta^{\rho}_{0}, \delta^{\rho}_{1}, \delta^{\rho}_{2},  \delta^{\rho}_{3}, x^{\rho} \}$. The conformal weight term in Eq.~(\ref{flat space extra symmetry}) is non-zero only if $w$ is a dilatation conformal Killing vector. Also, Eq.~(\ref{flat space extra symmetry}) continues describing a symmetry transformation if we replace $w$ with any of the six Killing vectors of the Lorentz algebra in Minkowski spacetime (this is not true however in the case of the conformal Killing vectors of special conformal transformations).}

{Last, we observe that the transformation~(\ref{flat space extra symmetry}) is a product of two transformations. Unlike in $dS_{4}$, in Minkowski spacetime, each of the two transformations present in the product~(\ref{flat space extra symmetry}) is also a symmetry. {To be specific, the flat-space equations~(\ref{flat space eqn}) are invariant under the replacement $\Psi_{\mu_{1}...\mu_{r}} \rightarrow \gamma^{5}   \Psi_{\mu_{1}...\mu_{r}}$ (infinitesimal axial rotations), as well as under $\Psi_{\mu_{1}...\mu_{r}} \rightarrow \mathbb{L}_{{w}}{\Psi}_{\mu_{1}...\mu_{r}} +\frac{3-2r}{8} \, \partial^{\alpha}w_{\alpha}\Psi_{\mu_{1}...\mu_{r}}$ (infinitesimal translations and scale transformations).}}
%%%%%%%%%%%%%%%%%%%%%%%%%%%%%%%%%%%%%%%%%%%%%%%%%%%%%%%%%%%%%%%%%%%%%%%%%%%%%%%%%%%%%%%%%%%%%%%%%%%%%%%%%%%%%%%%

%%%%%%%%%%%%%%%%%%%%%%%%%%%%%%%%%%%%%%%%%%%%%%%%%%%%%%%%%%%%%%%%%%%%%%%%%%%%%%%%%%%%%%%%%%%%%%%%%
$$  \textbf{Further discussions}  $$

{The main result of the present paper, i.e. the fact that strictly massless spin-$s \geq 3/2$ fermionic gauge potentials on $dS_{4}$ have $so(4,2)$ symmetry, is a new interesting feature of field theory on $dS_{4}$ (and possibly on $AdS_{4}$, although this has not been verified yet). Such a $so(4,2)$ symmetry at the level of gauge potentials does not appear on Minkowski spacetime~\cite{Conf_Bekaert}. However, at the level of the gauge-invariant field strengths, (bosonic and fermionic) strictly massless theories have $so(4,2)$ symmetry on both $(A)dS_{4}$ and Minkowski spacetimes~\cite{Conf_Bekaert}. Interestingly, our result does not contradict the no-go theorem of Ref.~\cite{Conf_Bekaert}, according to which there cannot be $so(4,2)$ symmetry at the level of strictly massless gauge potentials on $(A)dS_{4}$, because our gauge potentials are complex (Dirac) tensor-spinors.}

In Ref.~\cite{Vasiliev1}, using the unfolded formalism, Vasiliev presented a $sp(8, \mathbb{R})$ invariant formulation of free massless fields (gauge potentials) of any spin in $AdS_{4}$ and showed that the free field equations are invariant under $o(4,2)$ (see also Ref.~\cite{Vasiliev2}). Although further study is required, it is likely that the dS version of Vasiliev's conformal invariance~\cite{Vasiliev1} is related to the conformal-like symmetries we presented in this paper.

Last, it is worth recalling that unitary superconformal field theories on $dS_{4}$ are known to exist~\cite{dS revisited}. In view of our newly discovered conformal-like symmetries for strictly massless fermions, it is interesting to look for new (and possibly unitary) supersymmetric theories on $dS_{4}$ that include strictly massless fermions of any spin $s \geq 3/2$ - see also Ref.~\cite{Letsios_unconventional}.~\footnote{Recent interesting discussions on $dS_{2}$ supergravity can be found in Ref.~\cite{Beatrix}.}
%%%%%%%%%%%%%%%%%%%%%%%%%%%%%%%%%%%%%%%%%%%%%%%%%%%%%%%%%%%%%%%%%%%%%%%%%%%%%%%%%%%%%%%5
\acknowledgments
The author is grateful to Atsushi Higuchi for useful discussions and encouragement. The author also thanks Gizem {S}eng\"{o}r for discussions and comments, as well as Rakibur Rahman for communications. He would also like to thank Dionysios Anninos, F.F. John, Spyros Konitopoulos, Alan Rios Fukelman, Gizem {S}eng\"{o}r, and Guillermo `il professore' Silva for useful discussions and encouragement. Also, it is a pleasure to thank Xavier Bekaert and Mikhail Vasiliev for useful discussions. The author acknowledges financial support from the Department of Mathematics, University of York, and from the WW Smith Fund. Last, but not least, I would like to thank Alex for reminding me that there exist poetic qualities in life beyond poems, which was at the very least inspiring, for lack of a better wor(l)d.

\noindent

%%%%%%%%%%%%%%%%%%%%%%%%%%%%%%%%%%%%%%%%%%%%%%%%%%%%%%%%%%%%%%%%%5%%%%%%%%
\appendix 

\section{Deriving Eq.~(\ref{infinitesimal dS of physical modes}) by analytically continuing \texorpdfstring{$so(5)$}{so(5)} rotation generators and their matrix elements to \texorpdfstring{$so(4,1)$}{so(4,1)}} \label{Appendix matrix elements}
The aim of this Appendix is to explain how to use group-theoretic tools and analytic continuation techniques in order to derive the transformation properties of physical modes in Eq.~(\ref{infinitesimal dS of physical modes}).

\subsection{Background material for representations of \texorpdfstring{$so(5)$}{so(5)} and Gelfand-Tsetlin patterns}
The representations of the algebra $so(D+1)$ - with arbitrary $D$ - and the specification of the matrix elements of the generators have been studied by Gelfand and Tsetlin~\cite{Gelfand}.

The $D(D+1)/2$ generators $I_{AB} = -I_{BA}$ ($A,B=1,2,...,D+1$) of $so(D+1)$ satisfy the commutation relations
\begin{align}
   [I_{AB},I_{CD} ] = (\delta_{BC} I_{AD} +\delta_{A
   D}I_{BC})- (A\leftrightarrow B) .
\end{align}
 In Ref.~\cite{Gelfand}, the action of the $so(D+1)$ generators has been determined in the decomposition $so(D+1) \supset so(D)$. In particular, the representation space for a $so(D+1)$ representation is chosen to be the direct sum of the representation spaces of all representations of $so(D)$ that appear in the $so(D+1)$ representation. (If a representation of $so(D)$ appears in a representation of $so(D+1)$, then it appears with multiplicity one.) Similarly, the generators of $so(D)$ are determined in the decomposition $so(D) \supset so(D-1)$ and so forth. In other words, Gelfand and Tsetlin~\cite{Gelfand} determined a $so(D+1)$ representation in the decomposition $so(D+1) \supset so(D) \supset ... \supset so(2)$.

\noindent \textbf{Focusing on $\bm{so(5)}$}. We now specialise to $so(5)$ - since this is the non-compact partner of the dS algebra $so(4,1)$. Let us review some basic results obtained by Gelfand and Tsetlin~\cite{Gelfand} (with slightly modified notation).

A (unitary) irreducible representation of $so(5)$ is specified by the highest weight $\vec{s}=(s_{1},s_{2})$ with $s_{1} \geq s_{2} \geq 0$, where the numbers $s_{1}$ and $s_{2}$ are simultaneously integers or half-odd-integers. The 10 anti-hermitian generators $I_{AB} = -I_{BA}$ ($A,B=1,...,5$) act on a finite-dimensional vector space corresponding to a direct sum of $so(4)$ representation spaces (as described at the beginning of the Subsection). Let $v$ denote the orthonormal basis vectors in the $so(5)$ representation space. Each basis vector is uniquely labelled by a ``Gelfand-Tsetlin pattern'',~$\alpha$, as follows:
\begin{align}\label{Gelfand basis}
    v (\alpha) \equiv v \begin{pmatrix} s_{1} && s_{2}\\
                          f_{1} && f_{2}\\
                          p &&           \\
                          q&&                  
     \end{pmatrix} .
\end{align}
The labels $s_{1}, s_{2}$ are the same for all basis vectors, since they correspond to the highest weight specifying the $so(5)$ representation. The rest of the labels in Eq.~(\ref{Gelfand basis}) specify the content of the $so(5)$ representation concerning the chain of subalgebras $so(4) \supset so(3) \supset so(2)$. In particular, the labels $f_{1} , f_{2}$ correspond to a $so(4)$ highest weight $\vec{f} \equiv (f_{1} , f_{2})$ with 
$f_{1} \geq |f_{2}|$, where $f_{1}$ and $f_{2}$ are both integers or half-odd integers, while $f_{2}$ can be negative. The $so(3)$ weight $p \geq 0$ is an integer or half-odd integer. The full basis of the representation space is given by all $v(\alpha)$'s in eq.~(\ref{Gelfand basis}) - with fixed $s_{1}, s_{2}$ - satisfying:
\begin{align}\label{so(5)>so(4)>so(3)}
   &  s_{1} \geq f_{1} \geq s_{2} \geq |f_{2}|, \nonumber \\
  & f_{1} \geq p \geq |f_{2}|, \nonumber \\
  & p \geq q \geq -p.
\end{align}
The numbers $s_{1}, s_{2}, f_{1}, f_{2} , p$ and $q$ are all integers or half-odd integers. 

In order to obtain the desired transformation formulae~(\ref{infinitesimal dS of physical modes}) using analytic continuation, we need to study the action of the generator $I_{54}$ on the basis vectors~(\ref{Gelfand basis}). This is given by~\cite{Gelfand}:
\begin{align}\label{action of I54}
   - I_{54} \,\,v \begin{pmatrix} s_{1} && s_{2}\\
                          f_{1} && f_{2}\\
                          p &&           \\
                          q&&                  
     \end{pmatrix}  =& -\frac{1}{2}A({f_{1},f_{2}})~v \begin{pmatrix} s_{1} && s_{2}\\
                          f_{1}+1 && f_{2}\\
                          p &&           \\
                          q&&                  
     \end{pmatrix} -\frac{1}{2}B(f_{1},f_{2})\,~v \begin{pmatrix} s_{1} && s_{2}\\
                          f_{1} && f_{2}+1\\
                          p &&           \\
                          q&&                  
     \end{pmatrix}  \nonumber \\
     &+\frac{1}{2}A({f_{1}-1,f_{2}})~v \begin{pmatrix} s_{1} && s_{2}\\
                          f_{1}-1 && f_{2}\\
                          p &&           \\
                          q&&                  
     \end{pmatrix} +\frac{1}{2} B(f_{1},f_{2}-1)\,~v \begin{pmatrix} s_{1} && s_{2}\\
                          f_{1} && f_{2}-1\\
                          p &&           \\
                          q&&                  
     \end{pmatrix},
\end{align}
where 
\begin{align}\label{raising l coef}
    A(f_{1},f_{2}) =  \sqrt{\frac{(f_{1}-p+1)(f_{1}+p+2)(s_{1}-f_{1})(s_{1}+f_{1}+3)(f_{1}-s_{2}+1)(f_{1}+s_{2}+2)}{(f_{1}+f_{2}+1)(f_{1}+f_{2}+2) (f_{1}-f_{2}+1) (f_{1}-f_{2}+2)}}
\end{align}
and
\begin{align} \label{raising tilder coef}
    B(f_{1},f_{2}) =  \sqrt{\frac{(p-f_{2})(f_{2}+p+1)(s_{2}-f_{2})(s_{2}+f_{2}+1)(s_{1}-f_{2}+1)(s_{1}+f_{2}+2)}{(f_{1}+f_{2}+1)(f_{1}+f_{2}+2) (f_{1}-f_{2}) (f_{1}-f_{2}+1)}}.
\end{align}
(Our matrix elements differ from the matrix elements of Ref.~\cite{Gelfand} by a factor of $1/2$.)
Note that $A(f_{1},-f_{2})=A(f_{1}, f_{2})$ and $B(f_{1},f_{2})=B(f_{1}, -f_{2}-1)$.
%%%%%%%%%%%%%%%%%%%%%%%%%%%%%%%%%%%%%%%%%%%%%%%%%%%%%%%%%%%%%%%%%%%%%%%%%%%%%%%%%%%%%%%%%%%%%%%%%%%%%%%%%%%%%%%%%%%%%%%%%%%%%%%%%%%%%%%%%%%%%%%%%%%%%%%%%%%%%%%%%%%%%%%%%%%%%%
\subsection{Specialising to \texorpdfstring{$so(5)$}{so(5)} representations formed by tensor-spinor spherical harmonics on \texorpdfstring{$S^{4}$}{S}}
The line element of $S^{4}$ can be parametrised as
\begin{equation} \label{S^4_metric}
    ds_{S^{4}}^{2}=d\theta_{4}^{2}+\sin^{2}{\theta_{4}} \,d\Omega^{2},
\end{equation}
where $0 \leq \theta_{4}  \leq \pi$ and $d\Omega^{2}$ is the line element of $S^{3}$~(\ref{S^3_metric}). For later convenience, note that the line element~(\ref{S^4_metric}) can be analytically continued to the $dS_{4}$ line element~(\ref{dS_metric}) by making the replacement
\begin{align}\label{replacement anal cont}
    \theta_{4}  \rightarrow  x=\frac{\pi}{2} - it
\end{align}
- the variable $x$ has been already introduced in Eq.~(\ref{introducing x(t)=pi/2-it}).

Let $\slashed{\nabla} = \gamma^{\mu} \nabla_{\mu}$ be the Dirac operator on $S^{4}$, where $\gamma^{\mu}$ and $\nabla_{\mu}$ are the gamma matrices and covariant derivative, respectively, on $S^{4}$. We are interested in (totally symmetric) rank-$r$ tensor-spinor spherical harmonics $ \hat{\psi}^{(n;\,\tilde{r} ,\,\sigma \ell; \,{m}; k)}_{\mu_{1}...   \mu_{r}} (\theta_{4}, \bm{\theta_{3}})$ (with $\sigma = \pm$) on $S^{4}$ that satisfy~\cite{Homma}
\begin{align}\label{tens-spin sphr harmonics on S4}
  & \slashed{\nabla}  \hat{\psi}^{(n;\,\tilde{r} ,\,\sigma \ell; \,{m}; k)}_{\mu_{1}...   \mu_{r}}  = - i(n+2)\,  \hat{\psi}^{(n;\,\tilde{r} ,\,\sigma \ell; \,{m}; k)}_{\mu_{1}...   \mu_{r}},  \nonumber \\
   &\gamma^{\mu_{1}} \hat{\psi}^{(n;\,\tilde{r} ,\,\sigma \ell; \,{m}; k)}_{\mu_{1}...   \mu_{r}}  = \nabla^{\mu_{1}} \hat{\psi}^{(n;\,\tilde{r} ,\,\sigma \ell; \,{m}; k)}_{\mu_{1}...   \mu_{r}} =0,\hspace{5mm}(n=r,r+1,...),
\end{align}
where $n$ is the angular momentum quantum number on $S^{4}$~\footnote{There are also tensor-spinor spherical harmonics on $S^{4}$ that satisfy Eqs.~(\ref{tens-spin sphr harmonics on S4}) but with an opposite sign for the eigenvalue. We will not discuss these here as they form equivalent $so(5)$ representations with the tensor-spinor spherical harmonics in Eq.~(\ref{tens-spin sphr harmonics on S4}).}.
The representation-theoretic meaning of the labels $n,\sigma, \ell,\tilde{r}, m$ and $k$ will be discussed below. The hat has been used in order to indicate that the eigenmodes $\hat{\psi}^{(n;\,\tilde{r} ,\,\sigma \ell; \,{m}; k)}_{\mu_{1}...   \mu_{r}} (\theta_{4}, \bm{\theta_{3}})$ are normalised with respect to the standard inner product on $S^{4}$
\begin{align}\label{inner prod on S4}
   \int_{S^{4}} \sqrt{g_{S^{4}}}\, d\theta_{4} \,d \theta_{3}\, d \theta_{2} \, d\theta_{1} \,  \hat{\psi}^{(n';\,\tilde{r}' ,\,\sigma' \ell'; \,{m'}; k') \dagger}_{\mu_{1}...   \mu_{r}}\,\hat{\psi}^{(n;\,\tilde{r} ,\,\sigma \ell; \,{m}; k)\mu_{1}...   \mu_{r}} = \delta_{nn'} \delta_{\ell \ell'} \delta_{\sigma \sigma'} \delta_{\tilde{r}  \tilde{r}'}\delta_{mm'} \delta_{k k'},
\end{align}
where $g_{S^{4}}$ is the determinant of the $S^{4}$ metric. The indices $\mu_{1},...,\mu_{r}$ run from $\theta_{1}$ to $\theta_{4}$, while the indices $\tilde{\mu}_{1},...,\tilde{\mu}_{r}$ run from $\theta_{1}$ to $\theta_{3}$.

\noindent \textbf{Gelfand-Tsetlin patterns and tensor-spinor spherical harmonics.} The ten Killing vectors of $S^{4}$ act on the solution space of Eqs.~(\ref{tens-spin sphr harmonics on S4}) in terms of the Lie-Lorentz derivatives~(\ref{Lie_Lorentz}), and the latter generate a representation of $so(5)$ on this solution space. In particular, for each allowed value of $n$, the set of eigenmodes $\{ \hat{\psi}^{(n;\,\tilde{r} ,\,\sigma \ell; \,{m}; k)}_{\mu_{1}...   \mu_{r}} \}$ forms an irreducible representation of $so(5)$ with highest weight
\begin{align}
  \vec{s} = (s_{1}, s_{2})=\left( n+\frac{1}{2}, r+\frac{1}{2}   \right)  \hspace{5mm}
\end{align}
with $n=r,r+1,...$. Each eigenmode $\hat{\psi}^{(n;\,\tilde{r} ,\,\sigma \ell; \,{m}; k)}_{\mu_{1}...   \mu_{r}} (\theta_{4}, \bm{\theta_{3}})$ corresponds to a Gelfand-Tsetlin pattern (see Eq.~(\ref{Gelfand basis}))
\begin{align}\label{Gelfand basis tensor-spinors S^4}
   \alpha =  \begin{pmatrix}n+\tfrac{1}{2} && r+\tfrac{1}{2}\\
                         \ell+\tfrac{1}{2} &&  \sigma (\tilde{r} + \tfrac{1}{2})\\
                            m+\tfrac{1}{2} &&          \\
                          k+ \tfrac{1}{2}&&                  
     \end{pmatrix} .
\end{align}
The numbers $\ell,m$ and $k$ are the angular momentum quantum numbers on $S^{3}$, $S^{2}$ and $S^{1}$, respectively, and their allowed values are found from~(\ref{so(5)>so(4)>so(3)}). 

Based on the discussion in the previous paragraph, we can identify each eigenmode $\hat{\psi}^{(n;\,\tilde{r} ,\,\sigma \ell; \,{m}; k)}_{\mu_{1}...   \mu_{r}} (\theta_{4}, \bm{\theta_{3}})$ with a basis vector~(\ref{Gelfand basis}) labeled by the pattern~(\ref{Gelfand basis tensor-spinors S^4}). In particular, we make the identifications: 
\begin{align} \label{v(+)-> psi+ identify}
  v\begin{pmatrix}n+\tfrac{1}{2} && r+\tfrac{1}{2}\\
                         \ell+\tfrac{1}{2} &&   \tilde{r} + \tfrac{1}{2}\\
                            m+\tfrac{1}{2} &&          \\
                           k+ \tfrac{1}{2}&&                  
     \end{pmatrix}   \rightarrow  \hat{\psi}^{(n;\,\tilde{r} ,\,+ \ell; \,{m}; k)}_{\mu_{1}...   \mu_{r}}  
\end{align}
     and
     \begin{align}\label{v(-)-> psi- identify}
      v\begin{pmatrix} n+\tfrac{1}{2} &&r+\tfrac{1}{2}\\
                         \ell+\tfrac{1}{2} &&  -( \tilde{r} + \tfrac{1}{2})\\
                            m+\tfrac{1}{2} &&          \\
                          k+ \tfrac{1}{2}&&                  
     \end{pmatrix}   \rightarrow -i\,(-1)^{\tilde{r}} \hat{\psi}^{(n;\,\tilde{r} ,\,- \ell; \,{m}; k)}_{\mu_{1}...   \mu_{r}},   
     \end{align}
     where the phase factor $-i (-1)^{\tilde{r}}$ has been introduced for convenience.
%%%%%%%%%%%%%%%%%%%%%%%%%%%%%%%%%%%%%%%%%%%%%%%%%%%%%%%%%%%%%%%%%%%%%%%%%%%%%%%%%%%%%%%%%%%%%%%%%%%%%%%%%%%%%%%%%%%%%%%%%
\subsection{Transformation properties of tensor-spinor spherical harmonics on \texorpdfstring{$S^{4}$}{S} under \texorpdfstring{$so(5)$}{so}}
In this Subsection, we find the $so(5)$ transformation formulae for $\mathbb{L}_{\mathcal{S}}\hat{\psi}^{(n;\,\tilde{r}=r ,\,\pm \ell; \,{m}; k)}_{\mu_{1}...   \mu_{r}}$ that (after analytic continuation) will give rise to the $so(4,1)$ transformation formulae~(\ref{infinitesimal dS of physical modes}) for the physical modes of the strictly massless fermions on $dS_{4}$. Here $\mathbb{L}_{\mathcal{S}}$ is the Lie-Lorentz derivative on $S^{4}$ with respect to the Killing vector
\begin{align}\label{S4 rot before dS boost}
    \mathcal{S} = \mathcal{S}^{\mu} \partial_{\mu} = \cos{\theta_{3}}\, \frac{\partial}{\partial \theta_{4}} - \cot{\theta_{4}} \sin{\theta_{3}}\, \frac{\partial}{\partial {\theta_{3}}}.
\end{align}
This Killing vector corresponds to the $so(5)$ generator $I_{45} = -I_{54}$ in Eq.~(\ref{action of I54}) and by making the replacement~(\ref{replacement anal cont}) it is analytically continued as: $\mathcal{S}  \rightarrow i X$, where $X$ is the dS boost Killing vector~(\ref{dS boost}).

We focus on the eigenmodes $\hat{\psi}^{(n;\,\tilde{r}=r ,\,+ \ell; \,{m}; k)}_{\mu_{1}...   \mu_{r}}(\theta_{4}, \bm{\theta_{3}})$ and $\hat{\psi}^{(n;\,\tilde{r}=r ,\,- \ell; \,{m}; k)}_{\mu_{1}...   \mu_{r}}(\theta_{4}, \bm{\theta_{3}})$, as the former will be analytically continued to the physical modes $
 {\Psi}^{(phys ,\,+ \ell; \,m;k)}_{ {\mu}_{1}...{\mu}_{r}}(t,\bm{\theta}_{3})$~(\ref{physmodes_positive_spin_r+1/2_dS4}) and the latter to the physical modes $
 {\Psi}^{(phys ,\,- \ell; \,m;k)}_{ {\mu}_{1}...{\mu}_{r}}(t,\bm{\theta}_{3})$~(\ref{physmodes_negative_spin_r+1/2_dS4}). 
 We will also discuss in passing the eigenmodes
 $\hat{\psi}^{(n;\,\tilde{r}=r-1 ,\,\pm \ell; \,{m}; k)}_{\mu_{1}...   \mu_{r}}(\theta_{4}, \bm{\theta_{3}})$ as they will be analytically continued to the pure gauge modes ${\Psi}^{(pg ,\,\tilde{r}=r-1,\,\pm \ell; \,{m};k)}_{ {\mu}_{1}...{\mu}_{r}}(t,\bm{\theta}_{3}) $, which appear in the transformation formulae~(\ref{infinitesimal dS of physical modes}).
 
 Explicit expressions for the infinitesimal transformations $\mathbb{L}_{\mathcal{S}}\hat{\psi}^{(n;\,\tilde{r}=r ,\,\pm \ell; \,{m}; k)}_{\mu_{1}...   \mu_{r}}$ and \newline $\mathbb{L}_{\mathcal{S}}\hat{\psi}^{(n;\,\tilde{r}=r-1 ,\,\pm \ell; \,{m}; k)}_{\mu_{1}...   \mu_{r}}$ are immediately found from Eq.~(\ref{action of I54}) with the use of Eqs.~(\ref{v(+)-> psi+ identify}) and (\ref{v(-)-> psi- identify}). However, these transformation properties refer to normalised eigenmodes, while the desired dS transformation properties~(\ref{infinitesimal dS of physical modes}) refer to un-normalised eigenmodes. Therefore, we will first find the $so(5)$ transformation properties for the un-normalised eigenmodes on $S^{4}$ (the un-normalised eigenmodes will be defined below), and then perform analytic continuation to $dS_{4}$.

%%%%%%%%%%%%%%%%%%%%%%%%%%%%%%%%%%%%%%%%%%%%%%%%%%%%%%%%%%%%%%%%%%%%%%%%%%%%%%%%%%%%%%%%%%%%%%%%%
$$\textbf{Some useful expressions for eigenmodes}~\textbf{on}~\bm{S^{3}} $$
For later convenience, let us present some expressions for certain tensor-spinor spherical harmonics on $S^{3}$~[see Eqs.~(\ref{tensor-spinor+eigen_S3}) and (\ref{tensor-spinor--eigen_S3})]. These expressions can be easily obtained using the method of separation of variables as has been explained in Refs.~\cite{Camporesi, CHH, Letsios_announce}. Below, we use the notation $\bm{\theta_{3}} = (\theta_{3}, \theta_{2}, \theta_{1}) = (\theta_{3}, \bm{\theta_{2}})$. We only need the following expressions for our computations:

%%%%%%%%%%%%%%%%%%%%%%%%%%%%%%%%%%%%%%%%%%%%%%%%%%%%%%%%%%%%%%%%%%%%%%%%%%%
\noindent $\bullet$ \textbf{Rank-$\bm{r}$ eigenmodes $  \tilde{\psi}^{(\ell; {m};k)}_{\pm \tilde{\mu}_{1}... \tilde{\mu}_{r}}(\theta_{3}, \bm{\theta_{2}})$ on $\bm{S^{3}}$:} The component $\tilde{\psi}^{(\ell; {m};k)}_{\pm \theta_{3} \theta_{3}...\theta_{3}}(\theta_{3}, \bm{\theta_{2}})$ is a spinor on $S^{2}$. It is given by
\begin{align}
    \tilde{\psi}^{(\ell; {m};k)}_{\pm \underset{r~ \text{times}}{\theta_{3} \theta_{3}...\theta_{3}}}(\theta_{3}, \bm{\theta_{2}})=&~\frac{\tilde{c}(r,\ell;m)}{\sqrt{2}}\frac{1}{\sqrt{2}}(\bm{1}+i \tilde{\gamma}^{3})\Big\{\tilde{\phi}^{(r)}_{\ell m}(\theta_{3}) \pm i  \tilde{\psi}^{(r)}_{ \ell m}(\theta_{3})\tilde{\gamma}^{3}\Big\}  \tilde{\tilde{\psi}}^{(m;k)}_{-}(\bm{\theta_{2}}),
\end{align}
where $\tfrac{\tilde{c}(r,\ell;m)}{\sqrt{2}}$ is the normalisation factor, $\tilde{\tilde{\psi}}^{(m;k)}_{-}(\bm{\theta_{2}})$ are the spinor eigenfunctions of the Dirac operator $\tilde{\tilde{\slashed{\nabla}}}$ on $S^{2}$ satisfying 
$$\tilde{\tilde{\slashed{\nabla}}} \tilde{\tilde{\psi}}^{(m;k)}_{-} = -i(m+1) \tilde{\tilde{\psi}}^{(m;k)}_{-}, $$
while the spinors $\tilde{\tilde{\psi}}^{(m;k)}_{+} \equiv \tilde{\gamma}^{3}\,\tilde{\tilde{\psi}}^{(m;k)}_{-}$ satisfy
$$\tilde{\tilde{\slashed{\nabla}}} \tilde{\tilde{\psi}}^{(m;k)}_{+} = +i(m+1) \tilde{\tilde{\psi}}^{(m;k)}_{+}. $$
The functions $\tilde{\phi}^{(r)}_{ \ell m}(\theta_{3})$ and $\tilde{\psi}^{(r)}_{ \ell m}(\theta_{3})$ correspond to special cases of the following functions:
\begin{align}
    \tilde{\phi}^{(\tilde{a})}_{\ell m}(\theta_{3}) =&~\tilde{\kappa}_{\tilde{\phi}}(\ell, m) \left(\cos{\frac{\theta_{3}}{2}}\right)^{m+1-\tilde{a}}\left(\sin{\frac{\theta_{3}}{2}}\right)^{m-\tilde{a}} \nonumber \\ &\times  F\left(-\ell+m,\ell+m+3;m+\frac{3}{2};\sin^{2}\frac{\theta_{3}}{2}\right),\label{phitilde_a} 
\end{align} 
and
 \begin{align}
    \tilde{\psi}^{(\tilde{a})}_{\ell m}(\theta_{3})
    &=\,\tilde{\kappa}_{\tilde{\phi}}(\ell, m)\,\frac{\ell+\frac{3}{2}}{m+\frac{3}{2}}\left(\cos{\frac{\theta_{3}}{2}}\right)^{m-\tilde{a}}\left(\sin{\frac{\theta_{3}}{2}}\right)^{m+1-\tilde{a}}\nonumber\\  
    &\times  F\left(-\ell+m,\ell+m+3;m+\frac{5}{2};\sin^{2}\frac{\theta_{3}}{2}\right),\label{psitilde_a} 
    \end{align}
 where $\tilde{a}$ is an integer, while the factor $\tilde{\kappa}_{\tilde{\phi}}(\ell,m)$ is given by
  \begin{align}
   \tilde{\kappa}_{\tilde{\phi}}(\ell,m)= \frac{\Gamma(\ell+\frac{3}{2})}{\Gamma(\ell-m+1)\,\Gamma{(m+\frac{3}{2})}}.
  \end{align}
%%%%%%%%%%%%%%%%%%%%%%%%%%%%%%%%%%%%%%%%%%%%%%%%%%%%%%%%%%%%%%%%%%%%%%%%%%%%%%%%%%%%%%%%%%%%%%%%%%%

\noindent $\bullet$ \textbf{Rank-$(\bm{r-1})$ eigenmodes $  \tilde{\psi}^{(\ell; {m};k)}_{\pm \tilde{\mu}_{2}... \tilde{\mu}_{r}}(\theta_{3}, \bm{\theta_{2}})$ on $\bm{S^{3}}$:} The component $\tilde{\psi}^{(\ell; {m};k)}_{\pm \theta_{3} \theta_{3}...\theta_{3}}(\theta_{3}, \bm{\theta_{2}})$ is a spinor on $S^{2}$. It is given by
\begin{align}
    \tilde{\psi}^{(\ell; {m};k)}_{\pm \underset{r-1~ \text{times}}{\theta_{3} \theta_{3}...\theta_{3}}}(\theta_{3}, \bm{\theta_{2}})=&~\frac{\tilde{c}(r-1,\ell;m)}{\sqrt{2}}\frac{1}{\sqrt{2}}(\bm{1}+i \tilde{\gamma}^{3})\Big\{\tilde{\phi}^{(r-1)}_{\ell m}(\theta_{3}) \pm i  \tilde{\psi}^{(r-1)}_{ \ell m}(\theta_{3})\tilde{\gamma}^{3}\Big\}  \tilde{\tilde{\psi}}^{(m;k)}_{-}(\bm{\theta_{2}}),
\end{align}
where $\tfrac{\tilde{c}(r-1,\ell;m)}{\sqrt{2}}$ is the normalisation factor, while $\tilde{\phi}^{(r-1)}_{ \ell m}(\theta_{3})$ and $\tilde{\psi}^{(r-1)}_{ \ell m}(\theta_{3})$ are given by Eqs.~(\ref{phitilde_a}) and (\ref{psitilde_a}), respectively, with $\tilde{a}=r-1$.

%%%%%%%%%%%%%%%%%%%%%%%%%%%%%%%%%%%%%%%%%%%%%%%%%%%%%%%%%%%%%%%%%%%%%%%%%%%%%%%%%%%%%%%%%%%%%%%%%%%%%%%%%%%%%%%%%%%%%%%%%%%
$$\textbf{Expressions for the eigenmodes}~   \bm{\hat{\psi}^{(n ;\,\tilde{r}=r,\,\pm\ell; \,m;k)}_{\mu_{1} {\mu}_{2}...{\mu}_{r}}}~\textbf{on}~\bm{S^{4}} $$
%%%%%%%%%%%%%%%%%%%%%%%%%%%%%%%%%%%%%%%%%%%%%%%%%%%%%%%%%%%%%%%%%%%%%%%%%%%%%%%%%%%%%%%%%%
Working as in Section~\ref{Sec_mode solutions}, we separate variables for equations~(\ref{tens-spin sphr harmonics on S4}) on $S^{4}$. We find
\begin{align}\label{physmodes_positive_spin_r+1/2_sphere4}
 &  \hat{\psi}^{(n ;\,\tilde{r}=r,\,+\ell; \,m;k)}_{\theta_{4} {\mu}_{2}...{\mu}_{r}}(\theta_{4},\bm{\theta}_{3})= 0,\nonumber\\     &\hat{\psi}^{(n ;\,\tilde{r}=r,\,+\ell; \,m;k)}_{\tilde{\mu}_{1}...\tilde{\mu}_{r}}(\theta_{4},\bm{\theta}_{3})=\frac{c(r,n;\tilde{r}=r,\ell)}{\sqrt{2}}  \begin{pmatrix} i \psi^{(-r)}_{ n\ell}(\theta_{4}) \, \tilde{\psi}_{+\tilde{\mu}_{1}...\tilde{\mu}_{r}}^{(\ell; m;k)} (\bm{\theta_{3}})  \\ -   \phi^{(-r)}_{ n\ell}(\theta_{4}) \, \tilde{\psi}^{(\ell; m;k)}_{+\tilde{\mu}_{1}...\tilde{\mu}_{r}} (\bm{\theta_{3}}
    )  \end{pmatrix}
\end{align}
and
\begin{align}\label{physmodes_negative_spin_r+1/2_sphere4}
   &\hat{\psi}^{(n ;\,\tilde{r}=r,\,-\ell; \,m;k)}_{\theta_{4} {\mu}_{2}...{\mu}_{r}}(\theta_{4},\bm{\theta}_{3})= 0,\nonumber\\   & \hat{\psi}^{(n ;\,\tilde{r}=r,\,-\ell; \,m;k)}_{\tilde{\mu}_{1}...\tilde{\mu}_{r}}(\theta_{4},\bm{\theta}_{3})= \frac{c(r,n;\tilde{r}=r,  \ell)}{\sqrt{2}} \begin{pmatrix}  \phi^{(-r)}_{ n\ell}(\theta_{4}) \, \tilde{\psi}_{-\tilde{\mu}_{1}...\tilde{\mu}_{r}}^{(\ell; m;k)} (\bm{\theta_{3}})  \\ -  i \psi^{(-r)}_{ n\ell}(\theta_{4}) \, \tilde{\psi}^{(\ell; m;k)}_{-\tilde{\mu}_{1}...\tilde{\mu}_{r}} (\bm{\theta_{3}}
    )  \end{pmatrix},
\end{align}
where $\tfrac{c(r,n;\tilde{r}=r, \ell)}{\sqrt{2}}$ is a normalisation factor that will be determined below. The functions $\phi^{(-r)}_{ n\ell}(\theta_{4}) $ and $\psi^{(-r)}_{ n\ell}(\theta_{4})$ belong to the following family of functions:
  \begin{align}
    \phi^{(a)}_{n \ell}(\theta_{4})=&~\kappa_{\phi}(n,\ell)\,\left(\cos{\frac{\theta_{4}}{2}}\right)^{\ell+1-a}\left(\sin{\frac{\theta_{4}}{2}}\right)^{\ell-a}\nonumber \\
    &\times F\left(-n+\ell,n+\ell+4;\ell+2;\sin^{2}\frac{\theta_{4}}{2}\right),\label{phi_a} 
\end{align}
\begin{align}
    \psi^{(a)}_{n \ell}(\theta_{4})=&~\kappa_{\phi}(n,\ell)\frac{n
    +2}{\ell+2}  \left(\cos{\frac{\theta_{4}}{2}}\right)^{\ell-a} \left(\sin{\frac{\theta_{4}}{2}} \right)^{\ell+1-a} \nonumber\\ &\times F\left(-n+\ell,n+\ell+4;\ell+{3};\sin^{2}\frac{\theta_{4}}{2}\right),\label{psi_a}
\end{align}
where the factor $\kappa_{\phi}(n,\ell)$ is given by
  \begin{align}\label{normlsn_fac_of_Jacobi}
   \kappa_{\phi}(n,\ell)= \frac{\Gamma(n+2)}{\Gamma(n-\ell+1)\Gamma{(\ell+2)}}.
  \end{align}
Substituting the eigenmode~(\ref{physmodes_positive_spin_r+1/2_sphere4}) (or~(\ref{physmodes_negative_spin_r+1/2_sphere4})) into the inner product~(\ref{inner prod on S4}), and using the normalisation of the tensor-spinor eigenmodes on $S^{3}$~(\ref{normlzn_S3}), we find 
\begin{align}\label{norm_factor_rank-r_tilder=r_on_S4}
    \left|   \frac{c(r,n;\tilde{r}=r,\ell)}{\sqrt{2}} \right|^{2} = 2^{2r-3}\frac{\Gamma(n-\ell+1)\Gamma(4+n+\ell)}{\left| \Gamma(n+2)  \right|^{2}}.
\end{align}

\noindent \textbf{Introducing the un-normalised eigenmodes.}~Now, let us define the un-normalised eigenmodes ${\psi}^{(n ;\,\tilde{r},\,\pm\ell; \,m;k)}_{\mu_{1} {\mu}_{2}...{\mu}_{r}}(\theta_{4},\bm{\theta}_{3})$ (for any value of $\tilde{r}\in \{0,...,r\}$) as
\begin{align}
    {\psi}^{(n ;\,\tilde{r},\,\pm\ell; \,m;k)}_{\mu_{1} {\mu}_{2}...{\mu}_{r}}(\theta_{4},\bm{\theta}_{3}) \equiv \frac{\sqrt{2}}{c(r,n; \tilde{r},\ell)}\frac{1}{ \kappa_{\phi}(n,\ell)}\,\hat {\psi}^{(n ;\,\tilde{r},\,\pm\ell; \,m;k)}_{\mu_{1} {\mu}_{2}...{\mu}_{r}}(\theta_{4},\bm{\theta}_{3}),
\end{align}
 where the normalisation factors ${c(r,n; \tilde{r},\ell)}$ that are needed for our computations (and have not been defined yet) will be defined later.
(Recall that the un-normalised eigenmodes are the ones that will be analytically continued to $dS_{4}$.)

\noindent \textbf{Transformation of the un-normalised eigenmodes $ {\psi}^{(n ;\,\textcolor{red}{\tilde{r}=r},\,\pm\ell; \,m;k)}_{\mu_{1} {\mu}_{2}...{\mu}_{r}}$.}~The infinitesimal $so(5)$ transformation of the un-normalised modes $\mathbb{L}_{\mathcal{S}} {\psi}^{(n ;\,\tilde{r}=r,\,\pm\ell; \,m;k)}_{\mu_{1} {\mu}_{2}...{\mu}_{r}}$ can be straightforwardly found from the transformation of the normalised modes $\mathbb{L}_{\mathcal{S}} \hat{\psi}^{(n ;\,\tilde{r}=r,\,\pm\ell; \,m;k)}_{\mu_{1} {\mu}_{2}...{\mu}_{r}}$ (see the discussion at the beginning of this Subsection). We find in this manner
\begin{align}\label{infinitesimal so(5) of un-norm phys modes S^4}
   \mathbb{L}_{\mathcal{S}} {\psi}^{(n ;\,\tilde{r}=r,\,\pm\ell; \,m;k)}_{\mu_{1} {\mu}_{2}...{\mu}_{r}}=&-\frac{\kappa_{\phi}(n,\ell+1)}{2 \kappa_{\phi}(n,\ell)}\sqrt{\frac{(\ell-m+1)(\ell+m+3)}{(\ell+2)^{2}-r^{2}}} (n+\ell+4)\, {\psi}^{\left(n ;\,\tilde{r}=r,\,\pm(\ell+1); \,m;k\right)}_{\mu_{1} {\mu}_{2}...{\mu}_{r}}  \nonumber\\
   &+\frac{\kappa_{\phi}(n,\ell-1)}{2 \kappa_{\phi}(n,\ell)}\sqrt{\frac{(\ell-m)(\ell+m+2)}{(\ell+1)^{2}-r^{2}}} (n-\ell+1)\, {\psi}^{\left(n ;\,\tilde{r}=r,\,\pm(\ell-1); \,m;k\right)}_{\mu_{1} {\mu}_{2}...{\mu}_{r}}  \nonumber\\
   &+\frac{\sqrt{(n+2)^{2}-r^{2}}}{2}K_{\ell m} \frac{c(r,n; \tilde{r}=r-1  ,\ell)}{c(r,n;\tilde{r}=r ,   \ell)}\,{\psi}^{\left(n ;\,\tilde{r}=r-1,\,\pm\ell; \,m;k\right)}_{\mu_{1} {\mu}_{2}...{\mu}_{r}} ,
\end{align}
where 
\begin{align}\label{Klm}
    K_{\ell m}=\sqrt{  \frac{\left( (m+1)^{2}-r^{2}\right) \,(2r+1)}{\left((\ell+1)^{2}-r^{2}   \right)\,\left((\ell+2)^{2}-r^{2}   \right)  }     }.
\end{align}
Note that, under this $so(5)$ transformation, the modes $ {\psi}^{(n ;\,\tilde{r}=r,\,+\ell; \,m;k)}_{\mu_{1} {\mu}_{2}...{\mu}_{r}}$ do not mix with the modes $ {\psi}^{(n ;\,\tilde{r}=r,\,-\ell; \,m;k)}_{\mu_{1} {\mu}_{2}...{\mu}_{r}}$. This observation plays a key role when performing analytic continuation to $dS_{4}$, as it implies that the strictly massless fermions on $dS_{4}$ correspond to a direct sum of irreducible representations of $so(4,1)$ - see Eq.~(\ref{infinitesimal dS of physical modes}).

%%%%%%%%%%%%%%%%%%%%%%%%%%%%%%%%%%%%%%%%%%%%%%%%%%%%%%%%%%%%%%%%%%%%%%%%%%%%%%%%%%
 $$\textbf{Expressions for the eigenmodes}~   \bm{\hat{\psi}^{(n ;\,\tilde{r}=r-1,\,\pm\ell; \,m;k)}_{\mu_{1} {\mu}_{2}...{\mu}_{r}}} ~\textbf{on}~\bm{S^{4}}$$
By separating variables again for equations~(\ref{tens-spin sphr harmonics on S4}) we find
\begin{align}\label{tilde(r)=r-1_positive_spin_r+1/2_sphere4}
   &\hat{\psi}^{(n ;\,\tilde{r}=r-1,\,+\ell; \,m;k)}_{\theta_{4} \theta_{4} {\mu}_{3}...{\mu}_{r}}(\theta_{4},\bm{\theta}_{3})= 0,\nonumber\\  
   &\hat{\psi}^{(n ;\,\tilde{r}=r-1,\,+\ell; \,m;k)}_{\theta_{4}\tilde{\mu}_{2}...\tilde{\mu}_{r}}(\theta_{4},\bm{\theta}_{3})=\frac{c(r,n;  \tilde{r}=r-1 ,\ell)}{\sqrt{2}}  \begin{pmatrix} i \psi^{(-r+2)}_{ n\ell}(\theta_{4}) \, \tilde{\psi}_{+\tilde{\mu}_{2}...\tilde{\mu}_{r}}^{(\ell; m;k)} (\bm{\theta_{3}})  \\ -   \phi^{(-r+2)}_{ n\ell}(\theta_{4}) \, \tilde{\psi}^{(\ell; m;k)}_{+\tilde{\mu}_{2}...\tilde{\mu}_{r}} (\bm{\theta_{3}}
    )  \end{pmatrix}
\end{align}
and
\begin{align}\label{tilde(r)=r-1_negative_spin_r+1/2_sphere4}
   &\hat{\psi}^{(n ;\,\tilde{r}=r-1,\,-\ell; \,m;k)}_{\theta_{4} \theta_{4}{\mu}_{3}...{\mu}_{r}}(\theta_{4},\bm{\theta}_{3})= 0 \nonumber\\
   &\hat{\psi}^{(n ;\,\tilde{r}=r-1,\,-\ell; \,m;k)}_{\theta_{4}\tilde{\mu}_{2}...\tilde{\mu}_{r}}(\theta_{4},\bm{\theta}_{3})= \frac{c(r,n;  \tilde{r}=r-1 ,\ell)}{\sqrt{2}}  \begin{pmatrix}  \phi^{(-r+2)}_{ n\ell}(\theta_{4}) \, \tilde{\psi}_{-\tilde{\mu}_{2}...\tilde{\mu}_{r}}^{(\ell; m;k)} (\bm{\theta_{3}})  \\ -  i \psi^{(-r+2)}_{ n\ell}(\theta_{4}) \, \tilde{\psi}^{(\ell; m;k)}_{-\tilde{\mu}_{2}...\tilde{\mu}_{r}} (\bm{\theta_{3}}
    )  \end{pmatrix},
\end{align}
where $\tfrac{c(r,n;  \tilde{r}=r-1 ,\ell)}{\sqrt{2}} $ is the normalisation factor, while the functions $\phi^{(-r+2)}_{ n\ell}(\theta_{4})$ and $\psi^{(-r+2)}_{ n\ell}(\theta_{4})$ are given by Eqs.~(\ref{phi_a}) and (\ref{psi_a}), respectively, with $a=-r+2$. The components $ \hat{\psi}^{(n ;\,\tilde{r}=r-1,\,\pm\ell; \,m;k)}_{\tilde{\mu}_{1}...\tilde{\mu}_{r}}(\theta_{4},\bm{\theta}_{3})$ can be found using the TT conditions in Eq.~(\ref{tens-spin sphr harmonics on S4}).

Now that we know the expressions~(\ref{tilde(r)=r-1_positive_spin_r+1/2_sphere4}) and~(\ref{tilde(r)=r-1_negative_spin_r+1/2_sphere4}), we can perform the following calculation for later convenience. Letting $\mu_{1}=\theta_{4}$ and $\mu_{2}=...=\mu_{r}=\theta_{3}$ in $\mathbb{L}_{\mathcal{S}} {\psi}^{(n ;\,\tilde{r}=r,\,\pm\ell; \,m;k)}_{\mu_{1}...\mu_{r}}$ [Eq.~(\ref{infinitesimal so(5) of un-norm phys modes S^4})], we find
\begin{align}\label{infinitesimal so(5) of un-norm phys modes S^4 PLAY}
   \mathbb{L}_{\mathcal{S}} {\psi}^{(n ;\,\tilde{r}=r,\,\pm\ell; \,m;k)}_{\theta_{4} \theta_{3}...\theta_{3}}=\frac{\sqrt{(n+2)^{2}-r^{2}}}{2}K_{\ell m} \frac{c(r,n; \tilde{r}=r-1  ,\ell)}{c(r,n;\tilde{r}=r ,   \ell)}\,{\psi}^{\left(n ;\,\tilde{r}=r-1,\,\pm\ell; \,m;k\right)}_{\theta_{4} \theta_{3}...\theta_{3}} ,
\end{align}
while using the explicit expressions~(\ref{physmodes_positive_spin_r+1/2_sphere4}),~(\ref{physmodes_negative_spin_r+1/2_sphere4}), (\ref{tilde(r)=r-1_positive_spin_r+1/2_sphere4}) and (\ref{tilde(r)=r-1_negative_spin_r+1/2_sphere4}) we rewrite this equation as
\begin{align}\label{infinitesimal so(5) of un-norm phys modes S^4 PLAY2}
   \mathbb{L}_{\mathcal{S}} {\psi}^{(n ;\,\tilde{r}=r,\,\pm\ell; \,m;k)}_{\theta_{4} \theta_{3}...\theta_{3}}=\frac{1}{2}\frac{\tilde{c}(r,\ell;m)}{\tilde{c}(r-1,\ell;m)}\,{\psi}^{\left(n ;\,\tilde{r}=r-1,\,\pm\ell; \,m;k\right)}_{\theta_{4} \theta_{3}...\theta_{3}} .
\end{align}
Then, comparing Eqs.~(\ref{infinitesimal so(5) of un-norm phys modes S^4 PLAY}) and (\ref{infinitesimal so(5) of un-norm phys modes S^4 PLAY2}) we find
\begin{align}\label{useful proportionality relation}
   c(r,n; \tilde{r}=r-1  ,\ell)   \propto \frac{1}{\sqrt{(n+2)^{2}-r^{2}}} .
\end{align}
(We have used that $K_{\ell m}$ is given by Eq.~(\ref{Klm}), while $c(r,n;\tilde{r}=r ,   \ell)$ is given by Eq.~(\ref{norm_factor_rank-r_tilder=r_on_S4}).)
%%%%%%%%%%%%%%%%%%%%%%%%%%%%%%%%%%%%%%%%%%%%%%%%%%%%%%%%%%%%%%%%%%%%%%%%%%%%%%%%%%%%%%%%%%%%%%%%5

\noindent \textbf{Transformation of the un-normalised eigenmodes $ {\psi}^{(n ;\, \textcolor{red}{\tilde{r}=r-1},\,\pm\ell; \,m;k)}_{\mu_{1} {\mu}_{2}...{\mu}_{r}}$.}~Again, the infinitesimal $so(5)$ transformation of the un-normalised modes $\mathbb{L}_{\mathcal{S}} {\psi}^{(n ;\,\tilde{r}=r-1,\,\pm\ell; \,m;k)}_{\mu_{1} {\mu}_{2}...{\mu}_{r}}$ can be straightforwardly found from the transformation of the normalised modes $\mathbb{L}_{\mathcal{S}} \hat{\psi}^{(n ;\,\tilde{r}=r-1,\,\pm\ell; \,m;k)}_{\mu_{1} {\mu}_{2}...{\mu}_{r}}$ (see the discussion at the beginning of this Subsection). We find 
\begin{align}\label{infinitesimal so(5) of un-norm tilde(r)=r-1 modes S^4}
   \mathbb{L}_{\mathcal{S}} {\psi}^{(n ;\,\tilde{r}=r-1,\,\pm\ell; \,m;k)}_{\mu_{1} {\mu}_{2}...{\mu}_{r}}=&
   &-\frac{\sqrt{(n+2)^{2}-r^{2}}}{2}K_{\ell m} \frac{c(r,n; \tilde{r}=r  ,\ell)}{c(r,n;\tilde{r}=r-1 ,   \ell)}\,{\psi}^{\left(n ;\,\tilde{r}=r,\,\pm\ell; \,m;k\right)}_{\mu_{1} {\mu}_{2}...{\mu}_{r}}+... \,,
\end{align}
where `$...$' includes eigenmodes that are orthogonal to both ${\psi}^{\left(n ;\,\tilde{r}=r,\,\pm\ell; \,m;k\right)}_{\mu_{1} {\mu}_{2}...{\mu}_{r}}$ and ${\psi}^{\left(n ;\,\tilde{r}=r-1,\,\pm\ell; \,m;k\right)}_{\mu_{1} {\mu}_{2}...{\mu}_{r}}$.
%%%%%%%%%%%%%%%%%%%%%%%%%%%%%%%%%%%%%%%%%%%%%%%
\subsection{Performing analytic continuation}
Let us analytically continue the tensor-spinor spherical harmonics~(\ref{tens-spin sphr harmonics on S4}) on $S^{4}$ in order to obtain tensor-spinors satisfying Eqs.~(\ref{Dirac_eqn_fermion_dS}) and (\ref{TT_conditions_fermions_dS}) on $dS_{4}$. By making the replacements $\theta_{4} \rightarrow x(t)=  \pi/2 - it$ [see Eq.~(\ref{replacement anal cont})] and
\begin{align}
    n \rightarrow -2 - iM,
\end{align}
we analytically continue the un-normalised tensor-spinor spherical harmonics on $S^{4}$ to tensor-spinors on $dS_{4}$ as $${\psi}^{(n;\,\tilde{r} ,\,\sigma \ell; \,{m}; k)}_{\mu_{1}...   \mu_{r}}(\theta_{4}, \bm{\theta_{3}}) \rightarrow {\psi}^{(-2-iM;\,\tilde{r} ,\,\sigma \ell; \,{m}; k)}_{\mu_{1}...   \mu_{r}}(x(t), \bm{\theta_{3}}).$$ The analytically continued tensor-spinors satisfy Eqs.~(\ref{Dirac_eqn_fermion_dS}) and (\ref{TT_conditions_fermions_dS}) on $dS_{4}$, which we rewrite here again for convenience
\begin{align}\label{tens-spin sphr harmonics anal cont S4->dS}
  & \slashed{\nabla}  {\psi}^{(-2-iM;\,\tilde{r} ,\,\sigma \ell; \,{m}; k)}_{\mu_{1}...   \mu_{r}}  = - M\,  {\psi}^{(-2-iM;\,\tilde{r} ,\,\sigma \ell; \,{m}; k)}_{\mu_{1}...   \mu_{r}},  \nonumber \\
   &\gamma^{\mu_{1}} {\psi}^{(-2-iM;\,\tilde{r} ,\,\sigma \ell; \,{m}; k)}_{\mu_{1}...   \mu_{r}}  = \nabla^{\mu_{1}} {\psi}^{(-2-iM;\,\tilde{r} ,\,\sigma \ell; \,{m}; k)}_{\mu_{1}...   \mu_{r}} =0\hspace{5mm}.
\end{align}
Let us focus on imaginary values of the mass parameter $M$. For these values of $M$, a dS invariant (and time-independent) scalar product is given by~(\ref{axial_scalar prod}).

By applying the aforementioned analytic continuation techniques to the $so(5)$ transformation formulae~(\ref{infinitesimal so(5) of un-norm phys modes S^4}) and (\ref{infinitesimal so(5) of un-norm tilde(r)=r-1 modes S^4}), we find
\begin{align}\label{infinitesimal analcont-so(5) of un-norm phys modes S^4}
   \mathbb{L}_{X}& {\psi}^{(-2-iM ;\,\tilde{r}=r,\,\pm\ell; \,m;k)}_{\mu_{1} {\mu}_{2}...{\mu}_{r}}\nonumber\\
   =&~i\frac{\kappa_{\phi}(-2-iM,\ell+1)}{2 \kappa_{\phi}(-2-iM,\ell)}\sqrt{\frac{(\ell-m+1)(\ell+m+3)}{(\ell+2)^{2}-r^{2}}} (-iM+\ell+2)\, {\psi}^{\left(-2-iM ;\,\tilde{r}=r,\,\pm(\ell+1); \,m;k\right)}_{\mu_{1} {\mu}_{2}...{\mu}_{r}}  \nonumber\\
   &-i\frac{\kappa_{\phi}(-2-iM,\ell-1)}{2 \kappa_{\phi}(-2-iM,\ell)}\sqrt{\frac{(\ell-m)(\ell+m+2)}{(\ell+1)^{2}-r^{2}}} (-iM-\ell-1)\, {\psi}^{\left(-2-iM ;\,\tilde{r}=r,\,\pm(\ell-1); \,m;k\right)}_{\mu_{1} {\mu}_{2}...{\mu}_{r}}  \nonumber\\
   &-i\frac{\sqrt{-M^{2}-r^{2}}}{2}K_{\ell m} \frac{c(r,-2-iM; \tilde{r}=r-1  ,\ell)}{c(r,-2-iM;\tilde{r}=r ,   \ell)}\,{\psi}^{\left(-2-iM ;\,\tilde{r}=r-1,\,\pm\ell; \,m;k\right)}_{\mu_{1} {\mu}_{2}...{\mu}_{r}} ,
\end{align}
and
\begin{align}\label{infinitesimal analcont-so(5) of un-norm tilde(r)=r-1 modes S^4}
   \mathbb{L}_{X} &{\psi}^{(-2-iM ;\,\tilde{r}=r-1,\,\pm\ell; \,m;k)}_{\mu_{1} {\mu}_{2}...{\mu}_{r}}\nonumber\\
   &=i\frac{\sqrt{-M^{2}-r^{2}}}{2}K_{\ell m} \frac{c(r,-2-iM; \tilde{r}=r  ,\ell)}{c(r,-2-iM;\tilde{r}=r-1 ,   \ell)}\,{\psi}^{\left(-2-iM ;\,\tilde{r}=r,\,\pm\ell; \,m;k\right)}_{\mu_{1} {\mu}_{2}...{\mu}_{r}}+... \,,
\end{align}
while the analytically continued version of Eq.~(\ref{useful proportionality relation}) gives
\begin{align}
   c(r,-2-iM; \tilde{r}=r-1  ,\ell)   \propto \frac{1}{\sqrt{-M^{2}-r^{2}}} .
\end{align}

Recall that we focus on imaginary values of $M$. For convenience we assume that $-M^{2}>r^{2}$ [the value $-M^{2}=r^{2}$ corresponds to the strictly massless case~(\ref{sm_mass_parameter})].
Using the dS invariance~(\ref{dS invariance of axial inner}) of the scalar product~(\ref{axial_scalar prod}), we  have 
\begin{align}
   & \braket{  \mathbb{L}_{X} {\psi}^{(-2-iM ;\,\textcolor{red}{\tilde{r}=r},\,\pm\ell; \,m;k)}   |  {\psi}^{(-2-iM ;\, \textcolor{red}{\tilde{r}=r-1},\,\pm\ell; \,m;k)}} \nonumber\\
    &+\braket{  {\psi}^{(-2-iM ;\,\textcolor{red}{\tilde{r}=r},\,\pm\ell; \,m;k)} |\mathbb{L}_{X} {\psi}^{(-2-iM ;\,\textcolor{red}{\tilde{r}=r-1},\,\pm\ell; \,m;k)}  }=0.
\end{align}
Then, using the transformation formulae~(\ref{infinitesimal analcont-so(5) of un-norm phys modes S^4}) and (\ref{infinitesimal analcont-so(5) of un-norm tilde(r)=r-1 modes S^4}), we find 
\begin{align}
   & \braket{   {\psi}^{(-2-iM ;\,\textcolor{red}{\tilde{r}=r-1},\,\pm\ell; \,m;k)}   |  {\psi}^{(-2-iM ;\, \textcolor{red}{\tilde{r}=r-1},\,\pm\ell; \,m;k)}} \nonumber\\
    &=-\left| \frac{c(r,-2-iM; \tilde{r}=r  ,\ell)}{c(r,-2-iM;\tilde{r}=r-1 ,   \ell)}  \right|^{2}\,\braket{  {\psi}^{(-2-iM ;\,\textcolor{red}{\tilde{r}=r},\,\pm\ell; \,m;k)} | {\psi}^{(-2-iM ;\,\textcolor{red}{\tilde{r}=r},\,\pm\ell; \,m;k)}  }\\
    & \propto \sqrt{-M^{2}-r^{2}}^{2}.
\end{align}
From this equation, we understand that the analytically continued eigenmodes ${\psi}^{(-2-iM ;\,\textcolor{red}{\tilde{r}=r-1},\,\pm\ell; \,m;k)}_{\mu_{1}...\mu_{r}} $ have zero norm in the strictly massless limit ($M^{2}=-r^{2}$). In other words, they become pure gauge modes~(\ref{pure gauge modes +-}) in this limit, i.e. ${\psi}^{(-2+r ;\,\textcolor{red}{\tilde{r}=r-1},\,\pm\ell; \,m;k)}_{\mu_{1}...\mu_{r}}(x(t), \bm{\theta_{3}}) =  {\Psi}^{(pg ,\,\tilde{r}=r-1,\,\pm \ell; {m};k)}_{ {\mu}_{1}...{\mu}_{r}}(t,\bm{\theta}_{3}) $.

\noindent \textbf{Specialising to the strictly massless case and, finally, deriving Eq.~(\ref{infinitesimal dS of physical modes}).} Now we tune the mass parameter to the strictly massless value $M=i r$~(\ref{sm_mass_parameter}). The physical modes are
$ {\psi}^{(-2+r;\,\tilde{r} ,\,\pm \ell; \,{m}; k)}_{\mu_{1}...   \mu_{r}}(x(t), \bm{\theta_{3}}) \equiv {\Psi}^{(phys ,\,\pm \ell; \,m;k)}_{{\mu}_{1}...{\mu}_{r}} (t, \bm{\theta_{3}}) $ [see Eqs.~(\ref{physmodes_negative_spin_r+1/2_dS4}) and (\ref{physmodes_positive_spin_r+1/2_dS4})]. The infinitesimal dS transformation of these modes is found by letting $M=ir$ in Eq.~(\ref{infinitesimal analcont-so(5) of un-norm phys modes S^4}). By doing so, we straightforwardly arrive at Eq.~(\ref{infinitesimal dS of physical modes}), as required.

%%%%%%%%%%%%%%%%%%%%%%%%%%%%%%%%%%%%%%%%%%%%%%%%%%%%%%%%%%%%%%%%%%%%%%%%%%%%%%%%%%%%%%%%%%%%%%%%%%%%%%%%5
\section{Details for the computation of the commutator~(\ref{[hidden, hidden]}) between two conformal-like transformations}\label{Appe commutator}
 We wish to calculate $[T_{W}, T_{V}] \Psi_{\mu_{1}... \mu_{r}}$ in order to arrive at Eq.~(\ref{[hidden, hidden]}). For convenience, we split each of the conformal-like transformations in the commutator in two parts as in Eq.~(\ref{hidden= A+B}), i.e. $T_{W}\Psi_{\mu_{1}...\mu_{r}}=\Delta_{W}\Psi_{\mu_{1}...\mu_{r}}+P_{W}\Psi_{\mu_{1}...\mu_{r}}$ and $T_{V}\Psi_{\mu_{1}...\mu_{r}}=\Delta_{V}\Psi_{\mu_{1}...\mu_{r}}+P_{V}\Psi_{\mu_{1}...\mu_{r}}$. Then, we split $ [T_{W}, T_{V}] \Psi_{\mu_{1}... \mu_{r}}$ into three parts as
 \begin{align}\label{[hidden, hidden] in parts}
     [T_{W}, T_{V}] \Psi_{\mu_{1}... \mu_{r}} = [\Delta_{W}, \Delta_{V}]\Psi_{\mu_{1}... \mu_{r}} +\Big([\Delta_{W}, P_{V}]  -[\Delta_{V}, P_{W}]\Big) \Psi_{\mu_{1}... \mu_{r}}+[P_{W}, P_{V}] \Psi_{\mu_{1}... \mu_{r}}.
 \end{align}
Let us now calculate each of the three parts in this equation. (Recall that we denote the Lie bracket between two vectors as $[W,V]^{\mu}= \mathcal{L}_{W}V^{\mu}$.)
 
\noindent \textbf{Calculating} $\bm{[\Delta_{W}, \Delta_{V}]\Psi_{\mu_{1}... \mu_{r}}}.$~Using Eqs.~(\ref{V=nabla phi}) and (\ref{properties of phi 1}), we find (after a long calculation):
\begin{align}
   [\Delta_{W}, \Delta_{V}]\Psi_{\mu_{1}... \mu_{r}}&=~\mathbb{L}_{[W,V]}\Psi_{\mu_{1}... \mu_{r}}-2ir\left(  \nabla_{(\mu_{1}} +\frac{i}{2} \gamma_{(\mu_{1}}  \right)\gamma^{\lambda} \Psi^{\rho}_{\hspace{1mm} \mu_{2}...\mu_{r})}\, \nabla_{\lambda}[W,V]_{\rho} \nonumber \\
   &-2ir\, \nabla_{\lambda}[W,V]_{\rho} \left(   \gamma_{(\mu_{1}} K^{\lambda \rho}_{\hspace{4mm}| \mu_{2}...\mu_{r})}+ \gamma^{\rho} K^{\hspace{4mm}\lambda}_{(\mu_{1} \hspace{3mm}| \mu_{2}...\mu_{r})} + \gamma^{\lambda} K^{\rho}_{\hspace{2mm}(\mu_{1}|  \mu_{2}...\mu_{r})}  \right),
\end{align}
where we have used that any Killing vector $\xi$ (such as $[W,V]$) satisfies~\cite{gravitation}
\begin{align}
\nabla_{\mu_{1}}\nabla_{\lambda}\xi_{\rho}=R_{\rho \lambda \mu_{1}\sigma} \xi^{\sigma},
\end{align}
while we have also introduced the rank-$(r+1)$ tensor-spinor
\begin{align}\label{field-strength ishh}
   K_{\lambda \rho| \mu_{2}...\mu_{r}}=-   K_{ \rho \lambda| \mu_{2}...\mu_{r}}= K_{\lambda \rho| (\mu_{2}...\mu_{r})}=\left(  \nabla_{[\lambda}+\frac{ir}{2} \gamma_{[\lambda} \right)\Psi_{\rho] \mu_{2}...\mu_{r}},
\end{align}
which is anti-symmetric in its first two indices and symmetric in its last $r-1$ indices. (For $r=1$, this tensor-spinor coincides with the rank-2 anti-symmetric gauge-invariant field strength tensor-spinor $\left(  \nabla_{[\lambda}+\frac{i}{2} \gamma_{[\lambda} \right)\Psi_{\rho]}$, while for $r\geq 2$, $ K_{\lambda \rho| \mu_{2}...\mu_{r}}$ is not gauge-invariant.) Note that because of the field equations~(\ref{Dirac_eqn_fermion_dS_sm}) and~(\ref{TT_conditions_fermions_dS_sm}), the tensor-spinor~(\ref{field-strength ishh}) satisfies
\begin{align}\label{field-strength ishh gamma-trace}
   \gamma^{\lambda}K_{\lambda \rho| \mu_{2}...\mu_{r}}=0.
\end{align}
Now we will show that
\begin{align}\label{field-strength ishh Bianchi}
       \gamma_{\mu_{1}} K^{\lambda \rho}_{\hspace{4mm}| \mu_{2}...\mu_{r}}+ \gamma^{\rho} K^{\hspace{4mm}\lambda}_{\mu_{1} \hspace{3mm}| \mu_{2}...\mu_{r}} + \gamma^{\lambda} K^{\rho}_{\hspace{2mm}\mu_{1}|  \mu_{2}...\mu_{r}}  =0.
\end{align}
It is convenient to proceed by defining
\begin{align}\label{dual field-strength ishh defntn}
    ^{\star}K^{\alpha \beta}_{\hspace{4mm}| \mu_{2}...\mu_{r}}\equiv \frac{1}{2} \epsilon^{\alpha \beta \lambda \rho}K_{\lambda \rho| \mu_{2}...\mu_{r}},
\end{align}
which satisfies
\begin{align}\label{dual field-strength ishh Bianchi}
       \gamma_{\mu_{1}} \,^{*}K^{\alpha \beta}_{\hspace{4mm}| \mu_{2}...\mu_{r}}+ \gamma^{\beta} \,^{*}K^{\hspace{4mm}\alpha}_{\mu_{1} \hspace{3mm}| \mu_{2}...\mu_{r}} + \gamma^{\alpha} \,^{*}K^{\beta}_{\hspace{2mm}\mu_{1}|  \mu_{2}...\mu_{r}}  =0
\end{align}
(this is easy to show by contracting with $\epsilon_{\gamma \delta \alpha \beta}$ and using well-known properties of the totally anti-symmetric tensor). Then, using $\epsilon^{\alpha \beta \lambda \rho} = i \gamma^{5} \gamma^{[\alpha}  \gamma^{\beta} \gamma^{\lambda} \gamma^{\rho]}$ [see Eq.~(\ref{def_gamma5})] and the gamma-tracelessness property~(\ref{field-strength ishh gamma-trace}), we find that Eq.~(\ref{dual field-strength ishh defntn}) becomes
\begin{align}
    ^{\star}K^{\alpha \beta}_{\hspace{4mm}| \mu_{2}...\mu_{r}}=-i\,\gamma^{5}K^{\alpha \beta}_{\hspace{4mm}| \mu_{2}...\mu_{r}}.
\end{align}
Substituting this into Eq.~(\ref{dual field-strength ishh Bianchi}), we immediately derive Eq.~(\ref{field-strength ishh Bianchi}), and thus, we have
\begin{align}\label{[Delta_W, Delta_V]}
   [\Delta_{W}, \Delta_{V}]\Psi_{\mu_{1}... \mu_{r}}&=~\mathbb{L}_{[W,V]}\Psi_{\mu_{1}... \mu_{r}}-2ir\left(  \nabla_{(\mu_{1}} +\frac{i}{2} \gamma_{(\mu_{1}}  \right)\gamma^{\lambda} \Psi^{\rho}_{\hspace{1mm} \mu_{2}...\mu_{r})}\, \nabla_{\lambda}[W,V]_{\rho} .
\end{align}
%%%%%%%%%%%%%%%%%%%%%%%%%%%%%%%%%%%%%%%%%%%%%%%%%%%%%%%%%%%%%%%

\noindent \textbf{Calculating} $\bm{\Big([\Delta_{W}, P_{V}]  -[\Delta_{V}, P_{W}]\Big) \Psi_{\mu_{1}... \mu_{r}}}.$ We find
\begin{align}\label{[Delta_W, P_V]-()}
    \Big([\Delta_{W}, P_{V}]&  -[\Delta_{V}, P_{W}]\Big) \Psi_{\mu_{1}... \mu_{r}}=-\frac{2r}{2r+1}\nonumber\\&\times \left(\nabla_{(\mu_{1}}+\frac{i}{2} \gamma_{(\mu_{1}}   \right)\left[ 2 [W,V]^{\rho}   \Psi_{\mu_{2}...\mu_{r})\rho} -i(2r+1)\,  \nabla_{\lambda}   [W,V]_{\rho}\,   \gamma^{\lambda}   \Psi^{\rho}_{\mu_{2}...\mu_{r})}\right].
\end{align}
%%%%%%%%%%%%%%%%%%%%%%%%%%%%%%%%%%%%%%%%%%%%%%%%%%%%%%%%%%
\noindent \textbf{Calculating} $\bm{[P_{W}, P_{V}]  \Psi_{\mu_{1}... \mu_{r}}}.$ We find
\begin{align}\label{[P_W, P_V]}
    [P_{W}, P_{V}]& \Psi_{\mu_{1}... \mu_{r}}=\frac{4r}{(2r+1)^{2}}\nonumber\\&\times \left(\nabla_{(\mu_{1}}+\frac{i}{2} \gamma_{(\mu_{1}}   \right)\left[ r\,[W,V]^{\rho}   \Psi_{\mu_{2}...\mu_{r})\rho} +\,  \nabla_{\rho}   [W,V]_{\lambda}\,   (\nabla^{\rho}-\frac{i}{2}\gamma^{\rho})  \Psi^{\lambda}_{\mu_{2}...\mu_{r})}\right].
\end{align}

\noindent \textbf{Finally,} adding Eqs.~(\ref{[Delta_W, Delta_V]}), (\ref{[Delta_W, P_V]-()}) and (\ref{[P_W, P_V]}) by parts, we arrive at Eq.~(\ref{[hidden, hidden]}), as required.

% The \nocite command causes all entries in a bibliography to be printed out
% whether or not they are actually referenced in the text. This is appropriate
% for the sample file to show the different styles of references, but authors
% most likely will not want to use it.
\nocite{*}

%\bibliography{apssamp}% Produces the bibliography via BibTeX.
%%%%%%%%%%%%%%%%%%%%%%%%%%%%%%%%%%%%%%%%%%%%%%%%%%%%%%%%%%%%%%%%%%%%%%%%%%%%%%%%%%%%%%%%%%%%%%%%%%%%%%%%%%%%%%%%%%%%%%%%%%%%%%%%%%%%%%%%%%%%%%%5555555555555555555555555555%%%
%apsrev4-2.bst 2019-01-14 (MD) hand-edited version of apsrev4-1.bst
%Control: key (0)
%Control: author (8) initials jnrlst
%Control: editor formatted (1) identically to author
%Control: production of article title (0) allowed
%Control: page (0) single
%Control: year (1) truncated
%Control: production of eprint (0) enabled
\providecommand{\noopsort}[1]{}\providecommand{\singleletter}[1]{#1}%

\end{document}